\documentclass{cernyrep}
\usepackage{graphicx}
\usepackage{pstricks, pstricks-add, multido,pst-coil,pst-solides3d,pst-grad,pst-optic}
\usepackage{slantsc}
\usepackage{psfrag}
\usepackage{pst-text}
\usepackage{float}
\usepackage{subfig}
\allowdisplaybreaks[0]
\usepackage[colorlinks=true, linkcolor=black, citecolor=black, urlcolor=blue]{hyperref}
\usepackage{fancyhdr}
\fancyhfoffset{4 mm}
\fancypagestyle{ARTTITLE}{%
\fancyhf{} 
\lhead{\small{Proceedings of the 2018 CERN--Accelerator--School course on \it{Beam Instrumentation}, Tuusula, (Finland)}}
\lfoot{Available online at \url{https://cas.web.cern.ch/previous-schools}}
\rfoot{\thepage\hspace*{3mm}}

}

\graphicspath{{./fig/}}

\def\refFig{Fig.~\ref}
\def\refTbl{Table~\ref}
\def\refEqu{Eq.~\eqref}
\def\Equation#1#2{\begin{equation}\label{#1}{#2}\end{equation}}
\def\Figure#1#2#3#4#5{\begin{figure}#5\centering\includegraphics[scale=#1]{#2}\caption{\label{#3}#4}\end{figure}}

\newcommand{\qed}{\nobreak \ifvmode \relax \else
      \ifdim\lastskip<1.5em \hskip-\lastskip
      \hskip1.5em plus0em minus0.5em \fi \nobreak
      \vrule height0.75em width0.5em depth0.25em\fi}
\newcommand{\mywidth}{0.7\linewidth}
\usepackage{varwidth}
\usepackage{xcolor}

\DeclareCaptionFormat{captionformat}{\fontsize{10}{10}\selectfont#1#2#3}
\begin{document}
\title{RF Measurement Techniques}

\author{M.~Wendt}

\institute{CERN, Geneva, Switzerland}

\begin{abstract}
For the characterization of components, systems and signals in the range of 
microwave and radio-frequencies (RF) specific equipment and dedicated 
measurement instruments are used. 
In this article the fundamentals of RF signal processing and measurement techniques are discussed.
It gives complementary background information for the introduction to \emph{RF Measurement Techniques} 
and the \emph{Practical RF Course}, which are part of the
Advanced Accelerator Physics training program of the CERN Accelerator School (CAS)
and have also been presented at the CAS 2018 Special Topic Course in Beam Instrumentation.

\end{abstract}

\maketitle 
\thispagestyle{ARTTITLE}
\section{A note to the history of RF signal receiving and measurement techniques}
In the early days of radio-frequency (RF) engineering the available instrumentation 
for measurements
was rather limited. 
Besides elements acting on the heat developed by RF power (bi-metal contacts and
resistors with a very high temperature coefficient) only point/contact diodes, 
and to some extent vacuum tubes, were available as signal detectors. 
For several decades the slotted measurement line, see Section~\ref{VSWRsect}, 
was the only commonly used instrument to measure impedances 
and complex reflection coefficients. 
Around 1960 the tedious work with these coaxial and waveguide measurement lines became
considerably simplified with the availability of the vector network analyzer. At the same time the first
sampling oscilloscopes with 1~GHz bandwidth arrived on the market. This was possible due to progress in
solid-state (semiconductor) technology and advances in microwave elements (microstrip lines). Reliable,
stable and easily controllable microwave sources are the backbone of spectrum and network analyzers, as
well as sensitive (low-noise) receivers. The following sections focus on signal receiving devices
such as spectrum analyzers. An overview of network analysis is given later in Section~\ref{NetAnaSect}.

\section{Basic definitions, elements and concepts}
Before discussing key RF measurement devices, a brief overview of the most important components used in these devices and the related basic concepts are presented.
\subsection{Decibel}
Since the unit decibel (dB) is frequently used in RF engineering, a short introduction and definition of the terms are given. The decibel is a unit used to express relative differences between quantities, e.g.\ of signal power. It is expressed as the base-10 logarithm of the ratio of the powers between two signals:
\begin{equation}
 P\text{ [dB]} = 10 \cdot \text{log}(P/P_{0}).
\label{dbpower}
\end{equation}
It is also common to express the signal amplitude in dB. Since power is proportional to the square of the signal amplitude, a voltage ratio in dB is expressed as:
\begin{equation}
V\text{ [dB]} = 20 \cdot \text{log}(V/V_{0}).
\label{dbvoltage}
\end{equation}
In Eqs. (\ref{dbpower}) and (\ref{dbvoltage}), $P_{0}$ and $V_{0}$ are the reference power and voltage,
respectively. A given value in dB is the same for power ratios as for voltage ratios. It is important to
note that there are no `power dB' or `voltage dB' as dB values always express a ratio.
Conversely, the absolute power and voltage can be obtained from dB values by
\begin{eqnarray}
P = P_{0} \cdot 10^{\frac{P\text{ [dB]}}{10}}, \\
V = V_{0} \cdot 10^{\frac{V\text{ [dB]}}{20}}.
\label{conversion}
\end{eqnarray}
The advantage using a logarithmic scale as unit of the measurement is twofold:
\begin{itemize}
	\item [i)] typical RF signal powers tends to span several orders of magnitude; and
	\item [ii)] signal attenuation losses and gains can simply computed by subtraction and addition.
\end{itemize}
Table \ref{dB} helps to familiarize with signal ratios and the associated dB values.
\begin{table}[t]
\caption{Overview of common dB values and their conversion into power and voltage ratios}
\begin{center}
\begin{tabular}{ccc}
\hline
\hline
 & \bfseries Power ratio & \bfseries Voltage ratio \\
\hline
$-$20 dB & 0.01 & 0.1 \\
$-$10 dB & 0.1 & 0.32 \\
$-$6 dB & 0.25 & 0.5 \\
$-$3 dB & 0.50 & 0.71 \\
$-$1 dB & 0.74 & 0.89 \\
0 dB & 1 & 1 \\
1 dB & 1.26 & 1.12 \\
3 dB & 2.00 & 1.41 \\
6 dB & 4 & 2 \\
10 dB & 10 & 3.16 \\
20 dB & 100 & 10 \\
$n \cdot 10$ dB & 10$^{n}$ & 10$^{n\text{/2}}$ \\
\hline
\hline
\end{tabular}
\end{center}
\label{dB}
\end{table}

Absolute levels are expressed using a specific reference value, these dB systems are not based on SI units. Strictly speaking, the reference value should be included in parentheses when giving a dB value, e.g. +3 dB (1 W) indicates 3 dB at $P_{0} = 1$ W, thus 2 W. However, it is more common to add some typical reference values as letters after the unit, e.g.\ dBm defines dB using a reference level of $P_{0} = 1$ mW.
Thus, 0 dBm correspond to $-$30 dBW, where dBW indicates a reference level of $P_{0} = 1$ W. Often a reference impedance of 50 $\Omega$ is assumed.
Other common units are:
\begin{itemize}
	\item  [i)] dBmV for small voltages with $V_{0}$ = 1~mV; and
	\item  [ii)] dBmV/m for the electric field strength radiated from an antenna with reference field strength $E_{0} = 1$ mV/m.
\end{itemize}
\begin{figure}[h]%
\centering
 \includegraphics[width=\mywidth]{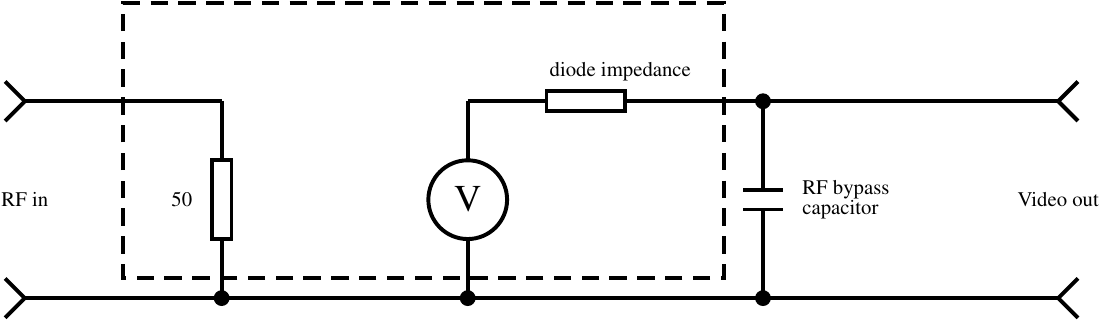}
\caption{Simplified equivalent circuit of a diode detector (w/o parasitic elements)}%
\label{circdiode}%
\end{figure}

\subsection{The RF diode}
\label{RFdiodesec}
\begin{figure}[t]%
\centering
\includegraphics[width=50mm]{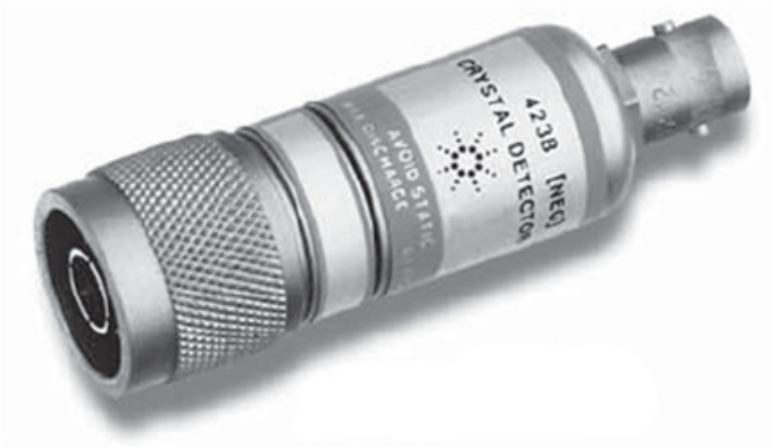}%
\caption{A typical \textit{Schottky} diode. The RF input of this detector diode is on the left and the video output on the right (courtesy \textit{Agilent}).}%
\label{diode1}%
\end{figure}

\begin{figure}[t]%
\centering
\includegraphics[width=70mm]{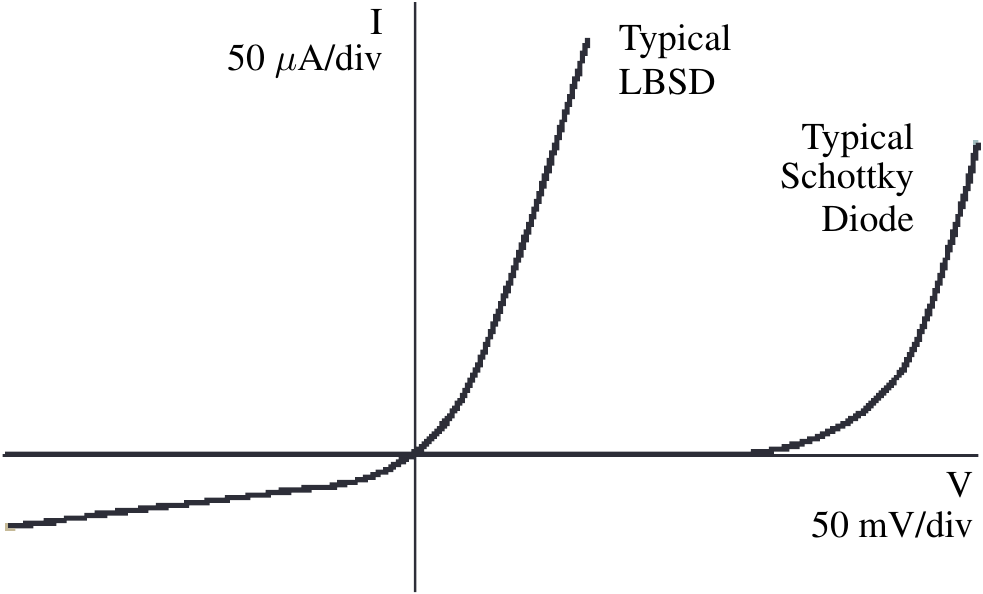}%
\caption{Current as a function of voltage for different diode types (LBSD = low barrier \textit{Schottky} diode)}%
\label{kenn}%
\end{figure}

\begin{figure}[t]%
\centering
\includegraphics[width=60mm]{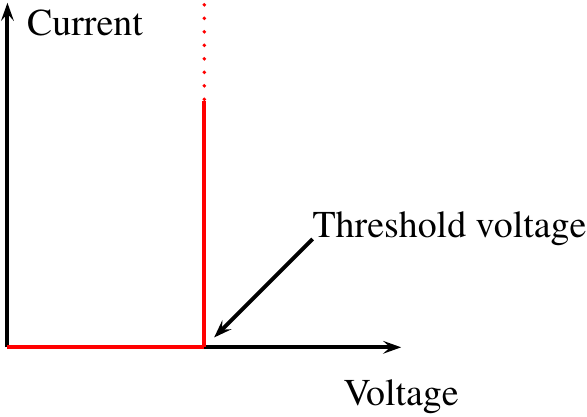}%
\caption{The current--voltage relation of an ideal commutator with threshold voltage}%
\label{comm}%
\end{figure}

One of the most important elements, even today inside the most sophisticated RF measurement devices is the fast RF diode or \textit{Schottky} diode. The basic metal--semiconductor junction has an intrinsically very fast switching time of well below a picosecond, provided that the geometric size and hence the junction capacitance of the diode has sufficiently small dimensions. However, the unavoidable, and voltage-dependent junction capacity will lead to limitations of the maximum operating frequency. 
The simplified equivalent circuit of such a diode is depicted in Fig. \ref{circdiode} and
an example of a commonly used \textit{Schottky} diode is shown in Fig. \ref{diode1}.
 of the most important properties of any diode is its IV-characteristic, 
which is the relation of the current passing the diode as a function of the applied voltage~\cite{vendelin}.
This relation is depicted graphically for two different types of diodes in Fig. \ref{kenn}.
It shows, the diode is a non-ideal commutator (in contrary to that shown in Fig. \ref{comm}) for small signals. Note that it is not possible to apply large signals, since this kind of diode would burn out.
Although there exist versions with rather large power handling capability of \textit{Schottky} diodes, these can stand more than 9\,kV and several tens of amperes, they are not suitable in microwave applications due to their large junction capacity.
\begin{figure}[t]%
\centering
 \includegraphics[width=80mm]{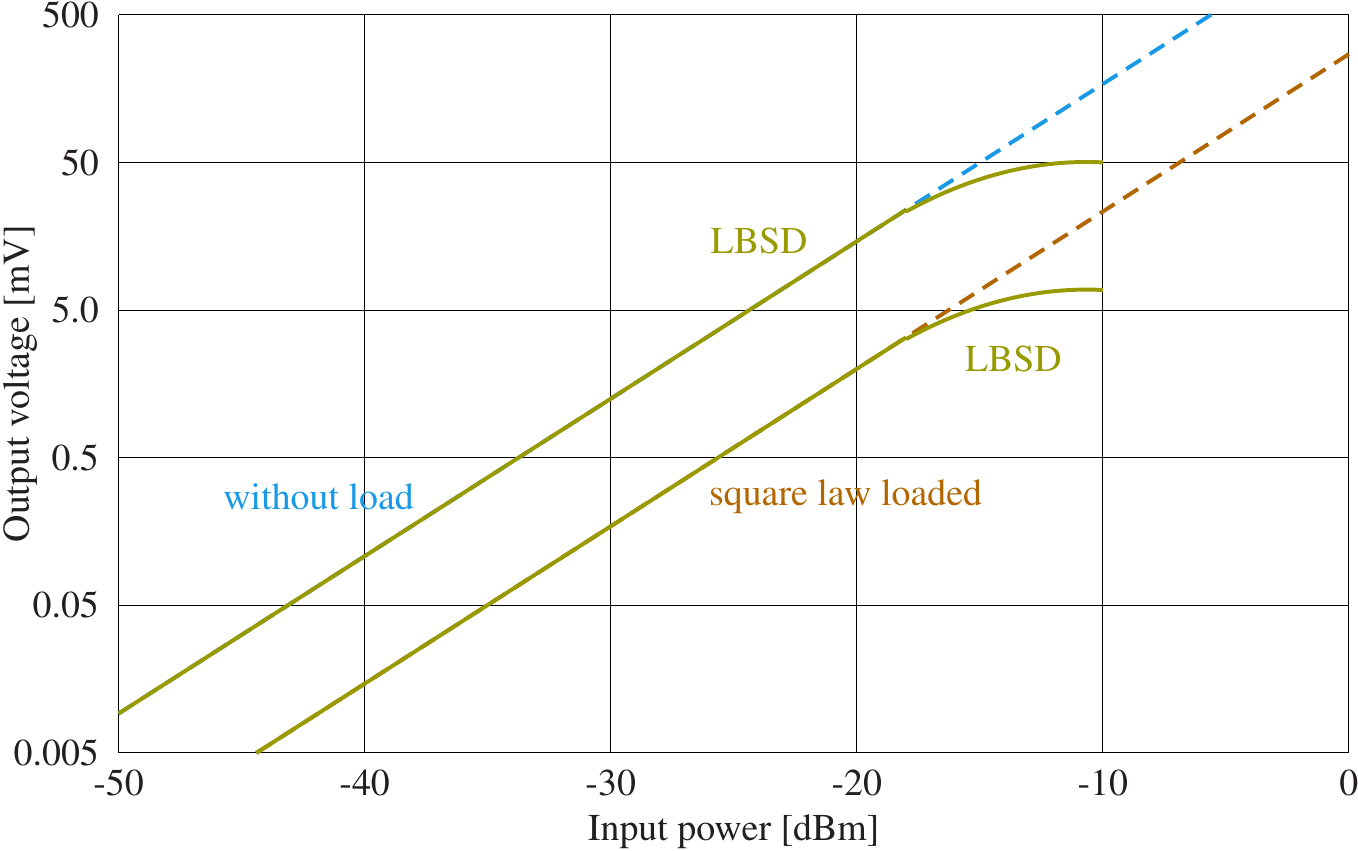}%
\caption{Relation between input power and output voltage}%
\label{squarelaw}%
\end{figure}
%
The region where the output voltage is proportional to the input power is called the square-law region (Fig. \ref{squarelaw}).
In this region the input power is proportional to the square of the input voltage and the output signal is proportional to the input power, hence the name square-law region.

The transition between the linear region and the square-law region is typically 
between $-$10 and $-$20 dBm (Fig. \ref{squarelaw}). 
For a more detailed description, see~\cite{src:oxford}.

There are some fundamental limitations when using diodes as detectors. The output signal of a diode (essentially DC or modulated DC if the RF is amplitude modulated) does not contain any phase information. In addition, the sensitivity of a diode limits the input level range to about $-$60 dB at best, which is not sufficient for many applications.

The minimum detectable power level of a RF diode is specified by the `tangential sensitivity', which typically amounts to $-$50 to $-$55 dBm for 10 MHz video bandwidth at the detector output \cite{thumm}.

To overcome these limitations, a more sophisticated method to utilize the RF diode is required. This method is presented in the next section.

\subsection{Mixer}
\label{Mixersec}
\begin{figure}[t]%
\centering
\includegraphics[width=80mm]{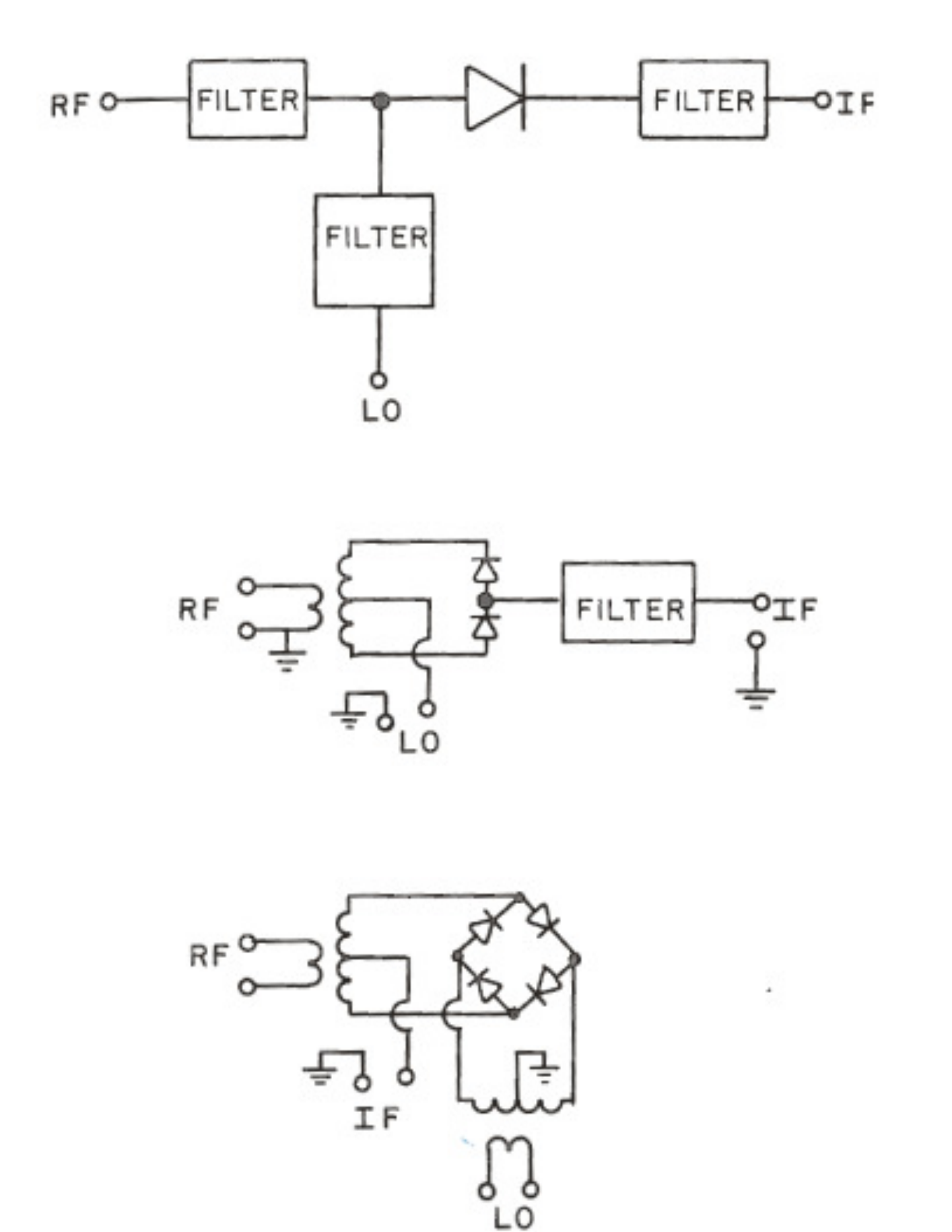}%
\caption{Examples of different mixer configurations}%
\label{mixers}%
\end{figure}

\begin{figure}[tb]%
\centering
\includegraphics[width=100mm]{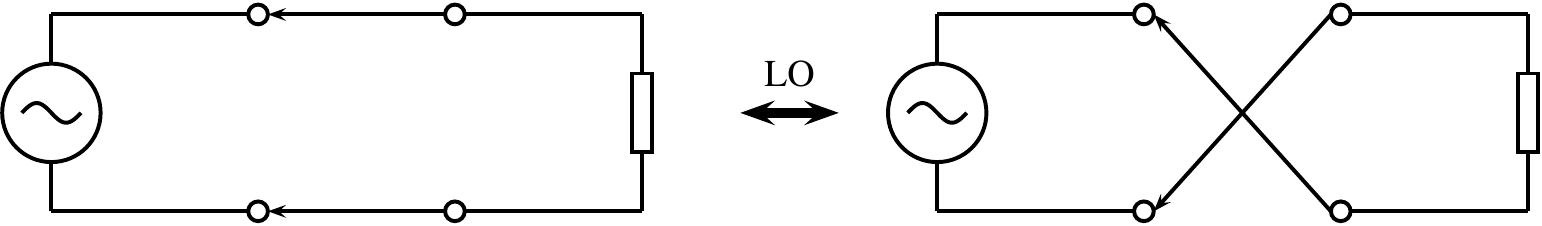}%
\caption{Two circuit configurations interchanging with the frequency of the LO where the switches represent the diodes.}%
\label{mixprinc}%
%
\centering
\includegraphics[width=\mywidth]{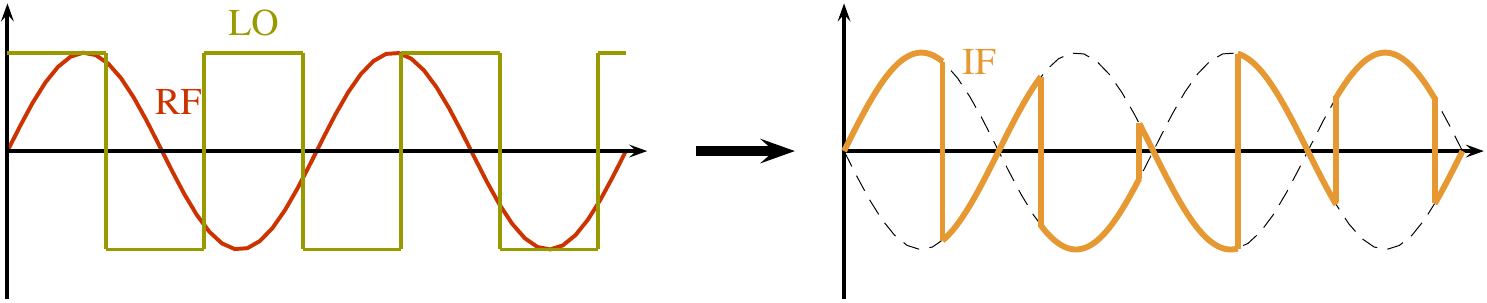}%
\caption{Time-
domain response of a mixer}%
\label{mixerresponse}%
\end{figure}

To include the detection of very small RF signals a device with a linear response over a wide range of signal levels (from 0 dBm (= 1 mW) down to the thermal noise = $-$174 dBm/Hz = 4$\cdot 10^{-21}$ W/Hz) is highly preferred. A RF mixer provides these features by using one, two or four diodes in different configurations (Fig. \ref{mixers}). 
A mixer is essentially a frequency multiplier with a very high dynamic range, 
implementing in it's simplest form the function
\begin{equation}
f_{1}(t) \cdot f_{2}(t) \text{\hspace{0.2cm}with } f_{1}(t) = \text{RF signal\hspace{0.2cm} and } f_{2}(t) = \text{ local oscillator (LO) signal}\hspace{0.2cm}
\label{mixer1}
\end{equation}
or more explicitly, for two sinusoidal signals 
with amplitudes $a_{i}$ and frequencies $f_{i}$ ($i = 1, 2$),
\begin{equation}
a_{1}\cos(2\pi f_{1}t + \varphi) \cdot a_{2} \cos (2\pi f_{2}t) = \frac{1}{2}a_{1}a_{2}\left[ \cos ((f_{1} + f_{2})t + \varphi)
+  \cos ((f_{1} - f_{2})t + \varphi)\right].
\label{mixer2}
\end{equation}
Thus, we obtain a response at the intermediate-frequency (IF) port as sum and difference frequencies of the local oscillator ($\textrm{LO} = f_{1}$) and RF ($ = f_{2}$) signals.
Examples of different mixer configurations are shown in Fig. \ref{mixers},
they all use diodes to multiply the two applied signals, RF and LO. 
These diodes operate like a switch, controlled by the frequency of the LO signal (Fig. \ref{mixprinc}).
The response of a mixer in the time 
domain is depicted in Fig. \ref{mixerresponse}.
The output signal is always in the ``linear regime'', provided that the mixer is not saturated with respect to the RF input signal. Note, with respect to the LO signal the mixer has to be always in saturation to insure the diodes operate almost as an ideal switch. 
The phase of the RF signal is conserved in the output signal available at the IF output.
\subsection{Amplifier}
\begin{figure}[H]%
\centering
\includegraphics[width=\mywidth]{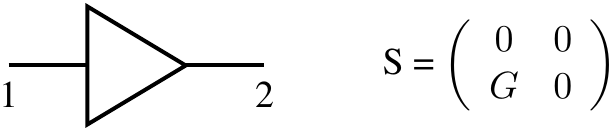}%
\caption{Circuit symbol and S-matrix of an ideal amplifier}%
\label{ampli}%
\end{figure}

\begin{figure}[H]%
\centering
\includegraphics[width=65mm]{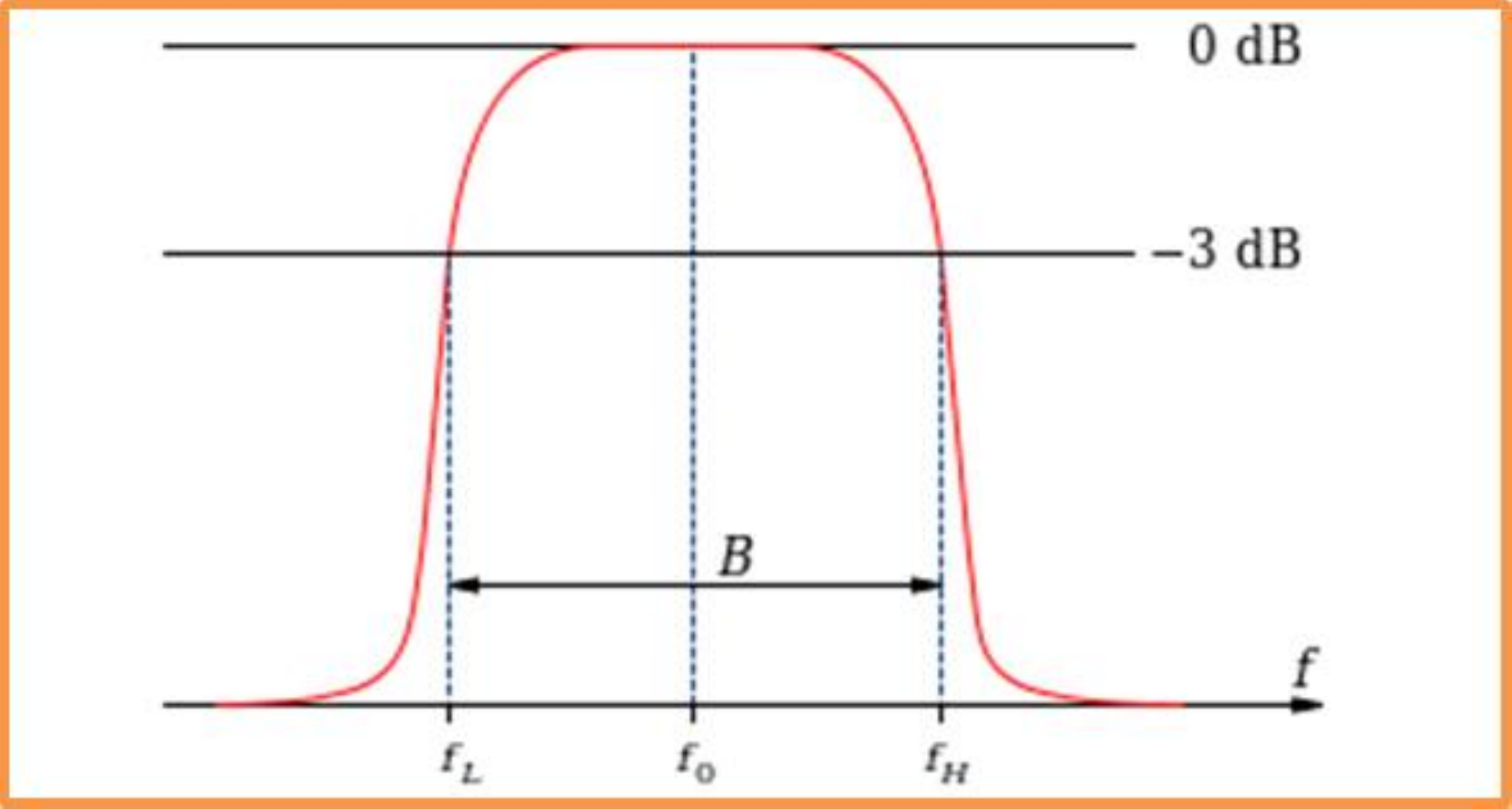}%
\caption{Definition of the bandwidth}%
\label{3dbbandwidth}%
\end{figure}

A linear amplifier, sometimes called ``gain stage'', 
auguments the input signal by a factor which is usually indicated in decibels (dB).
The ratio between the output and the input signals is called the transfer function and its magnitude -- the voltage gain $G$ -- is measured in dB and given as
\begin{equation}
G [\text{dB}] = 20 \cdot \frac{{V}_{\text{RFout}}}{{V}_{\text{RFin}}} \text{\hspace{0.2cm}or\hspace{0.2cm} } \frac{{V}_{\text{RFout}}}{{V}_{\text{RFin}}} = 20 \cdot \text{log} G [\text{lin}].
\label{gain}
\end{equation}
The circuit symbol of an amplifier is shown in Fig. \ref{ampli} together with its S-matrix.
%
The bandwidth of an amplifier specifies the frequency range where it is usually operated, 
see Fig.~\ref{3dbbandwidth}. 
This frequency range is defined by the $-$3~dB points\footnote{%
The $-$3~dB points are the values left and right of a reference value, 
typically the local maximum of the amplifier transfer function, and are 3 dB below that reference.} 
of the magnitude response with respect to its maximum or nominal transmission gain,
dividing the magnitude transfer function of the amplifier into a pass-band 
and a stop-band of equal transmitted power.

\begin{figure}[H]%
\centering
\includegraphics[width=100mm]{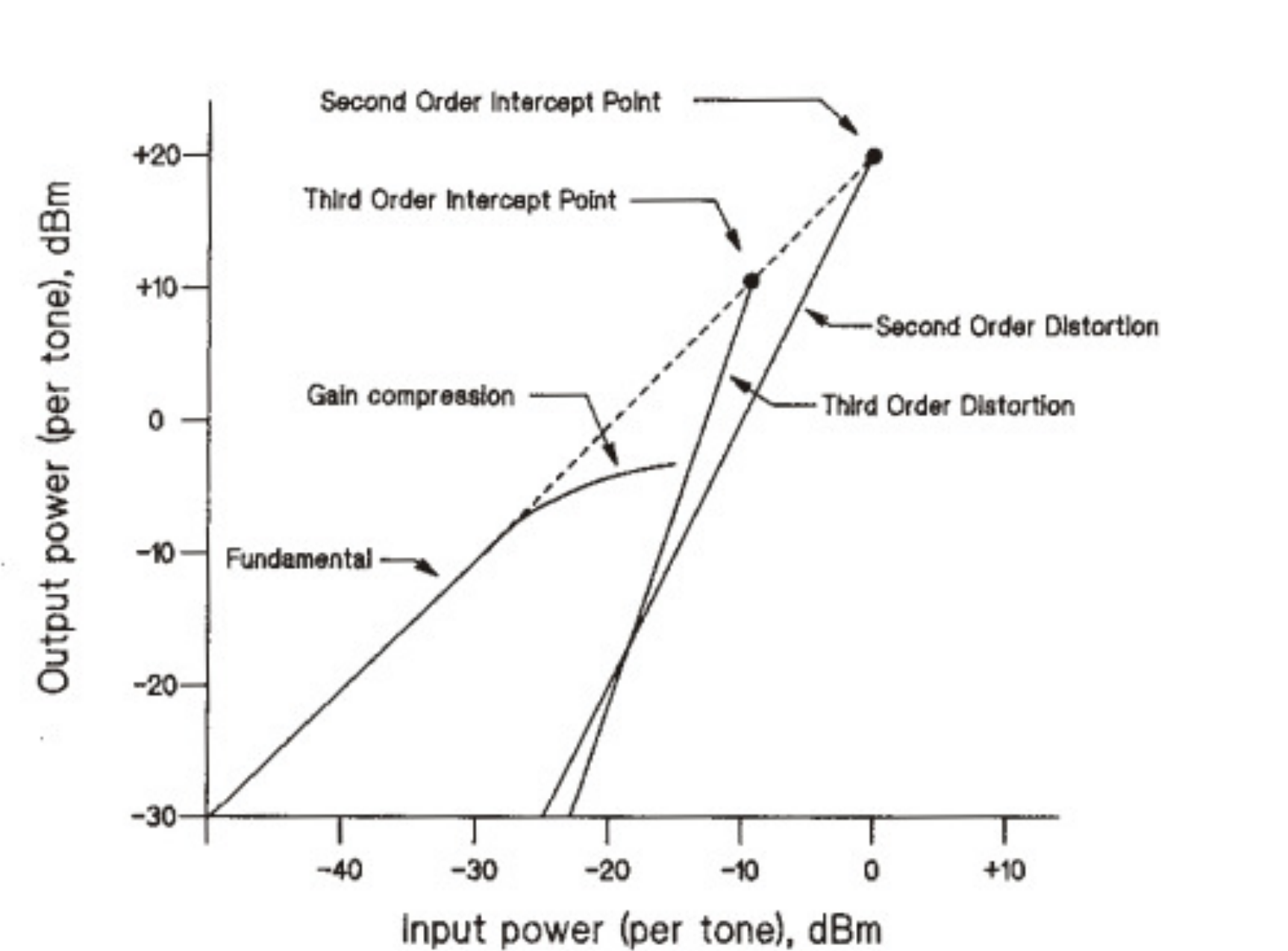}%
\caption{Example for the 1 dB compression point \cite{witte}}%
\label{1dB}
\end{figure}

For an ideal amplifier the output signal would always be proportional to the input signal. 
However, a real amplifier is non-linear, typically for larger signals the transfer characteristic 
deviates from its linear properties, which is validated for small-signal amplification. 
When increasing the output power of an amplifier, 
a point is reached where due to the non-linearities the small-signal gain 
is reduced by 1~dB (Fig. \ref{1dB}).
This output power level defines the so-called 1~dB compression point, which is an important measure of the output power capability, thus the dynamic range for the amplifier.

The transfer characteristic of an amplifier can be described in commonly used terms of RF engineering, i.e.\ the S-matrix, see Section~\ref{NetAnaSect}.
As implicitly contained in the S-matrix, both, amplitude and phase information of any spectral component are preserved when passing through an ideal amplifier. For a real amplifier the element $G = {S}_{21}$ (transmission from port 1 to port 2) is not a constant, but a complex function of frequency. Also the elements $S_{11}$ and $S_{22}$ are not zero.

\subsection{Interception points of non-linear devices}
Important characteristics of non-linear devices are the interception points. Here, only a brief overview is given, further information can be found in \cite{witte}.

\begin{figure}[t]%
\centering
\includegraphics[width=120mm]{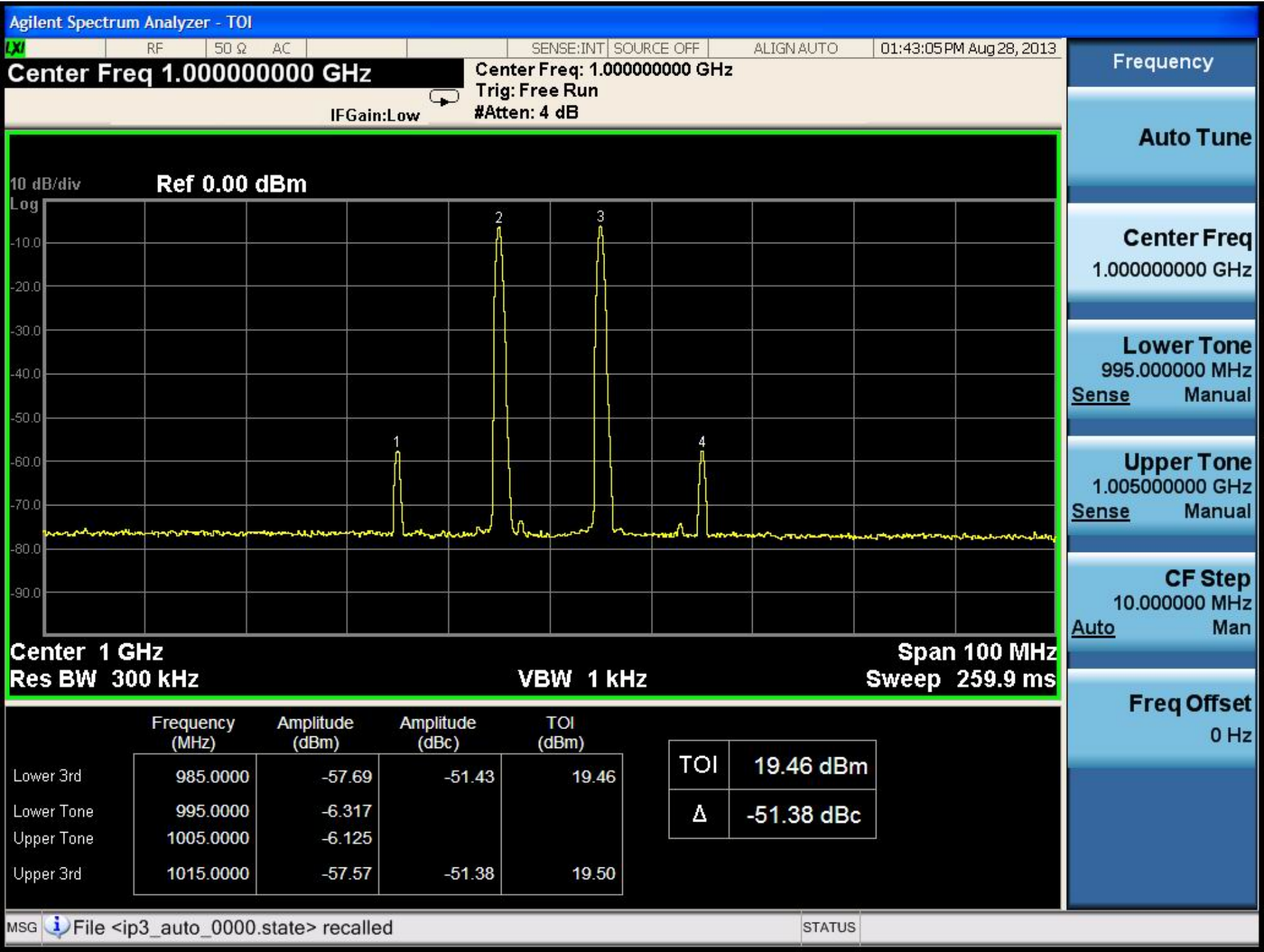}%
\caption{An example of automatized TOI measurement}%
\label{ip3}%
\end{figure}

The most relevant interception points is the interception point of third order (IP3 point). 
Its importance derives from its straightforward determination, 
plotting the input versus the output power on a logarithmic scale (Fig. \ref{1dB}). 
The IP3 point is usually not measured directly, but is extrapolated from the data, 
measured at much lower power levels in order to avoid overload 
or damage of the device under test (DUT). 
Applying two signals $(f_{1},f_{2} > f_{1})$ of closely spaced 
frequencies $\Delta f$ 
simultaneously to the DUT, the intermodulation products appear 
at +$\Delta f$ above $f_{2}$ and $-$$\Delta f$ below $f_{1}$. 
This method is called the third-order intermodulation (TOI). 
An example of an automatized TOI measurement is shown in Fig.~\ref{ip3}.

The transfer function of weakly non-linear devices can be approximated by a \textit{Taylor} expansion. Using $n$ higher order terms and plotting them together with an ideal linear device on a logarithmic scale results in two straight lines with different slopes ($x^{n} \stackrel{\text{log}}{\rightarrow} n \cdot \text{log }x$). Their intersection point is the intercept point of $n$th order. These points provide important information concerning the quality of non-linear devices.

In this context, the aforementioned 1 dB compression point of an amplifier is the intercept point of first order. For the method of measurements of the 1 dB compression point, see Section~\ref{1dBsect}.

Similar characterization techniques can also be applied for mixers, which, with respect to the LO signal, cannot be considered as weakly non-linear devices.
\subsection{The superheterodyne concept}
The word superheterodyne is composed of three parts: super (Latin: over), $\epsilon \tau \epsilon \rho \omega$ (hetero, Greek: different) and $\delta \upsilon \nu \alpha \mu \iota \sigma$ (dynamic, Greek: force), and can be translated as two forces superimposed\footnote{The direct translation (roughly) would be: another force becomes superimposed.}. Different abbreviations exist for the superheterodyne concept. In the USA it is often abbreviated by the simple word ``heterodyne'', and in Germany the shorter terms ``super'' or ``superhet'' are used. 
\begin{figure}[bht]%
\centering
\includegraphics[width=120mm]{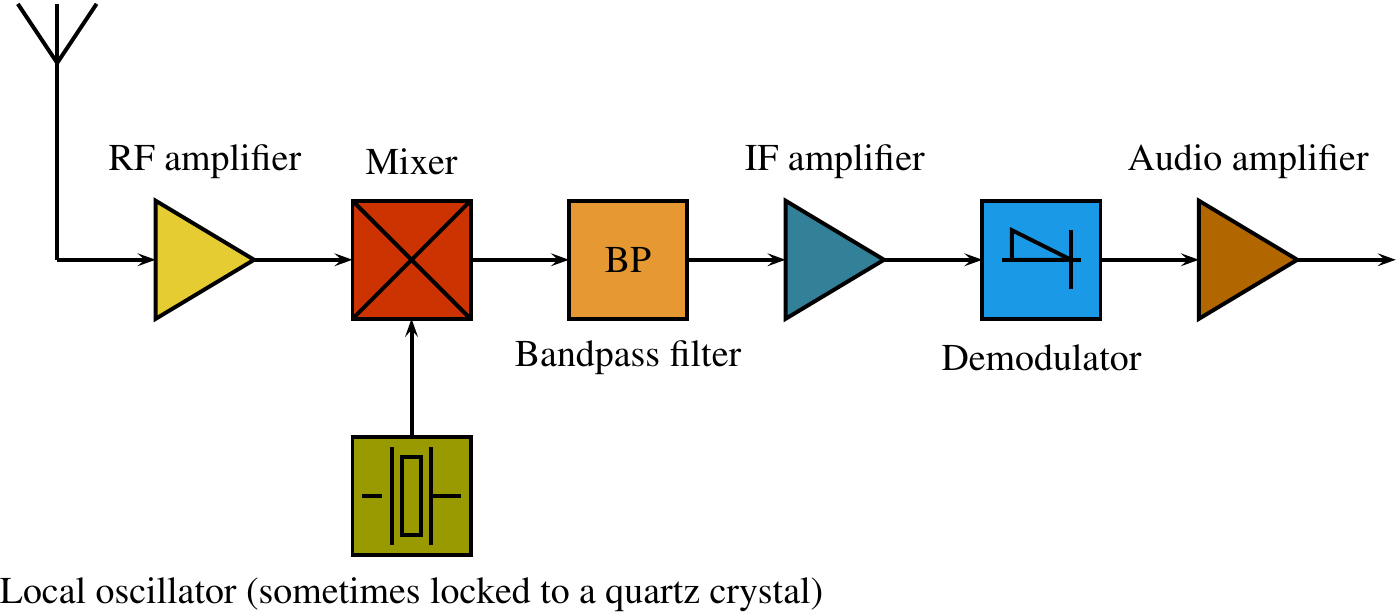}%
\caption{Schematic drawing of a superheterodyne radio receiver}%
\label{superhet}%
\end{figure}
%

A ``weak'' incident (RF) signal is subjected to non-linear superposition (i.e. mixing or multiplication) 
with a ``strong'` sine wave signal from a LO. 
At the mixer output sum and difference frequencies of the RF and LO signals appear. 
The LO signal can be tuned such that this IF output signal is always of same frequency, 
or stays within a very narrow frequency band. 
Therefore, a fixed-frequency bandpass with excellent transfer characteristics can be used, 
which is cheaper and easier to realize than a variable bandpass of the same performance. 
Also, gain-stages (amplifiers) operating at a lower IF frequency are of better quality and/or 
are more affordable. 
A well-known application of this principle is any simple radio receiver (Fig. \ref{superhet}).
\section{Spectrum analyser}
\label{SpecAnasec} 
RF spectrum analyzers can be found in virtually every control room of a modern particle accelerator. They are used for many aspects of beam diagnostics including Schottky signal acquisition and observation of RF signals. A spectrum analyzer is in principle very similar to a common superheterodyne broadcast receiver, except with respect to the choice of functions, change of parameters, and in general a more sophisticated, high quality design. It sweeps automatically through a specified frequency range, which corresponds to an automatic turning of the tuning knob on a radio. The signal is then displayed in the amplitude/frequency plane.
Originally, these kind of measurement instruments were setup manually 
and used a cathode ray tube (CRT) as display. 
Nowadays, with the availability of low-cost, powerful digital electronics for control and 
signal processing, basically every instrument can be remotely controlled. 
A microprocessor permits fast and reliable settings of the instrument, 
and an analog-digital-converter (ADC) in connection with digital signal processing hardware 
performs the acquisition and pre-processing of the measured signal values. 
The digital data processing enables extensive data treatment for error correction, 
complex calibration routines and self tests, which are a great improvement for 
RF signal measurements. 
However, the user of such sophisticated systems may not always be aware of the 
basic analogue signal path and processing, before the signals are digitized and 
prepared for user interaction. 
The basics of these analogue sections is discussed as follows.

In general, we distiguish two types of spectrum analyzers:
\begin{itemize}
	\item the scalar spectrum analyzer (SA) and
	\item the vector spectrum analyzer (VSA).
\end{itemize}
The SA provides only information of the amplitude of the applied signal, 
while the VSA provides information of the phase as well.
\subsection{Scalar spectrum analyzer}
A common oscilloscope displays a signal in the amplitude-vs.-time format (time domain). 
The SA follows a different approach and displays the RF signal in the frequency domain. 

%
\begin{figure}[hbt]%
\centering
\includegraphics[width=\mywidth]{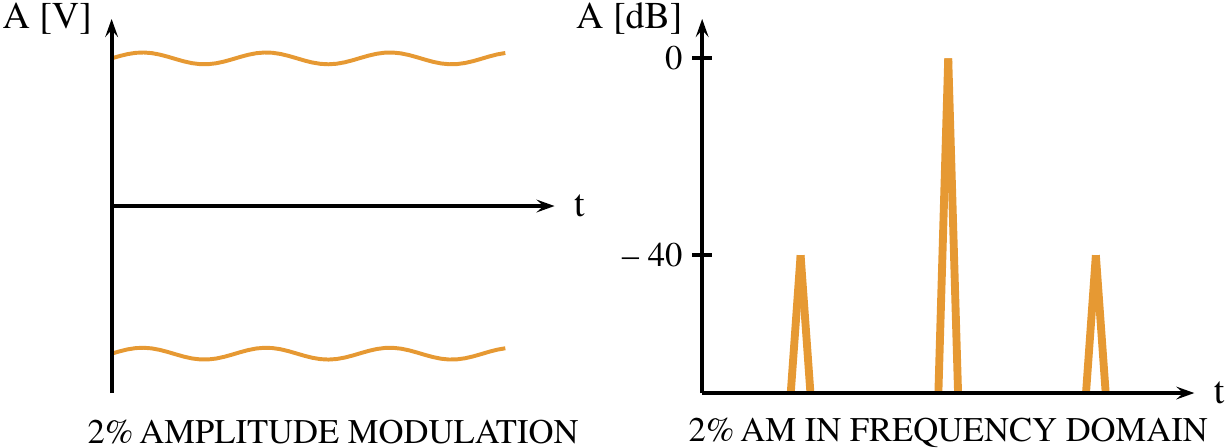}%
\caption{Example of amplitude modulation in time and frequency domains}%
\label{am}%
\end{figure}
%
One of the major advantages of the frequency-domain visualization lies in the 
higher sensitivity to perturbations of periodic signals. 
For example, a 2\% distortion of a sine-wave signal is already difficult to be observed 
on a the time domain display, but in the frequency domain on a logarithmic magnitude scale 
the related ``harmonics'' (Fig. \ref{am}) are clearly visible 
(here $-$40~dB below the main spectral line).
A very faint amplitude modulation (AM) of 10$^{-12}$ (power) on some sinusoidal signals 
would be completely invisible on a time domain trace, but can be displayed as two side 
harmonics 120 dB below the carrier in the frequency domain \cite{Schleifer}.

\begin{figure}[t]%
\centering
\includegraphics[width=\mywidth]{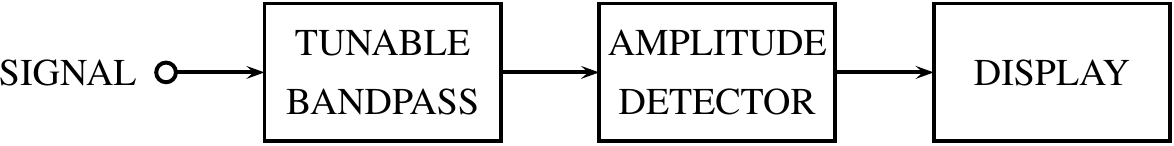}%
\caption{A tunable bandpass as a simple spectrum analyser (SA)}%
\label{bp}%
\end{figure}

\begin{figure}[t]%
\centering
\includegraphics[width=\mywidth]{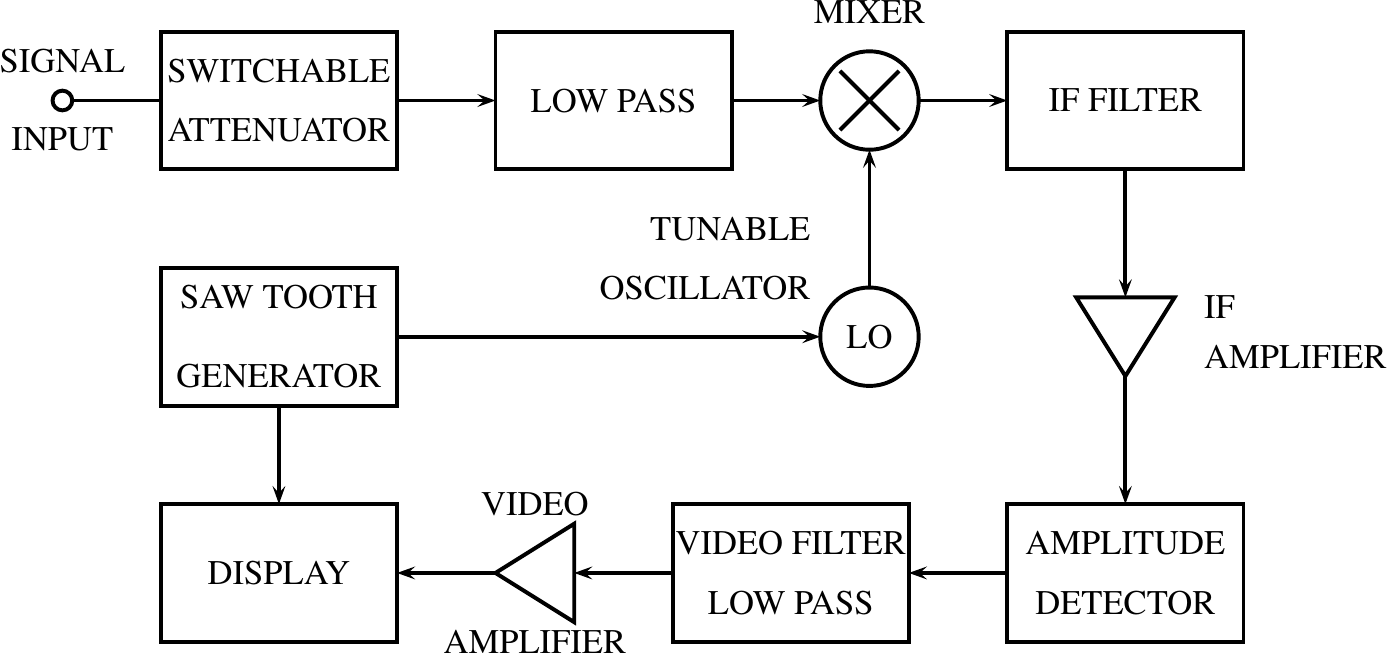}%
\caption{Block diagram of a spectrum analyzer}%
\label{sa}%
\end{figure}

In the following we consider only ``classical'' SAs, based on a swept tuned band-pass filter 
analysis (Fig. \ref{bp}), or utilizing the heterodyne receiver principle (Fig. \ref{sa}).

The simplest form of a swept frequency spectrum analyzer is based on a tunable bandpass. 
This may be a classical lumped element LC circuit or a YIG filter (YIG = yttrium iron garnet) 
for frequencies  $>$1~GHz. 
The LC filter exhibits poor tuning, stability and resolution. 
YIG filters are used in the microwave range (as preselectors) and for YIG oscillators. 
Their tuning range is about one decade, with $Q$ values exceeding 1000.

\begin{figure}[t]%
\centering
\includegraphics[width=0.4\linewidth]{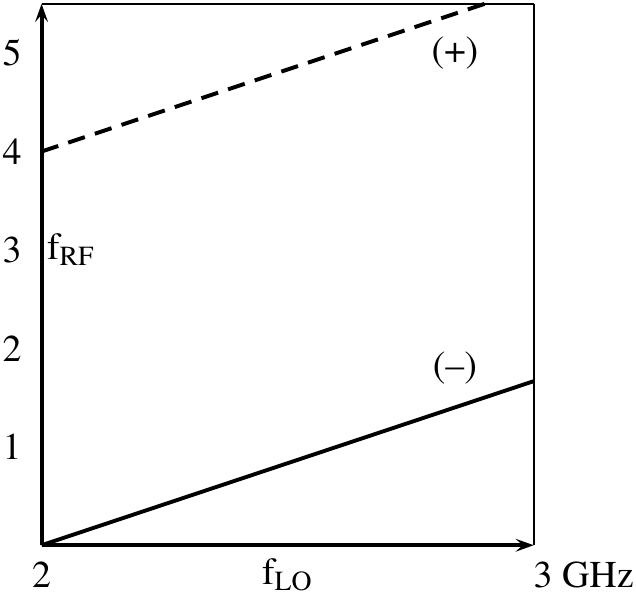}%
\caption{Frequency chart of the SA of Fig. \ref{sa}, $f_{\text{IF}}$ = 2~GHz}%
\label{fchart}%
\end{figure}

For superior performance, the superheterodyne principle is applied basically in all commercial spectrum analyzers (Fig. \ref{superhet}).
As already mentioned, the non-linear element (four-diode mixer or double-balanced mixer) delivers mixing products, like
\begin{equation}
f_{\text{signal}} = f_{\text{RF}} = f_{\text{LO}} \pm f_{\text{IF}}.
\label{eq:23a}
\end{equation}
Assuming an input frequency range $f_{\text{RF}}$ from 0 to 1~GHz for the spectrum analyzer shown in Fig. \ref{sa} and $f_{\text{LO}}$ ranging between 2 and 3~GHz, results in a frequency chart as shown in Fig. \ref{fchart}.

Obviously, for a wide range of input frequencies, while rejecting any image response, requires a sufficiently high IF. A similar situation occurs for AM- and FM-broadcast receivers (AM-IF = 455~kHz, FM-IF = 10.7~MHz). But, for a high IF (e.g. 2 GHz) a stable, narrowband IF filter is very challenging, therefore most SAs and high-quality
receivers use more than a single IF. Certain SAs have four different LOs, some fixed, some tunable. To perform a large tuning range, the first, and for fine tuning (e.g. 20~kHz range), the third LO are variable.

Multiple mixing stages may also be necessary when downconverting to a lower IF (required when using high-$Q$ quartz filters) to ensure a good image response suppression of the mixers.

It can be demonstrated that the frequency of the $n^{\text{th}}$ LO must be higher than the (say) 80~dB bandwidth (BW) of the $(n - 1)^{\text{th}}$ IF band-pass filter. A disadvantage of multiple mixing is the possible generation of intermodulation lines if amplitude levels in the conversion chain are not carefully controlled.

The requirements of a modern SA with respect to frequency generation and mixing are
\begin{itemize}
	\item high resolution,
	\item high stability (drift and phase noise),
	\item wide tuning range,
	\item no ambiguities
\end{itemize}
and, with respect to the amplitude response
\begin{itemize}
	\item large dynamic range ($>$$100$ dB),
	\item calibrated, stable amplitude response,
	\item low internal distortions.
\end{itemize}
It is important to notice that the bandwidth $\Delta f$ of the IF band-pass filter
is linked to sweep rate (or step width and rate when using a synthesizer):
\begin{equation}
\frac{\text{d}f}{\text{d}t} < (\Delta f)^{2}.
\label{eq:24a}
\end{equation}
In other words, the signal frequency has to remain stable within $\Delta T = 1/\Delta f$ for a given IF bandwidth $\Delta f$, which ensures steady-state conditions of the selected IF filter. 

On many instruments the proper relation between $\Delta f$ and the optimum sweep rate is selected automatically, but it can always be altered manually (setting of the resolution bandwidth).

Caution is advised when applying, but not necessarily displaying, two or more strong ($> 10\ \mbox{dBm}$) signals to the input. Third-order intermodulation products may appear (generated at the first mixer or amplifier) and could lead to misinterpretation of the signals to be analyzed.

Spectrum analyzers  usually have a rather poor noise figure of 20--40 dB, 
as they often do not use pre-amplifiers in front of the first mixer (dynamic range, linearity). 
But, with a good pre-amplifier, the noise figure can be reduced to almost that of the pre-amplifier. 
This configuration permits amplifier noise-figure measurements with a reasonable resolution 
of about 0.5 dB. 
The input of the amplifier to be tested is connected to the hot and cold terminations, 
and the two corresponding traces on the SA display 
are evaluated \cite{Schiek, Yip, Evans, Connor, Landstorfer}.

\subsection{Vector spectrum and fast Fourier transform analyzer}
\begin{figure}[t]%
\centering
\includegraphics[width=145mm]{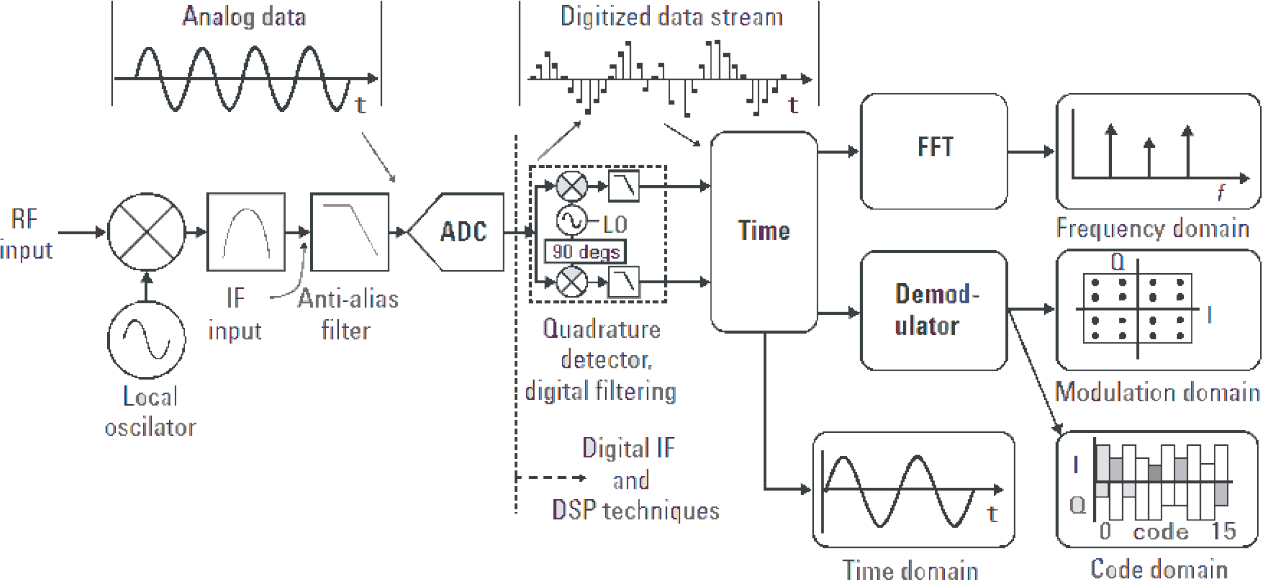}%
\caption{Block diagram of a vector spectrum analyser}%
\label{vsa}%
\end{figure}

The modern vector spectrum analyser (VSA) is essentially a combination of a two-channel 
digital oscilloscope  and a fast \textit{Fourier} transformation (FFT) based spectrum display. 
The incoming signal is down-converted, band-pass (BP) filtered, and passed to an analog-to-digital converter (ADC) (generalized Nyquist for BP signals; $f_{\text{sample}} = 2 \ \cdot $ BW).
Fig.~\ref{vsa} shows a typical, simplified schematic of a modern VSA.

The digitized signal is split into I (in-phase) and Q (quadrature, 90 degree offset) components with respect to the phase of some reference oscillator. Without this reference, the term ``vector'' would be meaningless for a spectral component.

One of the great advantages of a VSA, it easily allows to separate AM and FM components.

\begin{figure}[t]%
\centering
\includegraphics[width=\mywidth]{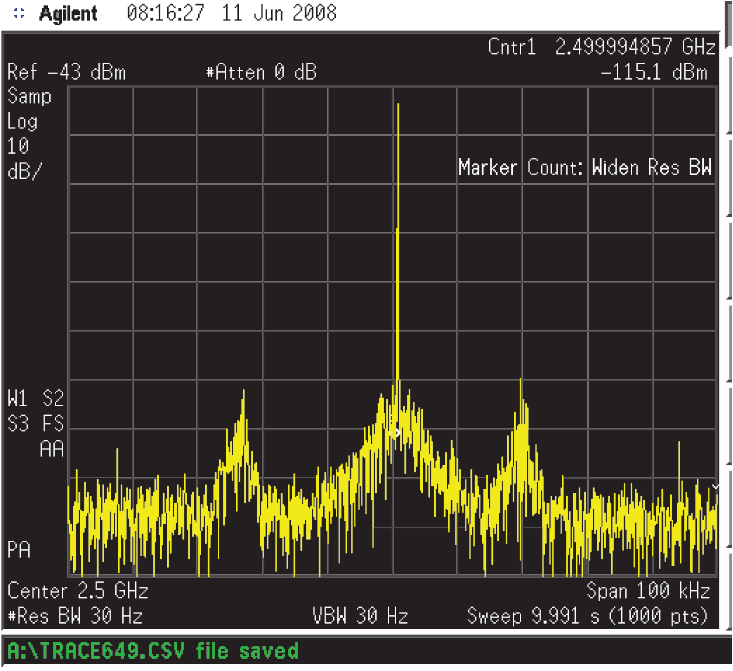}%
\caption{Single-sweep FFT display similar to a very slow scan on a swept spectrum analyser}%
\label{ec1}%
\end{figure}

\begin{figure}[tb]%
\centering
\includegraphics[width=\mywidth]{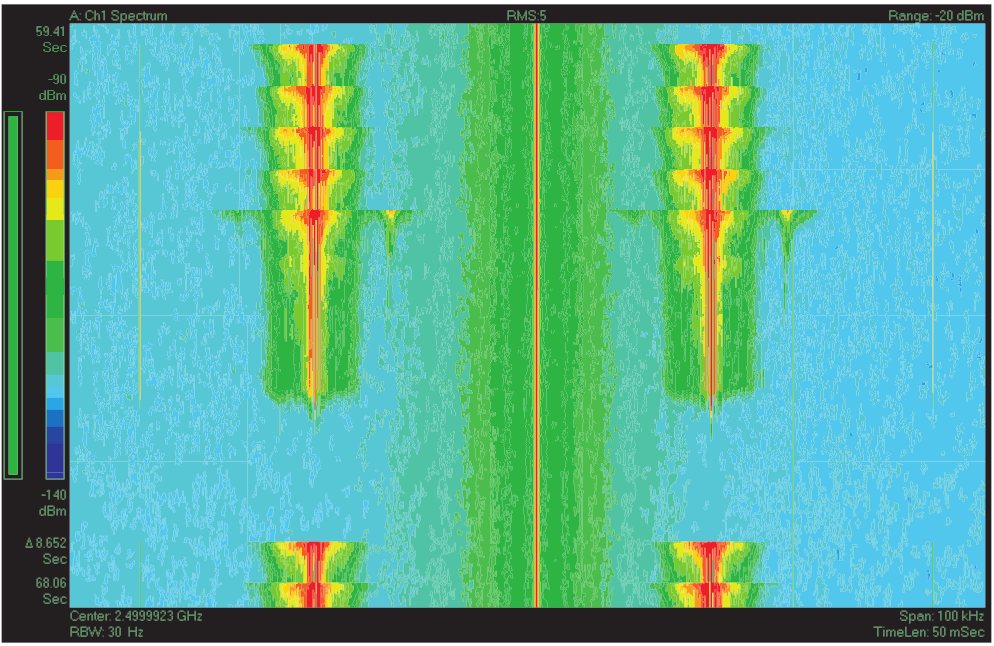}%
\caption{Spectrogram display containing about 200 traces as shown on the left-hand side in colour coding. Time runs from top to bottom.}%
\label{ec2}%
\end{figure}

An example of vector spectrum analyzer display and performance is given in Figs. \ref{ec1} and \ref{ec2}. Both figures were obtained during measurements of the electron cloud in the CERN Super Proton Synchrotron (SPS).

%

\section{Noise basics}
The concept of ``noise'' was originally studied for audible sound caused by statistical variations of the air pressure with a wide flat spectrum (white noise). It is now also used for electrical signals, with the noise ``floor'' determining the lower limit of the signal transmission. Typical noise sources are: \textit{Brownian} movement of charges (thermal noise), variations of the number of charges involved in the conduction (flicker noise) and quantum effects (\textit{Schottky} noise, shot noise). Thermal noise is only emitted by structures with electromagnetic losses, which, by reciprocity, also absorb power. Pure reactances do not emit noise (emissivity = 0).

Different categories of noise have been defined:
\begin{itemize}
	\item white, which has a flat spectrum,
	\item pink, being low-pass filtered and
	\item blue, being high-pass filtered.
\end{itemize}
In addition to the spectral distribution, the amplitude density distribution is also required in order to characterize a stochastic signal. For signals generated by superposition of many independent sources, the amplitude density has a \textit{Gaussian} distribution.
The noise power density delivered to a load by a black body is given by \textit{Planck's} formula:
\begin{equation}
\frac{N_{\text{L}}}{\Delta f} = hf\left(\text{e}^{hf/kT} - 1\right)^{-1},
\label{plank}
\end{equation}
where $N_{\text{L}}$ is the noise power delivered to the load, $h = 6.625 \cdot 10^{-34}$\,J\,s the \textit{Planck} constant and $k = 1.38056 \cdot 10^{-23}$\,J/K \textit{Boltzmann's} constant.

Equation (\ref{plank}) indicates a constant noise power density up to about 120 GHz (at 290 K) with 1\% error. Beyond, the power density decays and there is no ``ultraviolet catastrophe'', i.e. the total integrated noise power is finite.

The radiated power density of a black body is given as
\begin{equation}
W_{\text{r}}(f,T) = \frac{hf^{3}}{c^{2}\left[\text{e}^{hf/kT} - 1\right]}.
\label{radpower}
\end{equation}

\begin{figure}[tb]%
\centering
\includegraphics[width=\mywidth]{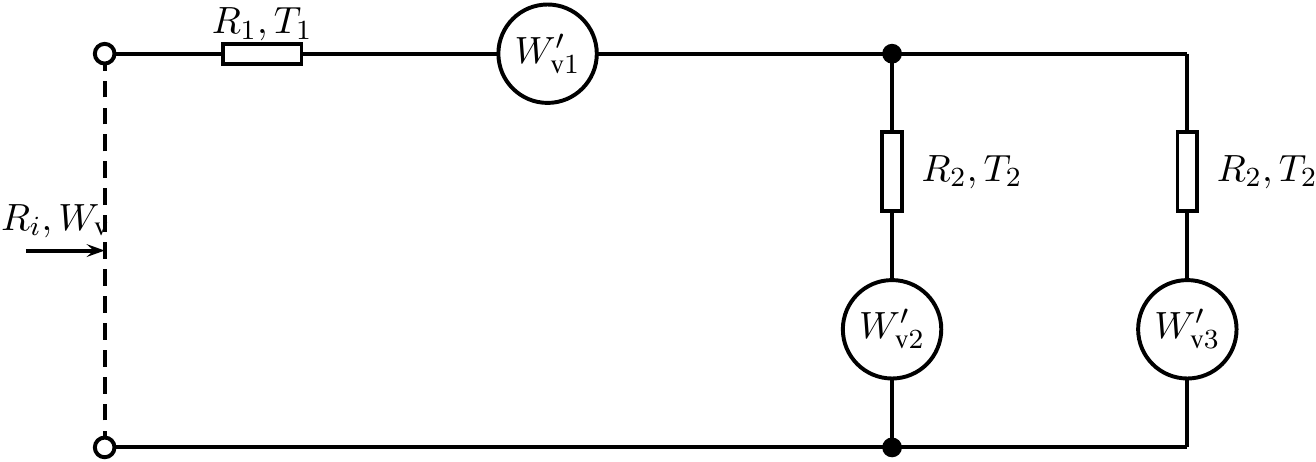}%
\caption{Noisy one-port with resistors of different temperatures \cite{Zinke, Schiek}}%
\label{noisyoneport}%
\end{figure}

\begin{figure}[t]%
\centering
\includegraphics[width=\mywidth]{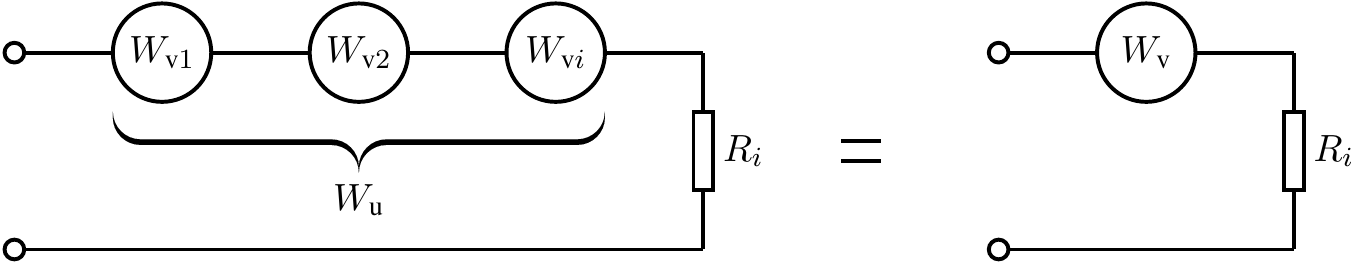}%
\caption{Equivalent sources for the circuit of Fig. \ref{noisyoneport}}%
\label{equivsources}%
\end{figure}

For $hf \ll kT$ the \textit{Rayleigh--Jeans} approximation of Eq. (\ref{plank}) holds:
\begin{equation}
N_{\text{L}} = kT \Delta f,
\label{rayleighjeans}
\end{equation}
where in this case $N_{\text{L}}$ is the power delivered to a matched load. 
The noise voltage $v(t)$ of a resistor $R$ with no load is given as
\begin{equation}
\overline{v^{2}(t)} = 4 kT R\Delta f
\label{nlnoise}
\end{equation}
and the short-circuit current $i(t)$ by
\begin{equation}
\overline{i^{2}(t)} = 4 \frac{kT \Delta f}{R} = 4 kT G\Delta f,
\label{scircurr}
\end{equation}
where $v(t)$ and $i(t)$ are stochastic signals, and $G$ is $1/R$. 
The linear averages $\overline{v(t)}, \overline{i(t)}$ vanishes, important are the quadratic averages $\overline{v^{2}(t)}, \overline{i^{2}(t)}$.
\begin{figure}[bht]%
\centering
\includegraphics[width=\mywidth]{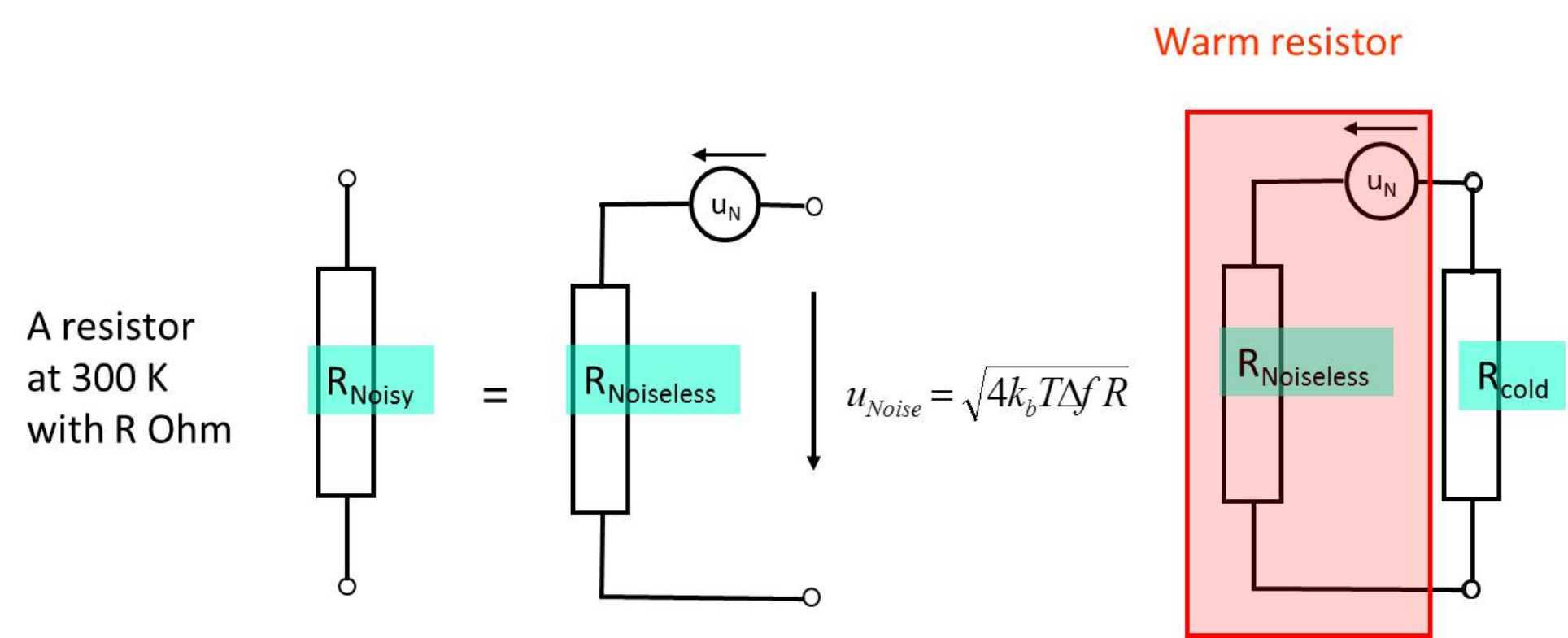}
\caption{Equivalent circuit of a noisy resistor terminated by a noiseless load }%
\label{resistor}%
\end{figure}
The available power (which is independent of $R$) is given by (see also Fig. \ref{resistor})
\begin{equation}
\frac{\overline{v^{2}(t)}}{4 R} = kT \Delta f.
\label{avpower}
\end{equation}
from which the spectral density function is defined as \cite{Schiek}
\begin{eqnarray}
\nonumber W_{\text{v}}(f) &=& 4kTR, \\
W_{\text{i}}(f) &=& 4kTG, \\
\nonumber \overline{v^{2}(t)} &=& \int_{f_{1}}^{f_{2}}W_{\text{v}}(f) \text{d}f.
\label{specdensfunc}
\end{eqnarray}
A noisy resistor may be composed of many elements (resistive network). Typically, it is made from a carbon grain structure, which has a homogeneous temperature. But if we consider a network of resistors with different temperatures, and hence with an inhomogeneous temperature distribution (Fig. \ref{noisyoneport}),
%
%
the spectral density function becomes
\begin{equation}
W_{\text{v}} = \sum_{j}W_{\text{v}j} = 4kT_{\text{n}}R_{\text{i}}, \\
\label{multires1}
\end{equation}
where $W_{\text{v}j}$ are the individual noise sources (Fig. \ref{equivsources}), $T_{\text{n}}$ is the total noise temperature, $R_{\text{i}}$ the total input impedance, and $\beta_{j}$ are coefficients indicating the fractional part of the input power dissipated in the resistor $R_{j}$. For simplicity it is assumed that all $W_{\text{v}j}$ are uncorrelated.
%
%

The relative contribution ($\beta_{j}$) of a lossy element to the total noise temperature is equal to the relative dissipated power multiplied by its temperature:
\begin{equation}
T_{\text{n}} = \beta_{1}T_{1} + \beta_{2}T_{2} + \beta_{3}T_{3} + \cdots = \sum_{j} \beta_{j}T_{j}
\label{relcont}
\end{equation}
A good example is the noise temperature of a satellite receiver, which is nothing else than a directional antenna. The noise temperature of free space amounts roughly to 3 K. The losses in the atmosphere, which is an air layer of 10 to 20 km height, causes a noise temperature at the antenna output of about 10 to 50 K. This is well below our room temperature of 290 K.

So far, only pure resistors have been considered. Looking at complex impedances, it is evident, losses occur only from dissipation in Re($Z$). The available noise power is independent of the magnitude of Re($Z$) with Re($Z$) $>$ 0. For Figs. \ref{noisyoneport} and \ref{equivsources}, Eq. (\ref{multires1}) still applies, except $R_{\text{i}}$ is replaced by Re($Z_{\text{i}}$). However, in complex impedance networks the spectral power density $W_{\text{v}}$ becomes frequency dependent \cite{Zinke}.

The rules mentioned above apply to passive structures. A forward-biased Schottky diode (external power supply) has a noise temperature of about $T_{0}$/2 + 10\%. A biased Schottky diode is not in thermodynamic equilibrium and only half of the carriers contribute to the noise \cite{Schiek}. But, it represents a real 50~$\Omega$ resistor when properly forward biased. For transistors, in particular field-effect transistors (FETs), the physical mechanisms are somewhat more complicated. Noise temperatures of 50 K have been observed for a FET at 290 K physical temperature.
\subsection{Noise-figure measurements with the spectrum analyzer}
Consider an ideal (noiseless) amplifier, terminated at its input (and output) with a load at 290 K with an available power gain ($G_{\text{a}}$). At the output we measure \cite{Yip, HP}:
\begin{equation}
P_{\text{a}} = kT_{0}\Delta fG_{\text{a}}.
\label{output}
\end{equation}

For $T_{0} = 290$ K (sometimes 300 K), we obtain $kT_{0}$ = $-$174 dBm/Hz 
($-$dBm = decibel below 1 mW). 
At the input we determine for a given signal $S_{\rm i}$ a certain signal-to-noise 
ratio $S_{\rm i}/{N}_{\rm i}$, and at the output $S_{\rm o}/{N}_{\rm o}$,  
from what the noise factor $F$ is defined as:
\begin{equation}
F = \frac{{S}_{\rm i}/{N}_{\rm i}}{{S}_{\rm o}/{N}_{\rm o}}
\label{noisefac}
\end{equation}%
and its logarithmic equivalent $\mathit{NF}$ follows as:
\begin{equation}
\mathit{NF} = 10 \log F
\label{noisefig}
\end{equation}

An ideal amplifier has $F = 1$ or $\mathit{NF} = 0$ dB. The noise temperature of this amplifier is 0 K, and signal and noise levels at the output are linearly increased by the gain. 
A real amplifier adds some noise, which leads to a decrease in ${S}_{\rm o}/{N}_{\rm o}$ due to the added noise $N_{\text{a}}$:
\begin{equation}
F = \frac{N_{\text{a}}+N_{\rm i} G_{\text{a}}}{N_{\rm i} G_{\text{a}}}=
\frac{{N}_{\text{a}} + kT_{0}\Delta f G_{\text{a}}}{kT_{0}\Delta f G_{\text{a}}}.
\label{adnoise}
\end{equation}

For a linear system the minimum noise factor amounts to $F_{\text{min}} = 1$ or $\mathit{NF_{\text{min}}}=0$ dB, however, for non-linear systems one may experience a noise factor $F < 1$.

 Noise factor and noise temperature are related by
 \begin{equation}
 T_{\text{e}} = \frac{N_{\text{a}}}{k\Delta f G_{\text{a}}} = T_0 (F-1)
 \label{noisetemp}
 \end{equation}
 with $T_{\text{e}}$ being the equivalent temperature of a source impedance into a perfect, noise-free device
 that would produce the same added noise $N_{\text{a}}$ \cite{HP}.

\begin{figure}[bht]%
\centering
\includegraphics[width=\mywidth]{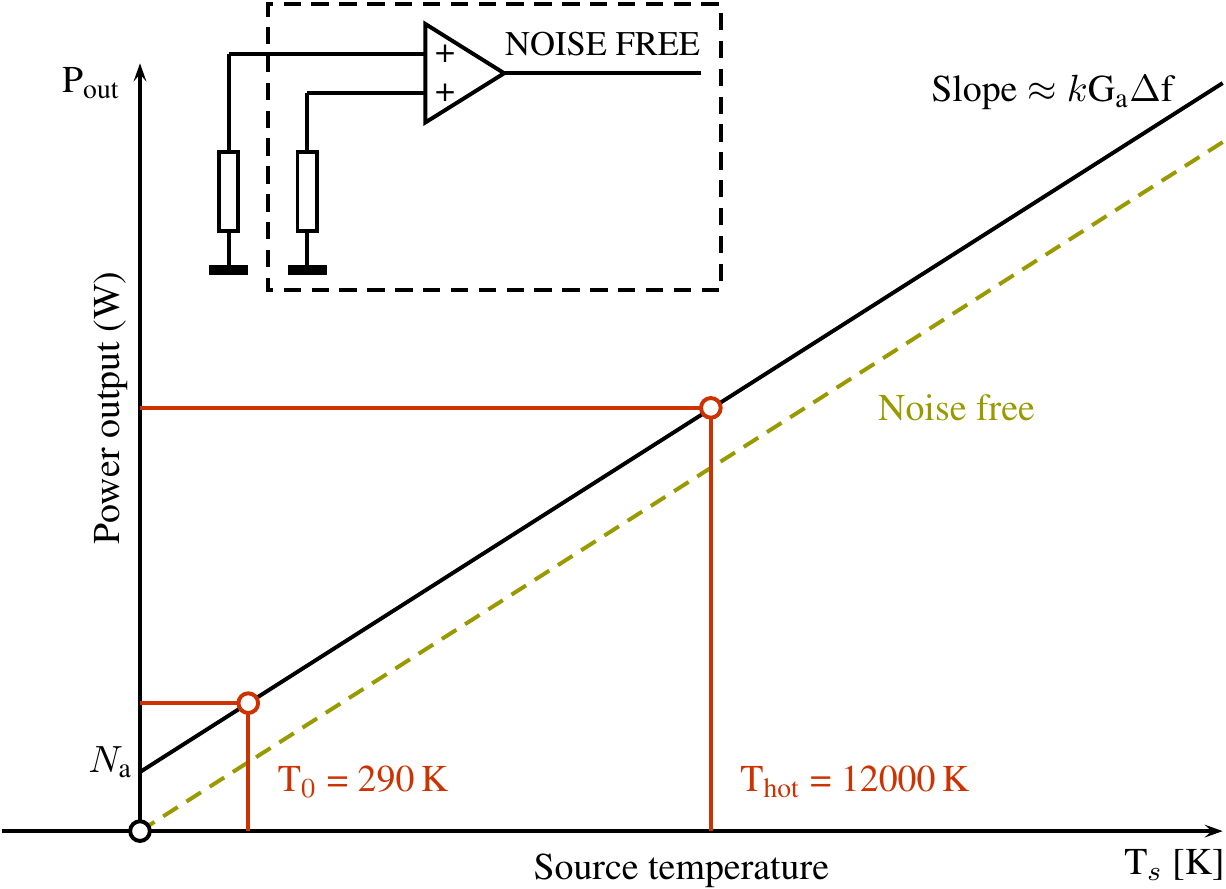}%
\caption{Relation between source noise temperature $T_{\text{s}}$ and output power $P_{\text{out}}$ for an ideal (noise-free) and a real amplifier \cite{Yip, HP}.}%
\label{tempoutput}%
\end{figure}
The so-called $Y$-factor method is a popular way to measure the noise figure.
It is based on a switchable noise source with two calibrated values $N_1$ and $N_2$ for the noise temperature, e.g.\ $T_{\text{c}}$ and $T_{\text{h}}$, corresponding to ``cold'' and ``hot''.
Usually a dedicated noise diode is used as noise source, switched between non-bias and bias operation to provide the two noise temperatures.
The calibrated noise level is defined as \emph{excess noise ratio} ($\mathit{ENR}$):
\begin{equation}
\mathit{ENR}_{\text{dB}} = 10 \log \left(\frac{T_{\text{h}}-T_{\text{c}}}{T_0}\right)
\label{exessnoisedb}
\end{equation}
For most noise figure calculations the  linear form is more useful:
\begin{equation}
\mathit{ENR} = 10^{\frac{\mathit{ENR}_{\text{dB}}}{10}}
\label{exessnoise}
\end{equation}


The noise source is connected to the amplifier or DUT to be analyzed, providing noise ``on'' ($N_2$) and ``off'' ($N_1$) conditions. 
The ratio of these noise powers is called the \emph{$Y$-factor}:
\begin{equation}
Y = \frac{N_2}{N_1}
\label{yfactor}
\end{equation}
$Y$-factor and $\mathit{ENR}$ can be used to determine the noise slope of the DUT, 
as illustrated in Fig.~\ref{tempoutput}.
The calibrated $\mathit{ENR}$ of the noise source represents a reference level for the input noise,
which allows the calculation of the internal (added) noise $N_{\text{a}}$ of the DUT:
\begin{equation}
N_{\text{a}} = k T_0 \Delta f G_1 \left( \frac{\mathit{ENR}}{Y-1}-1 \right)
\label{intnoise}
\end{equation}
The SA, operating in automatized \emph{noise figure mode}, controls the noise diode, i.e.\ switching between ``hot'' (on) and ``cold'' (off) states, acquiring the DUT output signal, and computes -- based on the calibrated $\mathit{ENR}$ -- the total \emph{system noise factor}
\begin{equation}
F_{\text{sys}} = \frac{\mathit{ENR}}{Y-1}
\label{sysnoise}
\end{equation}
which includes noise contributions from all parts of the system.
In case the ``cold'' noise temperature $T_{\text{c}}\neq T_0= 290$~K, Eq.~\ref{sysnoise} becomes
\begin{equation}
F_{\text{sys}} = \frac{\mathit{ENR-Y(T_{\text{c}}/T_0-1)}}{Y-1}
\label{sysnoisecorr}
\end{equation}
For low $\mathit{ENR}$ noise sources, $T_{\text{h}}<10\,T_{\text{c}}$, an alternative equation holds:
\begin{equation}
F_{\text{sys}} = \frac{\mathit{ENR(T_{\text{c}}/T_0)}}{Y-1}
\label{sysnoisealt}
\end{equation}

If $Y$ is close to 1, i.e.\ $F_{\text{sys}}\gg \mathit{ENR}$, the system noise factor ``masks'' the noise generated by the noise source, making an accurate measurement difficult or impossible.
Therefore the $Y$-factor method is limited to noise figure measurements with $\mathit{NF\approx 10}$~dB below the
$\mathit{ENR}$ of the noise source.

The literature explains a variety of other noise figure measurement methods \cite{Schiek,Connor,Landstorfer,Evans,Schiek2},
including the ``3~dB'' method \cite{HP} for the measurement of high noise figure devices, 
where the $Y$-factor method is limited. 

The noise figure of a cascade of amplifiers is given as \cite{Zinke, Schiek, Yip, HP, Schiek2}
\begin{equation}
F_{\text{total}} = F_{1} + \frac{F_{2} - 1}{G_{\text{a}1}} + \frac{F_{3} - 1}{G_{\text{a}1}G_{\text{a}2}} + \cdots .
\label{cascade}
\end{equation}
As Eq.~(\ref{cascade}) shows, the first amplifier in a cascade has a dominant effect on the total (system) noise figure, provided $G_{\text{a}1}$ is not too small and $F_{2}$ not too large. In order to select the best amplifier from a number of different units to be cascaded, the noise measure $M$
\begin{equation}
M = \frac{F - 1}{1 - (1/G_{\text{a}})}.
\label{noisemeasure}
\end{equation}
helps to select the optimal unit:\newline
The amplifier with the smallest $M$ should be selected as first unit in the cascade \cite{HP}.

\section{Introduction to network analysis and S-parameters}
\label{NetAnaSect}

One of the most common measurement tasks in the field of RF engineering is the analysis of circuits and
electrical networks. Such networks can be a simple one-port (two-pole), containing only a few passive
components (resistors, inductors and capacitors) or they may be complex units, consisting of passive, active and/or
non-linear components with several input and output ports.

A vector network analyzer (VNA) is
one of the most versatile and valuable pieces of measurement equipment used in a 
RF laboratory
or particle accelerator control room. 
The network analysis is performed by exciting
the device under test (DUT) with a well-defined input signal in terms of frequency and amplitude, 
and recording the response of the network, for each frequency step as complex value 
of the reflection and/or transmission coefficients.
These are the coefficients of the scattering parameters (S-Parameter), 
the properties to characterize a DUT at RF and microwave frequencies.
The best commercially available network analyzers can cover a frequency
range of ten (and more) orders of magnitude (from a few Hz to many GHz), 
with a resolution down to 0.1~Hz. 

In the following sections, scalar and vector network analyzers are introduced and measurement techniques for the
determination of S-parameters of networks are discussed.
S-parameters are basically defined only for linear networks. In the real world, many DUTs are at least
weakly non-linear (e.g.\ mixers, or active elements such as amplifiers). For the analysis of these devices
certain approximations or extensions of the definitions are required~\cite{src:xPar}.

Another interesting application is the determination of the beam transfer function (BTF), where the DUT is
a circulating particle beam in an accelerator.

\begin{figure}[t]
\begin{center}
\subfloat[ ]{
\includegraphics[width=0.35\textwidth]{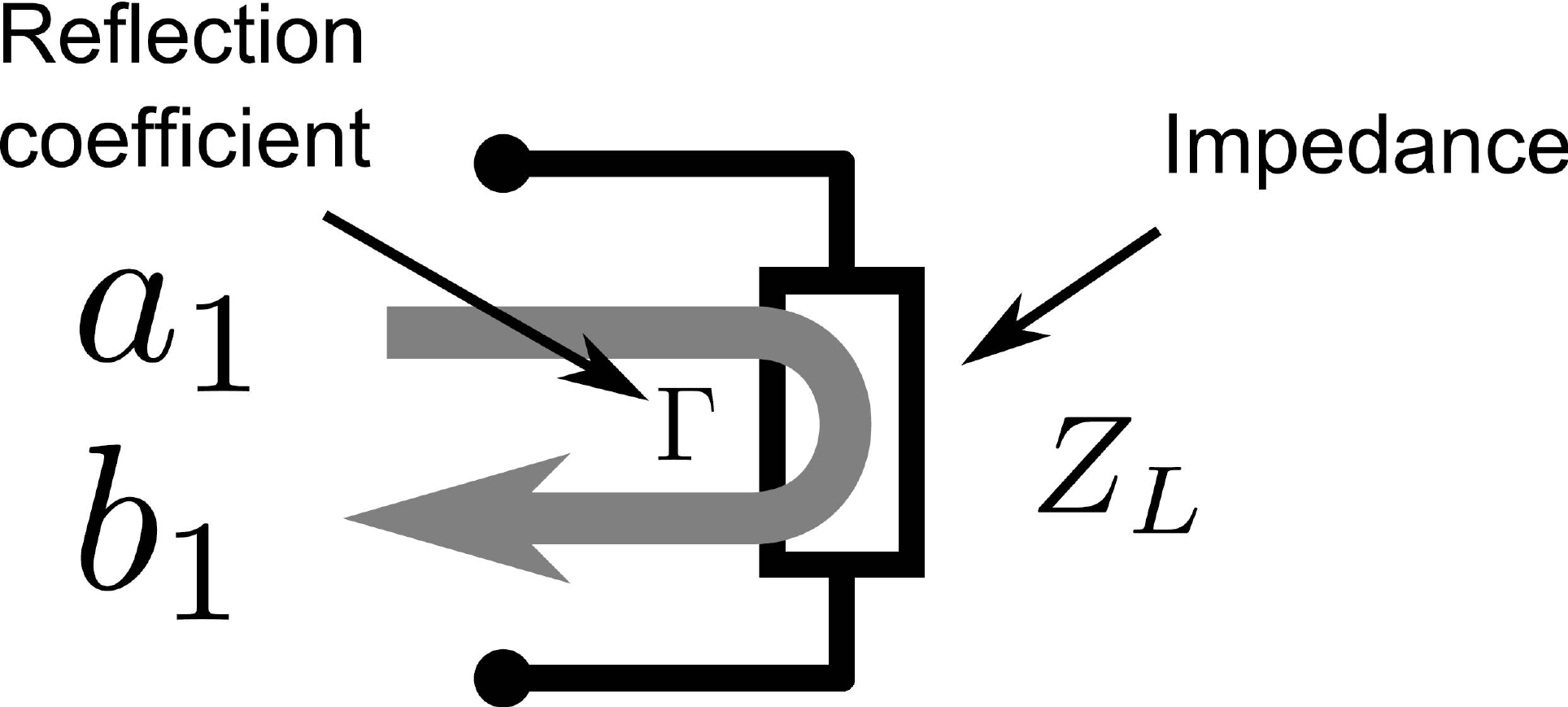}
\label{fig:1portWave}
}
\subfloat[ ]{
\includegraphics[width=0.3\textwidth]{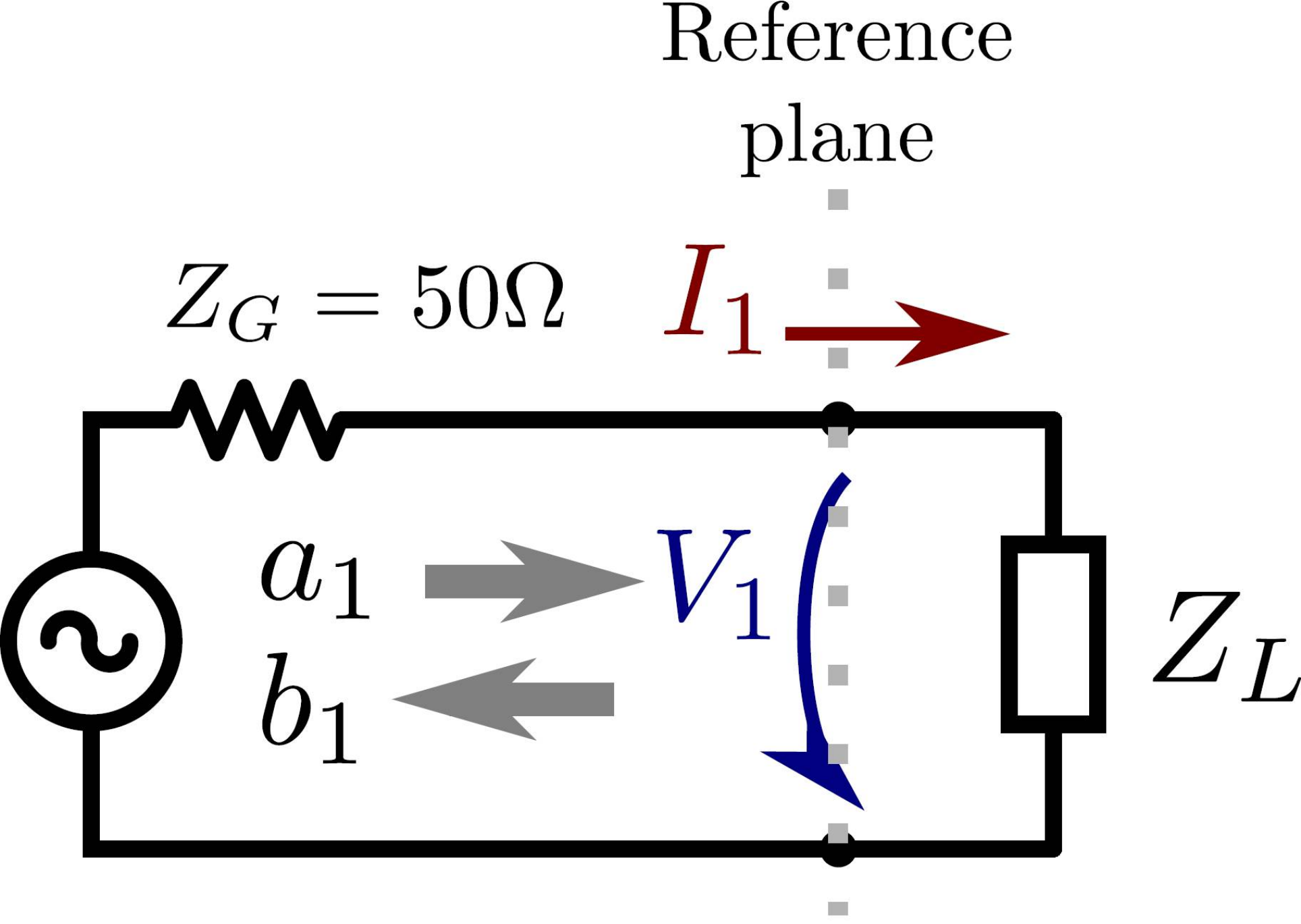}
\label{fig:1PortVI}
}
\caption{Wave quantities of a one-port (with two poles) and impedance $Z_L$:
(a) incident ($a_1$) and reflected ($b_1$) wave; (b) relation of $a_1$ and $b_1$ to $V_1$ and $I_1$.}
\end{center}
\end{figure}

\subsection{One-port networks}
In RF engineering, \emph{wave quantities} are preferred in favor currents 
or voltages for the characterization of RF circuits. 
We can distinguish between incident (a) and reflected waves (b).
The incident wave travels from a source to the DUT -- the reflected wave travels in the opposite direction.
This terminology is preferred, because in RF engineering the linear geometrical
dimensions of a circuit often are larger than 10\% of the corresponding free-space wavelength. 
Wave functions are defined in time and \emph{spacial} coordinates, and for this fact are preferred to voltages and currents, which typically are only defined in time.
This also requires the
definition of a reference plane, i.e.\ the physical location in space to which the measurement refers.
Without this reference plane, e.g.\ the phase of the reflection coefficient would be undefined, 
which would make vectorial measurements impossible. 
Of course, a mathematically correct description of the DUT in terms of voltages and currents still
holds, and also will return correct results, but working with wave quantities turns out to be much 
more convenient in practice. 
Both network discription methods -- if correctly applied -- have no fundamental limitation,
e.g. S-parameters can be used at very low frequencies and 
voltage and current descriptions can also be used at very high frequencies. 
Both methods are fully equivalent, for any frequency; the results are mutually convertible. 
This fact is expressed by conversion rules, namely S-parameters can be converted
into impedances and vice versa.

The interface of the DUT to the outside world is utilized by one or more \emph{pole pairs}, which are
commonly referred as \emph{ports}. A device with one pair of poles (as in \refFig{fig:1portWave}) is
defined as one-port, where one incident ($a_1$) and one reflected ($b_1$) wave can propagate simultaneously.
The index of the wave quantities represents the number of the port.

The wave quantities can be determined from the voltage and current at the port. They are
related to each other
\Equation{equ:waveVI}{a_1 = \frac{V_1 + I_1 Z_0}{2 \sqrt{Z_0}}, \quad \quad
b_1 = \frac{V_1 - I_1 Z_0}{2 \sqrt{Z_0}},}
where $V_1$ and $I_1$ represent the voltage and current
respectively at the port as depicted in \refFig{fig:1PortVI}. $Z_0$ is an arbitrary reference impedance
(often, but not necessarily always, the characteristic impedance $Z_0 = Z_{\rm G} = 50~\Omega$ of the system).

The wave quantities have the dimension of $\sqrt{W}$ (see~\cite{FritzSpara}).
This normalization is important for the conservation of energy. 
The power traveling towards the DUT
is calculated by $P_{\rm inc} = |a|^2$, the reflected power by $|b|^2$. 
It is important to note that this
definition is mainly used in the USA -- in European notation, 
the incident power is usually calculated by
$P_{\rm inc} = 0.5 |a|^2$. 
These conventions have no impact on the calculation of S-parameters and only need to be
considered when the absolute power is of interest.

The reflection coefficient $\Gamma$ represents the ratio between the incident wave and the reflected wave of a
specific port. It is defined as
\Equation{equ:gamma}{
\Gamma = \frac{b_1}{a_1}.}

By substitution with \refEqu{equ:waveVI}, we can find a relation between the complex 
(load) impedance $Z_L$ of a one-port and its complex reflection coefficient $\Gamma$:
\Equation{equ:zgamma}{
\Gamma = \frac{Z_L - Z_0}{Z_L + Z_0}.}


\subsection{Two-port networks}
For electrical networks with two ports (e.g. attenuators, amplifiers) we find more quantities 
to be measured. 
Besides the reflection coefficients for each port, the transmission in forward and reverse
directions also needs to be characterized. 
We now require the definition of the scattering parameters (S-parameters) for
two ports. 
The idea is to describe how the incident energy on one port is scattered by the network 
and exits through the other ports. 
All possible signal paths through a two-port are shown in \refFig{fig:2port2}. 
A two-port has four complex, frequency-dependent scattering parameters:
\Equation{}{
S_{11} = \frac{b_1}{a_1}, \quad\quad S_{12} = \frac{b_1}{a_2}, \quad\quad S_{21} = \frac{b_2}{a_1}, \quad\quad
S_{22} = \frac{b_2}{a_2}.}
Here $S_{11}$ and $S_{22}$ are equal to the reflection coefficients $\Gamma$ of their respective
ports -- but \emph{only} under the condition that the corresponding other port is terminated in its
characteristic impedance. 
$S_{21}$ and $S_{12}$ are the forward and reverse transmission coefficients,
respectively. 
The first index of the S-parameter defines at which port the outgoing wave is observed, 
the second index defines at which port the network is excited. 
This leads to the counterintuitive appearing situation, 
that for forward transmission the corresponding S-parameter is $S_{21}$, not $S_{12}$. 
The S-parameters are measured following exactly the same definition. 
The internal source of the network analyzer excites
an incident wave on port one, namely $a_1$. 
Now $b_1$ and $b_2$, the outgoing waves from the DUT, are
measured, which allows the determination of $S_{11}$ and $S_{21}$ (provided that port one 
and port two are terminated with their characteristic impedances).

\Figure{0.4}{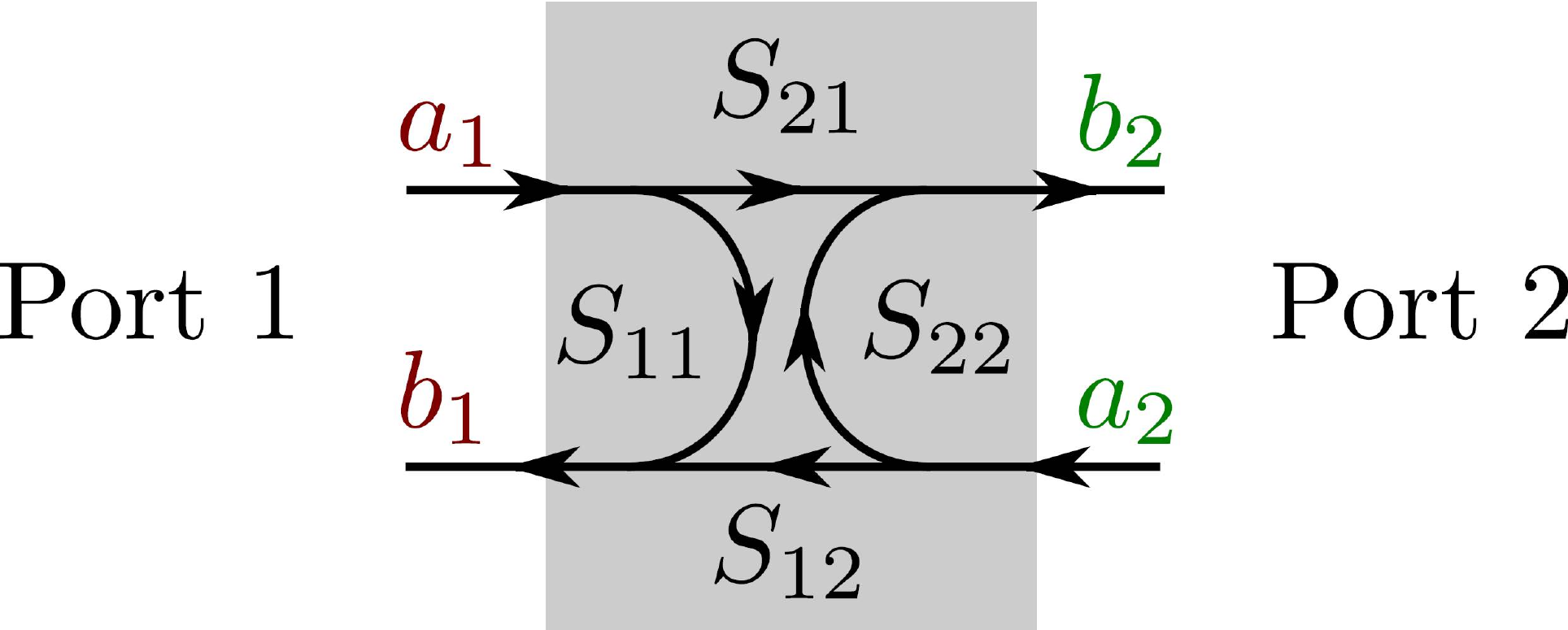}{fig:2port2}{All possible S-parameters of a two-port network}{[t]}

It is very important to \emph{always} terminate all ports of the DUT with their respective characteristic impedances.
In many situations this is $Z_0$, but there are cases where the characteristic impedance is different between port
one and port two, e.g.\ a transformer with a turns ratio of two, leading to an impedance transformation by
a factor of four. In this case the characteristic impedance would be for port one $50~\Omega$ and for port two
$12.5~\Omega$.

The termination prevents unwanted reflections and ensures the DUT is only excited by a single
incident wave. For practical S-parameter measurements this implies that any port of the DUT needs to be
connected to a matched load corresponding to the characteristic impedance of this port. This rule includes
in particular the port connected to the VNA output port, or in other words, the generator impedance has
also to match the impedance of the DUT. For example,
the analysis of a DUT with $25~\Omega$ characteristic impedance is not simply straightforward on a $50~\Omega$ network analyzer, unless special care is taken..
But, permitting a modern VNA, applying a
special calibration procedure allows the modification of the characteristic impedance of each VNA port to any value (within
a reasonable range from $> 5~\Omega$ to $< 500~\Omega$), and in this way to adapt to the requirements of the DUT.
However, the situation of the termination of ports becomes more complicated for the
characterization of beam elements, like beam pickups, kickers, and accelerating structures,
where strictly speaking the beam (waveguide) ports also need to be terminated in their
characteristic impedance.
Often simple solutions can be applied, like microwave absorbing foam, to avoid unwanted
reflections from open beam ports.

The S-parameters are an intrinsic property of the DUT and not a function of the incident power used for the
measurement (condition of linearity). Obviously, the S-parameters measured shall be independent of the
instrumentation used to perform the measurement.

Once all $n^2$ S-parameters for a given $n$-port network are measured, the properties of this network can be
described by a set of linear equations. For incident waves $a_1$ and $a_2$
of arbitrary phase and magnitude on a two-port, the outgoing or scattered waves $b_1$ and $b_2$ can be
determined
\begin{eqnarray}\label{equ:slinear}
b_1 = S_{11} a_1 + S_{12} a_2,\\
b_2 = S_{21} a_1 + S_{22} a_2.\nonumber
\end{eqnarray}

These equations can be written in matrix format, for convenience:
\begin{eqnarray}\label{equ:smatrix}
\vec{b} &= \textbf{S} \; \vec{a}\\
\left[
\begin{array}{c}
b_1\\
b_2
\end{array}
\right]&=
\left[
\begin{array}{cc}
S_{11} & S_{12}\\
S_{21} & S_{22}
\end{array}
\right]
\left[
\begin{array}{c}
a_1\\
a_2
\end{array}
\right].
\end{eqnarray}

The S-matrix is a linear model of the DUT. Its diagonal elements represent the reflection
coefficients of each port. The remaining elements characterize all possible signal transmission
paths between the ports. S-parameters are in general complex and a function of frequency. The set of
linear equations given by the S-matrix must be solved for a single frequency at a time. S-parameters are
typically acquired over a certain frequency range (span) for a number $N$ of discrete, equidistant frequency steps. With $N$ data
points, the system of equations has to be solved $N$ times.
A discussion of the general properties of the S-matrix can be found in~\cite{FritzSpara}.

\section{Scalar network analysis}
A scalar network analyzer measures only the amplitude, i.e.\ the magnitude of a -- reflected or transmitted -- signal, the
phase is not available. Consequently, only the absolute value (the magnitude) of the complex
S-parameters can be obtained. 
Today scalar network analyzers are basically obsolete, however, some key components and circuits are also found in VNAs, making this instrument a methodical way to introduce the concept of network analysis.

\Figure{0.7}{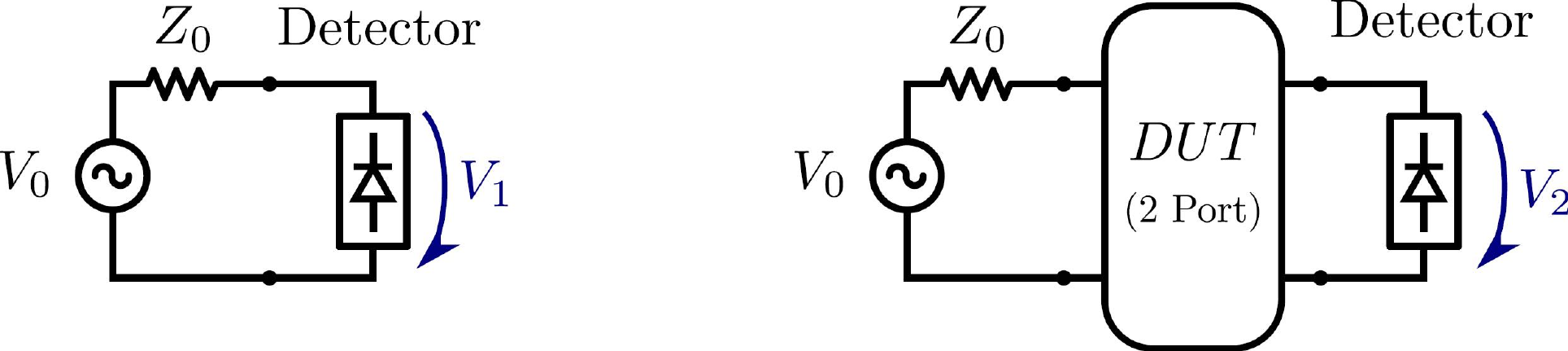}{fig:scalarSimple}{A simple measurement set-up for the scalar transmission coefficient ($|S_{21}|$)}{[t]}
A simple network analysis set-up, as it was used more than 50 years ago, is shown in
\refFig{fig:scalarSimple}.
The measurement is performed in two steps, in the first step (\refFig{fig:scalarSimple}, left) without the DUT to measure the power of the incident signal ($V_1$). Then the DUT is inserted (\refFig{fig:scalarSimple}, right), and $V_2$ is
measured. 
Following the magnitude of the transmission coefficient is calculated:
\Equation{equ:magTx}{|S_{21}| \propto \frac{V_2}{V_1}.}

To obtain the results in decibels, a logarithmic amplifier was connected to the output of the detector. 
It has a
logarithmic transfer function ($V_{\rm out} = \log V_{\rm in}$) and permits the display of a large dynamic range on a dB
scale. 
Furthermore, mathematical operations like multiplication or division, 
e.g.\ required for normalization in \refEqu{equ:magTx}, 
transforms simply into into an addition or subtraction, handled by operational amplifiers.

As detector any kind of device converting the input RF signal into a DC voltage is applicable, assuming its transfer function is ``reasonable''\footnotemark\ proportional to the RF power. There are basically three possibilities to achieve
this:
\footnotetext{With the term ``reasonable'' we point out the fact, that many detectors have a non-linear relation between
input power and output voltage.}

\begin{description}
\item[Rectifier]
A fast \textit{Schottky} diode and a low-pass filter are used to convert the input RF signal to a DC voltage.
Operating the diode in its square-law region ($P_{\rm in} < -10$ dBm) results in an output voltage proportional
to the RF power; see Section~\ref{RFdiodesec}.\newline 
Advantages: cheap, fast response (depending on $f_{\rm max}$ of the output filter). \newline
Limitations:
Commercially available RF power meters, based on \textit{Schottky} diodes, can operate from $-60$~dBm (limited by
tangential sensitivity) up to about +30~dBm (damage level). The non-linearity of the output signal versus input
power is compensated by electronic means (look-up table). Coaxial RF \textit{Schottky}
detectors are usually limited to maximum frequencies of approximately 100~GHz, 
essentially determined by the coaxial connector technology available. Usually an input matching network is required to match the input impedance of the
\textit{Schottky} diode to $Z_0 = 50~\Omega$.
\item[Thermal measurement]
Several types of detectors based on heating effects are available for the measurement of RF power. In a
bolometer (thermistor or barretter), the high temperature coefficient of the thermal conductivity of certain
metals or metal alloys is exploited. 
The temperature change $\Delta T$ of dissipated heat of the RF input signal is measured utilizing a 
DC-based temperature measurement, while applying a correction of the non-linearities.
Barretters utilize the positive temperature coefficient of metals like
tungsten and platinum. Thermistors consist of a metal oxide with a strong negative temperature coefficient.
Another class of RF power meters based on heating is the thermo-element, which takes advantage of the
thermo-electrical coefficient of a junction between two different metals. A well-known example is the Sb-Bi
junction, which has a temperature coefficient of about $10^{-4}$ V/K, which is one of the highest values
available for this kind of detector. Even larger values can be achieved using semiconductor--metal junctions,
where thermoelectric coefficients of $250~\mu$V/K have been achieved. For further details, see \cite{src:thumm}.
\item[Mixer]
Multiplying two sinusoidal signals with different frequencies results in signals of sum and
difference frequencies at the multiplier's output; see Section~\ref{Mixersec}. Technically this frequency mixing principle allows to convert a range of high-frequency signals to a much lower intermediate frequency (IF) band. Now the RF power measurement is performed in simpler ways at this IF.
\end{description}

\subsection{Automatic Gain Control (AGC)}
Often RF measurements are performed over a wide range of frequencies, requiring the signal strength, i.e.\ the amplitude $V_0$ of the source to be constant. 
This is usually achieved by an active feedback loop (\textit{levelling}), keeping $V_0$ constant,
independent of the operation frequency.
Any feedback loop requires a process variable which has to be detected and
controlled to a well defined set point, here the output signal level $V_1$. 
For the automatic gain control (AGC) loop in a NA e.g.\ a resistive
power divider can be used to provide this reference signal, while keeping inputs and outputs matched to $Z_0
= 50~\Omega$ (\refFig{fig:scalarFeedback}). For this example, the test signal arriving at the
DUT is reduced by 6 dB due to the insertion loss of the resistive power divider. However, the AGC feedback loop ensures
the stimulus signal applied to the DUT has always a constant, well defined power level over a wide frequency range.

\Figure{0.5}{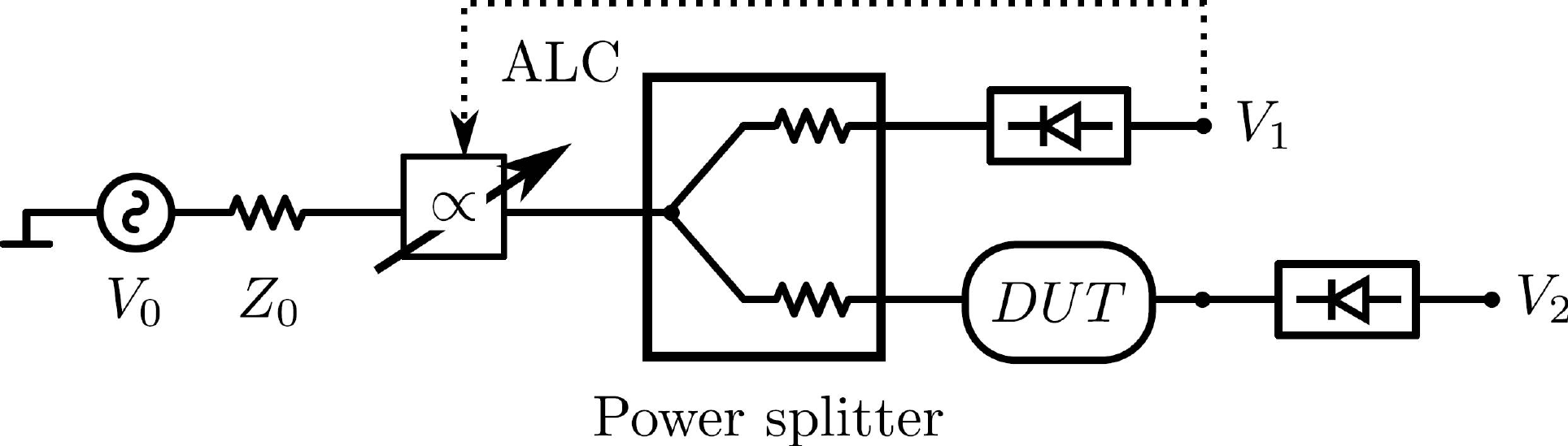}{fig:scalarFeedback}{Simplified circuit diagram of a typical automatic gain control}{[t]}

For the characterization of linear DUTs, only the ratio $V_2 / V_1$ is of interest, which is independent of the absolute value of $V_0$. In this case the S-parameter measurements do not require an AGC loop of the RF generator, but in
practice the gain control has many advantages, in particular for measurements on weakly non-linear
elements, such as amplifiers.

\Figure{0.6}{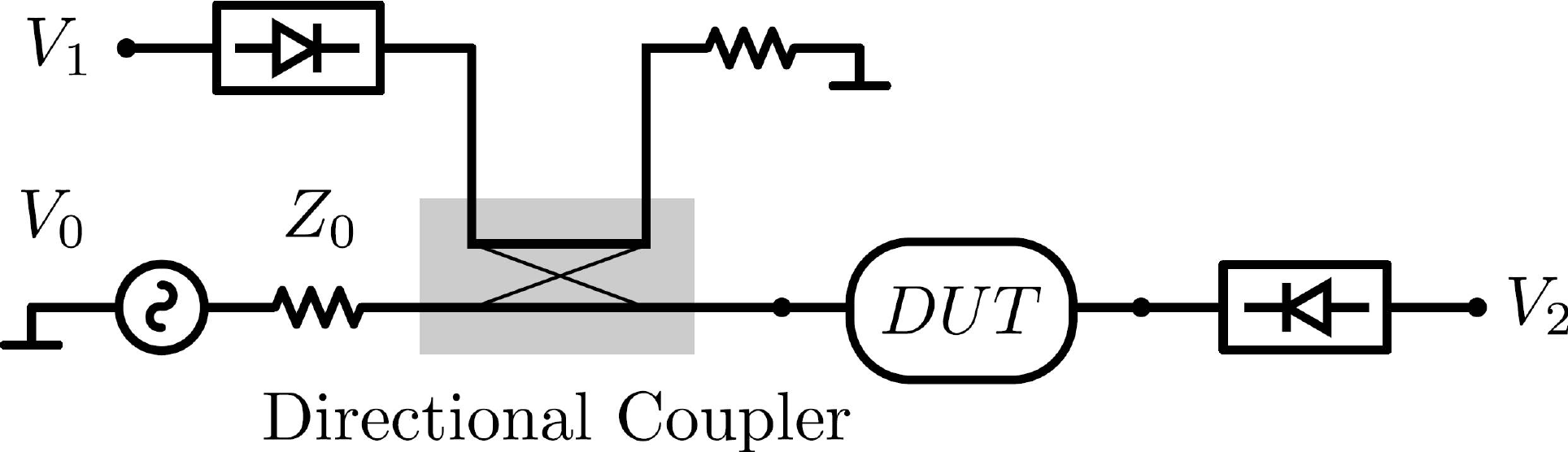}{fig:scalarFeedbackCoupler}{Feedback loop of a typical automatic gain control (AGC)}{[t]}

\subsection{Directional couplers}
Replacing the resistive power divider by a directional coupler reduces the insertion loss substantially, 
the principle is outlined in \refFig{fig:scalarFeedbackCoupler}. $V_1$ is an attenuated replica  -- defined by the
coupling factor -- of the forward-traveling wave, which is only used for as reference for the gain control.
Typically, directional couplers with a coupling coefficient of $-$20~dB are used for the purpose, they offer a transmission attenuation in the main branch of less than 0.3 dB. In contrast to the resistive power splitter, the transmission-line based directional coupler has a limited frequency range, and therefore other issues.

\Figure{0.7}{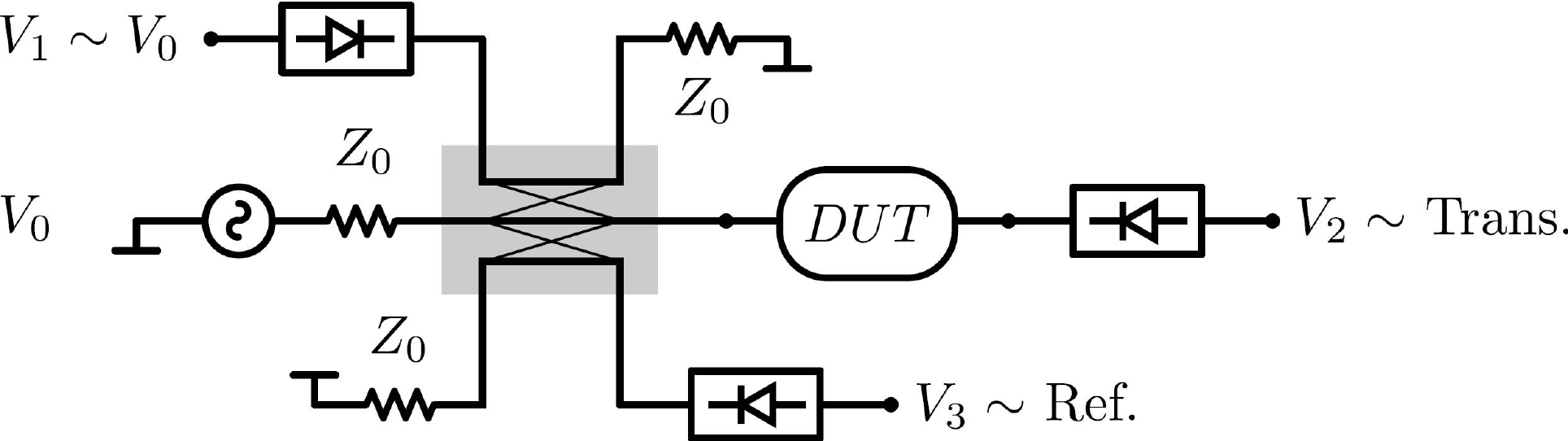}{fig:scalarBidirectional}{Dual directional coupler in a network analyzer}{[t]}

Modern network analyzers (both scalar and vectorial versions) measure the forward-transmission, as well as the reflection coefficient of a DUT simultaneously, without the need to manually re-connect DUT ports. Each port of the instrument is equipped with a dual directional coupler, providing simultaneously replicas
of the incident and reflected waves from the DUT, see \refFig{fig:scalarBidirectional}. These
directional couplers, in combination with some required switches and attenuators are commonly called \emph{test set}.
In the early days, network
analyzers consisted of separate building blocks, like S-parameter test set, frequency generator, display and
controller unit. All these elements had to be connected by many external cables. Modern instruments have
all those building blocks integrated in a single unit, including advanced computer controls with digital data acquisition and post-processing.

Based on \refFig{fig:scalarBidirectional}, the reflection and transmission coefficients are defined as
\Equation{equ:refFwd}{
|S_{11}| \propto \frac{V_3}{V_1}, \quad \quad |S_{21}| \propto \frac{V_2}{V_1}.}

From the ratio of the reflected wave to the incident wave ($S_{11}$), valuable quantities like standing wave
ratio (SWR), reflection coefficient, impedance, admittance as well as return loss
of the DUT are determined. From the ratio of the transmitted wave to the incident wave ($S_{21}$), gain resp.\ insertion loss, the transmission coefficient, the insertion phase, and
group delay of the DUT can be characterized.

\section{Vector measurements}
A vector network analyzer (VNA) is able to measure the magnitude \emph{and phase} of a complex S-parameter.
There are different hardware configurations which implement this kind of RF instrument, e.g.\ six-port reflectometers, certain RF bridge methods, or superheterodyne RF network analyzers. Here only the latter will be introduced.

\subsection{The modern vector network analyzer}

A modern VNA contains a RF generator which produces the signal stimulating the DUT. 
This signal is
usually generated by a synthesizer-type oscillator and is adjustable in very fine steps 
over a large frequency
range, in a programmable manner. 
Since all modern VNAs operate with analog and/or digital downconverters (mixing), 
the generation of a tracking LO frequency is also necessary. 
This tracking LO is typically generated by PLL
circuits and represents essentially a second oscillator following the 
main frequency with a specified frequency offset.

The observation (IF) band signal is typically processed digitally, allowing bandwidth settings over a wide range, e.g.\
1~Hz to 20~MHz and more.
In all stages of the signal path the vectorial nature of the signal is preserved, 
both phase and magnitude are processed, in the digital domain usually as I-Q (in-phase -- 
quadrature-phase) data, equivalent to real and imaginary parts.
Details on the internal signal processing of a VNA are found in \cite{src:fundVNA,
src:fundVNA2}. Note, similar to the spectrum analyzer, the sweep time and resolution bandwidth cannot be
adjusted independently.
A modern four-port vector network analyzer is shown in \refFig{fig:vectorPhoto}.

\Figure{3.0}{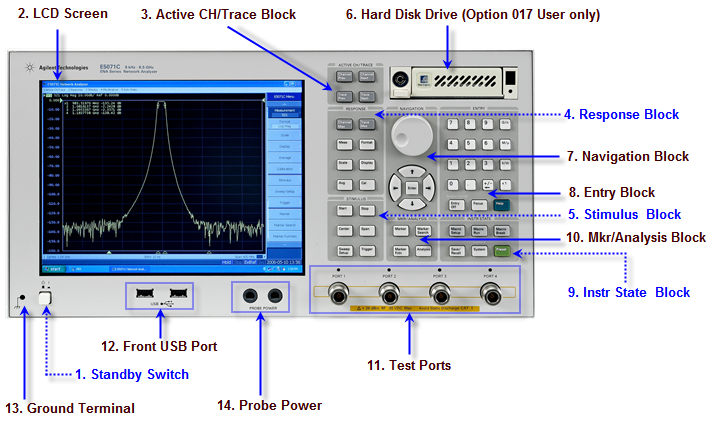}{fig:vectorPhoto}{A modern four-port VNA}{[t]}

Although complete network analysis of any $N$-port can be performed with a two-port VNA, a four-port unit is extremely
convenient for many measurement tasks. It permits a quick analysis, e.g. of a directional coupler or a
three-port circulator without the need for swapping cables, it also introduces virtual ports of balanced nature, and many other valuable features.

\subsection{Time-domain transformation (synthetic pulse technique)}
For any linear system, the frequency domain information (data) can be converted 
to the time domain by an inverse (fast) \textit{Fourier} transformation\footnotemark\ and vice versa, assuming the entire frequency vector data (magnitude and phase, or real and imaginary) is present. 
This is the basis of the synthetic pulse technique,
available on many modern VNAs. It was commercially introduced by Hewlett-Packard in the 1980s for network analyzer applications.

\footnotetext{More precisely: by a discrete \textit{Fourier} transformation (DFT). 
The fast \textit{Fourier} transformation (FFT) is just an optimized form of th eDFT, 
exploiting the symmetry of $2^n$ data samples, thus saving computation time. 
However, both algorithms will produce
the same result for the same input data.}

It renders the VNA even more versatile, allowing to display the impulse (\textit{Gaussian}) and/or step response of
the DUT, and to perform time-domain reflectometry (TDR) measurements. Typical applications of this
measurement techniques are:
\begin{enumerate}
  \item Localizing and evaluating discontinuities (faults) in transmission lines.
  \item Separating the scattering properties of sections of complicated RF networks by time-domain gating.
  \item Echo cancellation (in multipath environments).
  \item The synthetic pulse time-domain reflectometry can be very useful in
  trouble-shooting, e.g. of the accelerator beam-pipe. 
  By using waveguide modes 
  it  was successfully used to detect an obstacle in the LHC beam-pipe.
\end{enumerate}

The only constraint of the applicability of the synthetic pulse measurement technique, the DUT has to
be a \emph{linear} and \emph{time-invariant} (LTI) system.

\begin{figure}[t]
\begin{center}
\subfloat[ ]{
\includegraphics[width=0.5\textwidth]{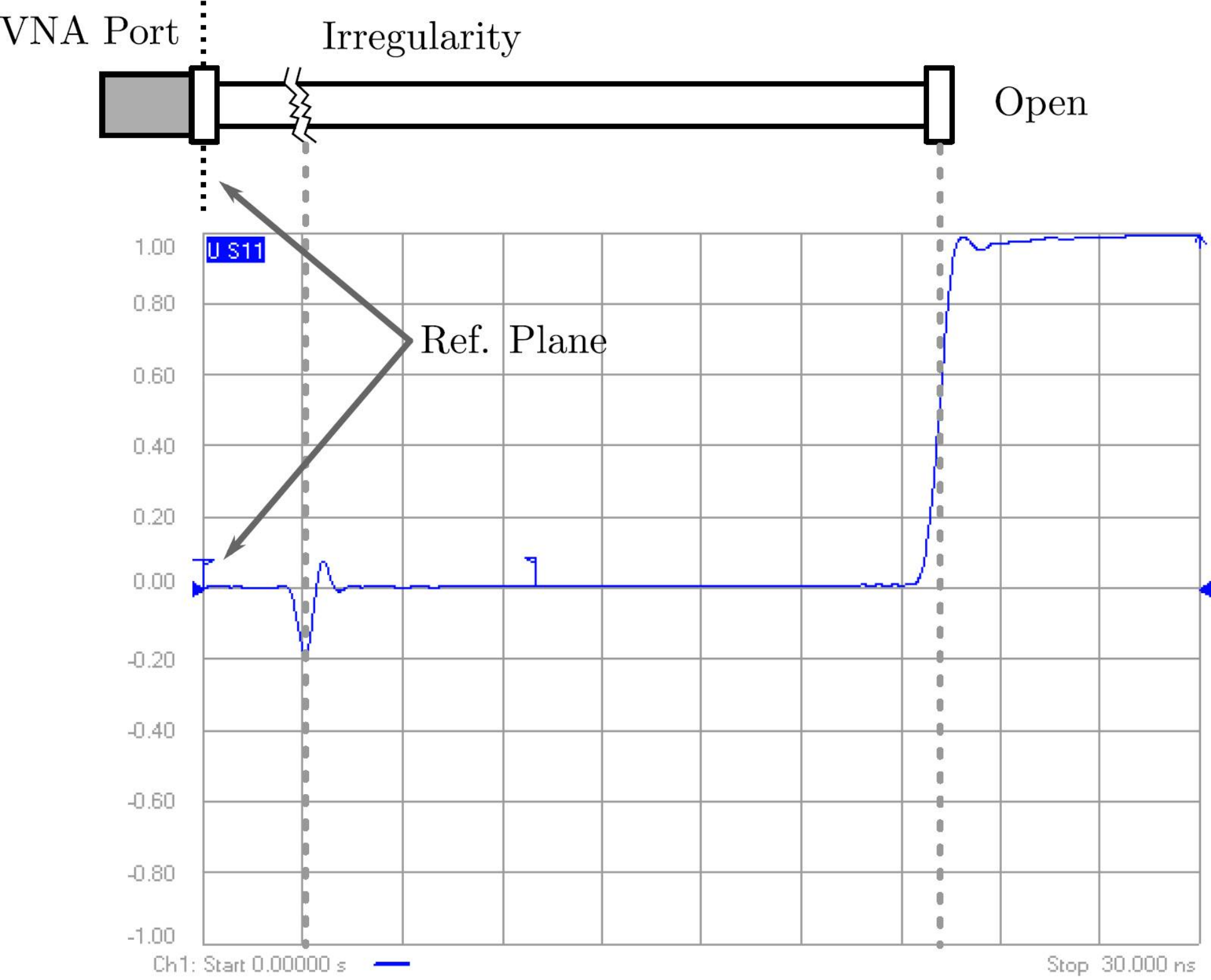}
\label{fig:iDFTcableS}
}
\subfloat[ ]{
\includegraphics[width=0.5\textwidth]{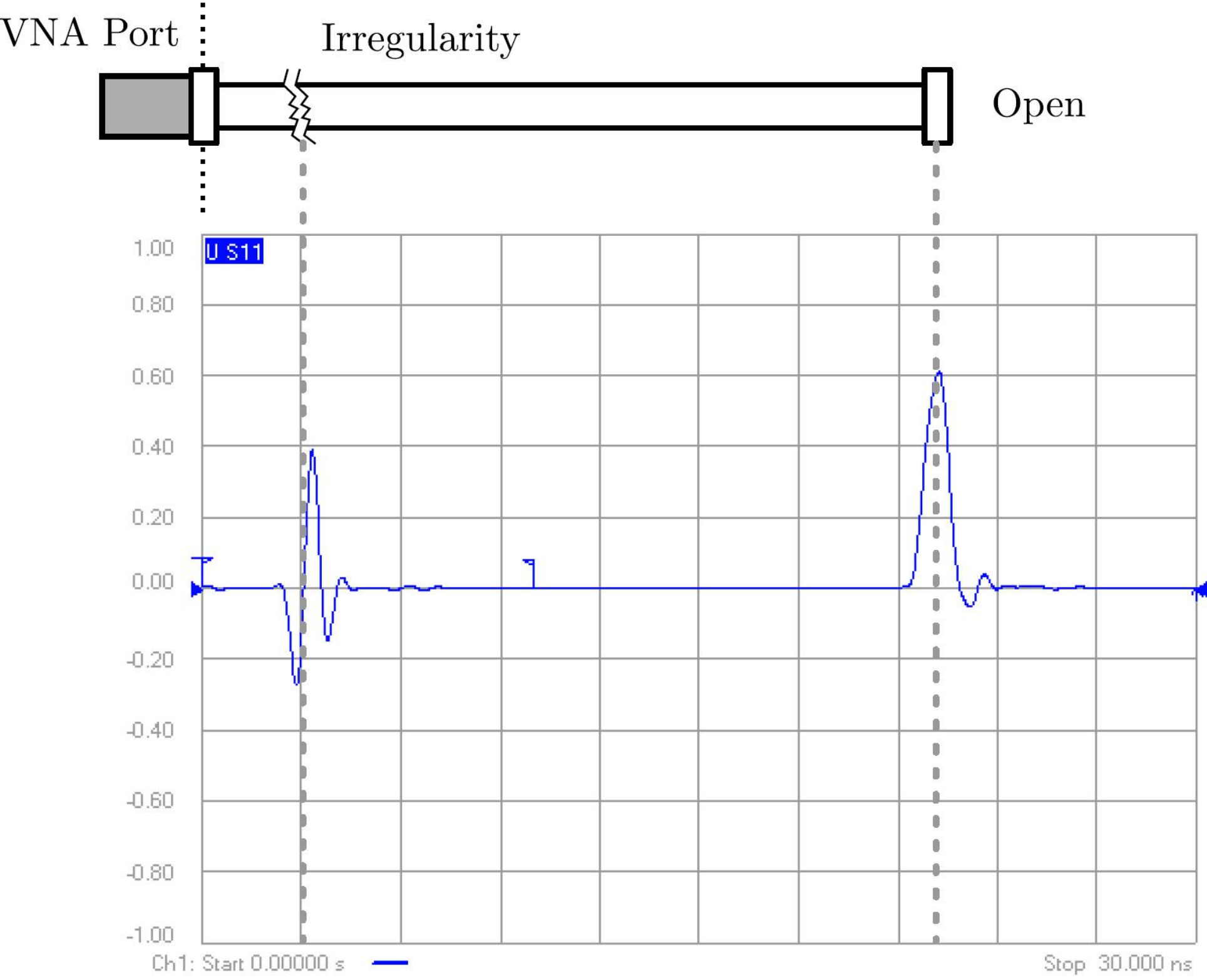}
\label{fig:iDFTcableI}
}
\caption{Synthetic pulse measurement with a VNA: (a) step response; (b) impulse response. \newline
The measured frequency data is converted by
an inverse discrete Fourier transformation (iDFT) 
to the time
domain. Now the synthetic impulse response of a transmission-line, here a coaxial cable, is displayed over time. The reflections of
the incident pulse on any irregularity or discontinuity, as well as the end of the cable are clearly identified. By measuring the time delay between the the reference plane and the location of the irregularity, or end of the cable (displayed as pulse or step in the reflection coefficient) the electrical length of the cable can be calculated.}
\label{fig:iDFTcable}
\end{center}
\end{figure}
A measurement example is shown in \refFig{fig:iDFTcable}. 
A transmission line with a given length
and some perturbation is connected to a calibrated VNA. 
The real part of the \textit{Fourier}-transformed reflection
coefficient ($S_{11}(\omega)$) is plotted versus time. The VNA permits the display of either, the synthetic step
(\refFig{fig:iDFTcableS}) or the impulse response (\refFig{fig:iDFTcableI}). The step is simply obtained by
(numerical) integration of the impulse response data.

The incident synthetic pulse is scattered from the discontinuity, but also from the open end of the cable. The
travel time for the pulse can be read on the horizontal axis on the time-domain display. In this example we
measure a delay of $t_d = 22$ ns until the open end of the cable becomes visible. This time accounts for the
impulse traveling towards the open end \emph{and back}; thus, the factor $1/2$ has to be taken into account
when calculating physical length $l$ of the transmission-line:
\Equation{equ:lineLength}{
	l = \frac{c}{\sqrt{\varepsilon_r}} \cdot \frac{1}{2} t_d.}
In this example the relative dielectric constant of the insulation in the coaxial cable is $\epsilon_r = 2.3$ (PTFE Teflon),
which returns a cable length of $l = 2.2$ m. The same method can be applied for obtaining the position of
any irregularity or discontinuity (deformation, bad connector) along the cable. Nearly all VNAs with time-domain option permit
the designation of the velocity factor ($1 / \sqrt{\varepsilon_r}$ for a homogeneously filled transmission line) and thus
convert travel time or electrical length to physical distance on the display.

Note that the step response shown in \refFig{fig:iDFTcableS} returns the local reflection factor versus time.
Along the cable it amounts to $\Gamma = 0$, except for the position of the irregularity, indicating a
well-matched $50~\Omega$ transmission line. At the end we notice a positive step to $\Gamma = 1$, indicating an
open circuit (see \refTbl{tbl:ref}).

\begin{table}[t]
\caption{Important values of the reflection coefficient}
\begin{center}
\begin{tabular}{lrr}
\hline
\hline
\bfseries DUT & ${\pmb Z_{\pmb L}}$ & ${\pmb\Gamma}$ \\
\hline
Open circuit 	&$\infty$	&+1\\
Short circuit 	&0			&--1\\
Matched load 	&$Z_0$&0\\
Load 			&$Z_0/2$	&--1/3\\
Load 			&$2 Z_0$	&1/3\\
\hline
\hline
\end{tabular}
\end{center}
\label{tbl:ref}
\end{table}

The reflected pulse in the impulse response trace (\refFig{fig:iDFTcableI}), related to the open end of the cable
does not reach unit amplitude due to fact of cable attenuation of the transmission line used
for this example -- a semi-rigid coaxial cable approximately 2 m length. The amplitude of this reflection from the open end
indicates the attenuation over twice the electrical length of the cable at the equivalent center frequency
($f_{\text{max}} = 3$ GHz, $f_{\text{centre}}= 1.5$ GHz) of the measurement.

For practical applications of the synthetic pulse technique, certain basic
properties of the discrete Fourier transform should be kept in mind, they are summarized in \refTbl{tbl:fft}.
For example, a long cable needs to be tested. 
This requires a long time window to ensure all multiple refections have decayed to zero,
which needs attention to ensure a sufficient narrow frequency sampling. 
The time interval $\Delta t$ is related to
$1/\Delta f$, and this reciprocal relation may cause issues if settings are kept in ``automatic'' mode.

\begin{table}[t]
\caption{Important characteristics of the FFT}
\begin{center}
\begin{tabular}{rcl}
\hline
\hline
\bfseries Time domain & & \bfseries Frequency domain\\
\hline
$T_{\text{max}}$ (time span)	&$\leftrightarrow$&	$\Delta f$ (frequency resolution)\\
$\Delta t$ (time resolution)	&$\leftrightarrow$&	$f_{\text{max}}$ (frequency span)\\
\hline
\hline
\end{tabular}
\end{center}\label{tbl:fft}
\end{table}

On the other hand, if a bad connector or cable damage needs to be located along a transmission-line, a high
resolution in time is required. Thus, the VNA has to measure over a wide frequency span ($f_{\text{max}}$).
Obviously, we would like to often use both, a high frequency span and a close spacing of the samples in the
frequency domain, but there are practical limitations: namely, the number of data points available. Usually in
modern instruments the number of data points available amounts to 60000 and, depending on the application,
compromises have to be accepted. 

Performing time-domain measurements with the vector network analyzer calls for two basic modes,
the ``low-pass'', or the ``band-pass'' mode to be selected. 

\subsubsection{Low-pass mode}
\begin{figure}[t]
\begin{center}
\subfloat[ ]{
\includegraphics[width=0.4\textwidth]{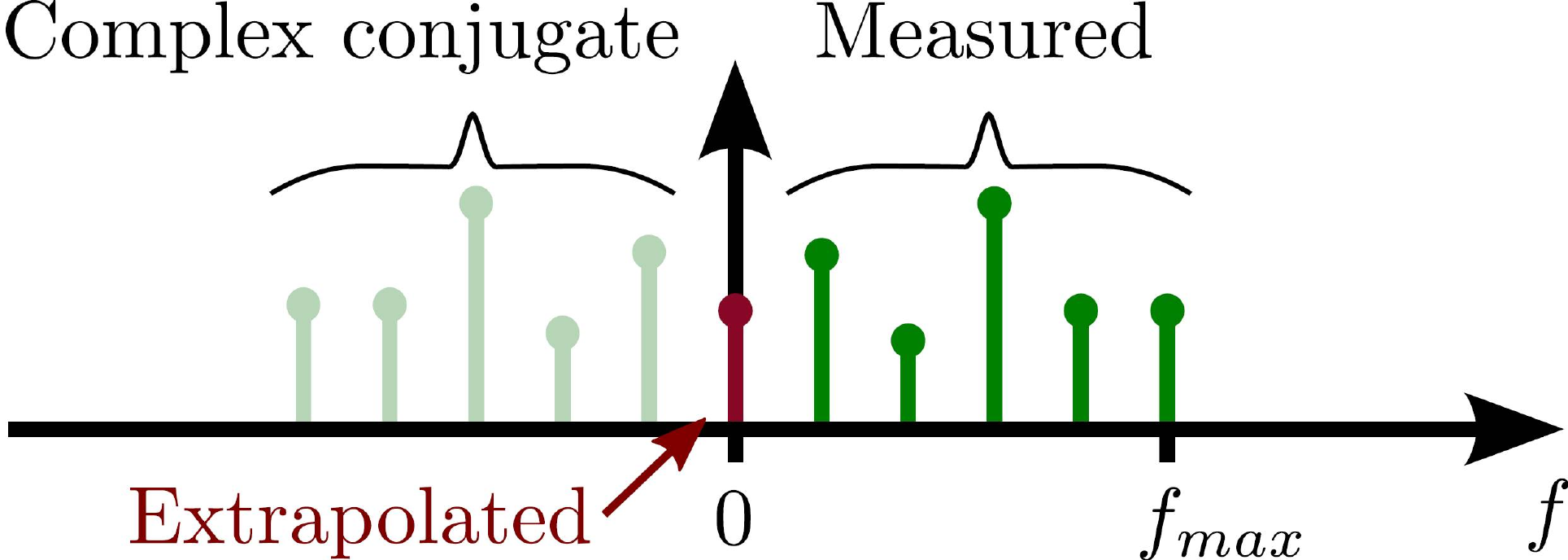}
\label{fig:lpm}
}
\hspace{15mm}
\subfloat[ ]{
\includegraphics[width=0.4\textwidth]{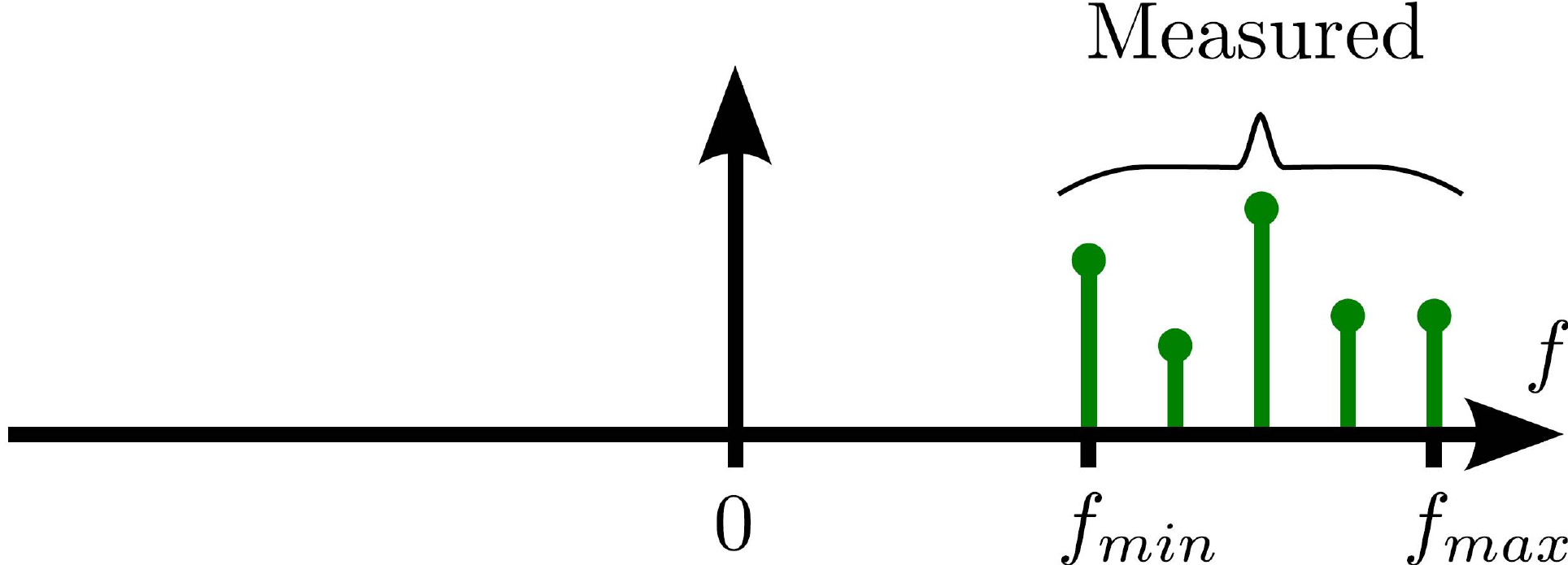}
\label{fig:bpm}
}
\caption{Sampling of frequency points for the different operating modes: (a) low-pass mode; (b) bandpass mode.}
\end{center}
\end{figure}

In low-pass mode the basic discrete \textit{Fourier} transformation algorithm is applied. This returns certain
constraints on the frequency-domain measurement data of the DUT (\refFig{fig:lpm}). The iDFT demands a
start frequency to always be 0 Hz (DC), and data is acquired in equidistant frequency steps between start and stop frequency.
Since most VNAs cannot measure at very low frequencies, the data points from DC to the minimum operation
frequency of the VNA are extrapolated mathematically. 
Data points for negative frequencies are derived from the measured
samples on the corresponding positive frequencies by complex conjugation. Compared to the bandpass mode, this
effectively doubles the number of data points available for the calculation of the time trace. For this
particular symmetry, the discrete \textit{Fourier} transformation returns a purely real-valued time trace. A
practical time domain reflectometry (TDR) measurement routine is setup as follows:
\begin{enumerate}
  \item The DUT is connected, the port and type of measurement are selected (transmission or reflection).
  \item The frequency range of interest and the number of data points are entered (this relates to the
  time domain by \refTbl{tbl:fft})
  \item After pushing the soft key, ``set frequency low-pass''\footnotemark, the instrument choses the exact
  sampling frequencies.
  \item Once the sampling points are defined, the VNA has to be calibrated (open, short, load for reflection
  measurements).
\end{enumerate}
\footnotetext{
This soft key may appear with slightly different naming, depending on the definitions of the manufacturer.}

In the low-pass mode, the trace appearing on the screen for time domain reflectometry (TDR) or time domain transmission (TDT)
is basically equivalent to what a real-time TDR or sampling oscilloscope display; see Section~\ref{ch:compare}.

\subsubsection{Band-pass mode}
In band-pass mode (\refFig{fig:bpm}) the spectral lines
(frequency-domain data points) no longer need to be equidistant, and extrapolated down to DC, 
they just need to cover the
frequency range of interest, e.g.\ from $f_{\rm min} = 1.2$ GHz to $f_{\rm max} = 1.5$ GHz. 
The start and stop
frequencies of the VNA can be chosen arbitrarily, which returns a high degree of flexibility and is especially
suited for the measurement of devices having a limited frequency range (example: waveguide-mode reflectometry).

The bandpass mode is the equivalent to a narrowband TDR (and also time-domain
transmission TDT) using the synthetic pulse technique. It permits the display of the impulse response only,
since no extrapolated information on a DC component is available. 
The measurement clearly identifies
position and size of perturbations along a transmission line, including waveguides.
Their characterization in terms of capacitive, inductive or resistive properties is possible, but not
straightforward~\cite{src:kruschdPaper}. Details on the
general properties and mathematical backgrounds of the low-pass and bandpass modes are found in
\cite{src:TDA, src:fundVNA2}.

\subsubsection{Windowing}
\begin{figure}[t]
\begin{center}
\subfloat[ ]{
\includegraphics[width=0.45\textwidth]{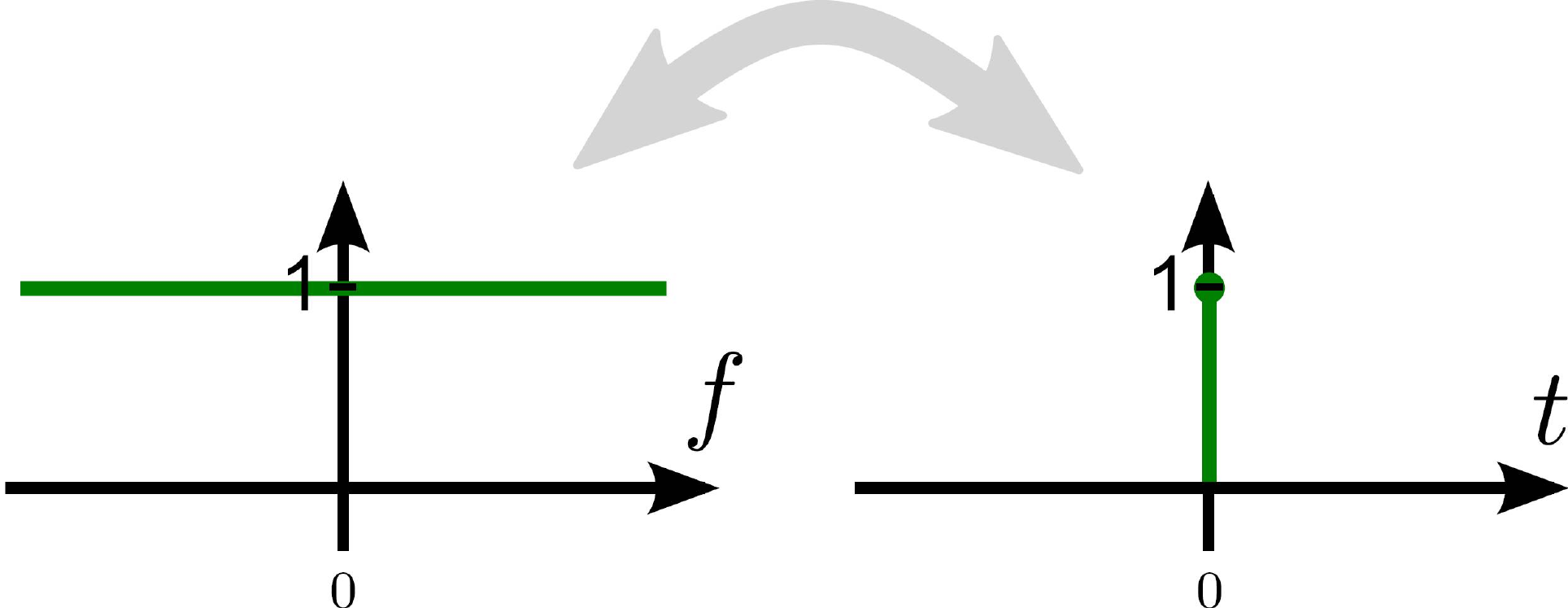}
\label{fig:spectLarge}
}
\hspace{11mm}
\subfloat[ ]{
\includegraphics[width=0.45\textwidth]{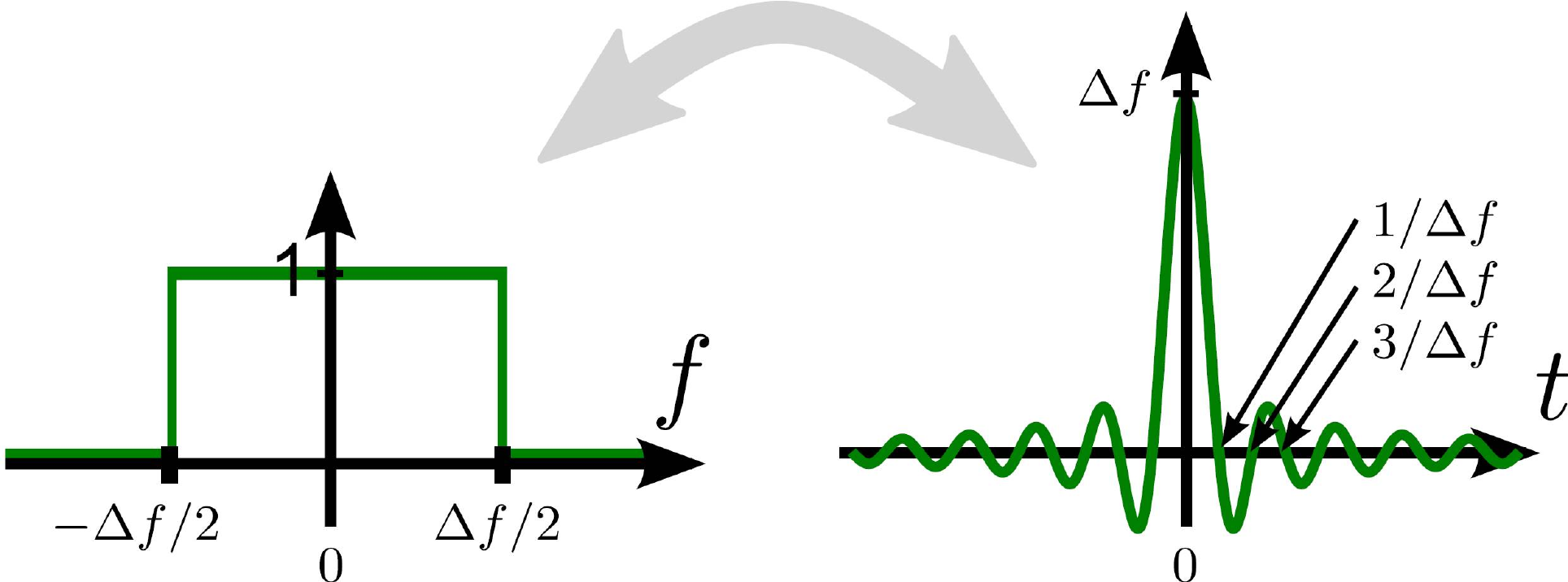}
\label{fig:spectSmall}
}
\caption{(a) Infinite frequency span. (b) Limited frequency span.
The limited frequency span $\Delta f$ \protect\footnotemark 
of the VNA leads to ``distortions'' of the time-domain synthetic pulse
measurement. The ideal response is convoluted with a $\mathit{sinc}$ function, 
which characteristics depend on $\Delta f$.}
\label{fig:rectSpect}
\end{center}
\end{figure}

As the VNA always samples a limited frequency spectrum, starting at $f_{\text{min}}$ and stopping at
$f_{\text{max}}$, the acquired spectrum is clipped by a rectangular envelope. 
Performing the iDFT, rectangular windowing artifacts show up in the time-domain data, 
as compared in \refFig{fig:rectSpect}.

\footnotetext{not to be confused with the previous definition of $\Delta f$ for the equidistant frequency samples}

An infinite spectrum of constant density (shown in \refFig{fig:spectLarge}) leads to a Dirac-pulse
function in the time domain. 
The Dirac pulse contains by definition all frequency components of equal
power. In \refFig{fig:spectSmall}, the spectrum is limited, for example, 
by the maximum operation frequency
of the VNA, or by some user settings. 
This can be expressed by multiplication of the ideal spectrum with a
rectangular function. 
The iDFT of a rectangular function of width $\Delta f$ leads to a $\mathit{sinc}$ function
(sometimes denoted as $\mathit{si}$ function) in the time domain. This relation is shown in \refEqu{equ:ftSinc} and
graphically in Fig.~\ref{fig:rectSpect}.
\begin{align}\label{equ:ftSinc}
 \begin{split}
	\text{Frequency domain} \quad &\Longleftrightarrow \quad \text{Time domain}\\
	\text{rect}\left(\frac{f}{\Delta f}\right) \quad &\Longleftrightarrow \quad \frac{\sin\left(\Delta f \pi
	t\right)}{\pi t} = \Delta f \cdot \text{sinc}\left(\Delta f \pi t\right).
	 \end{split}
\end{align}
To mitigate the effect of rectangular clipping of the spectrum in the time domain result, various weighting
functions are available. They smoothly filter (reduce) the amplitude of the spectrum around $f_{\rm min}$
and $f_{\rm max}$ in band-pass and low-pass mode. This helps to reduce the
strong sidelobes (ringing) in the time domain. 
However, the price to be paid is a
reduced pass-band, thus limiting the time resolution and the ability to distinguish between two
closely spaced impulses. The user has to select a reasonable trade off between the window weighting functions,
depending on the requirements of the particular measurement. 
The effect of some window functions on main and sidelobes
is shown in the frequency domain(!) on a logarithmic scale in \refFig{fig:fftWndw}.

\Figure{0.7}{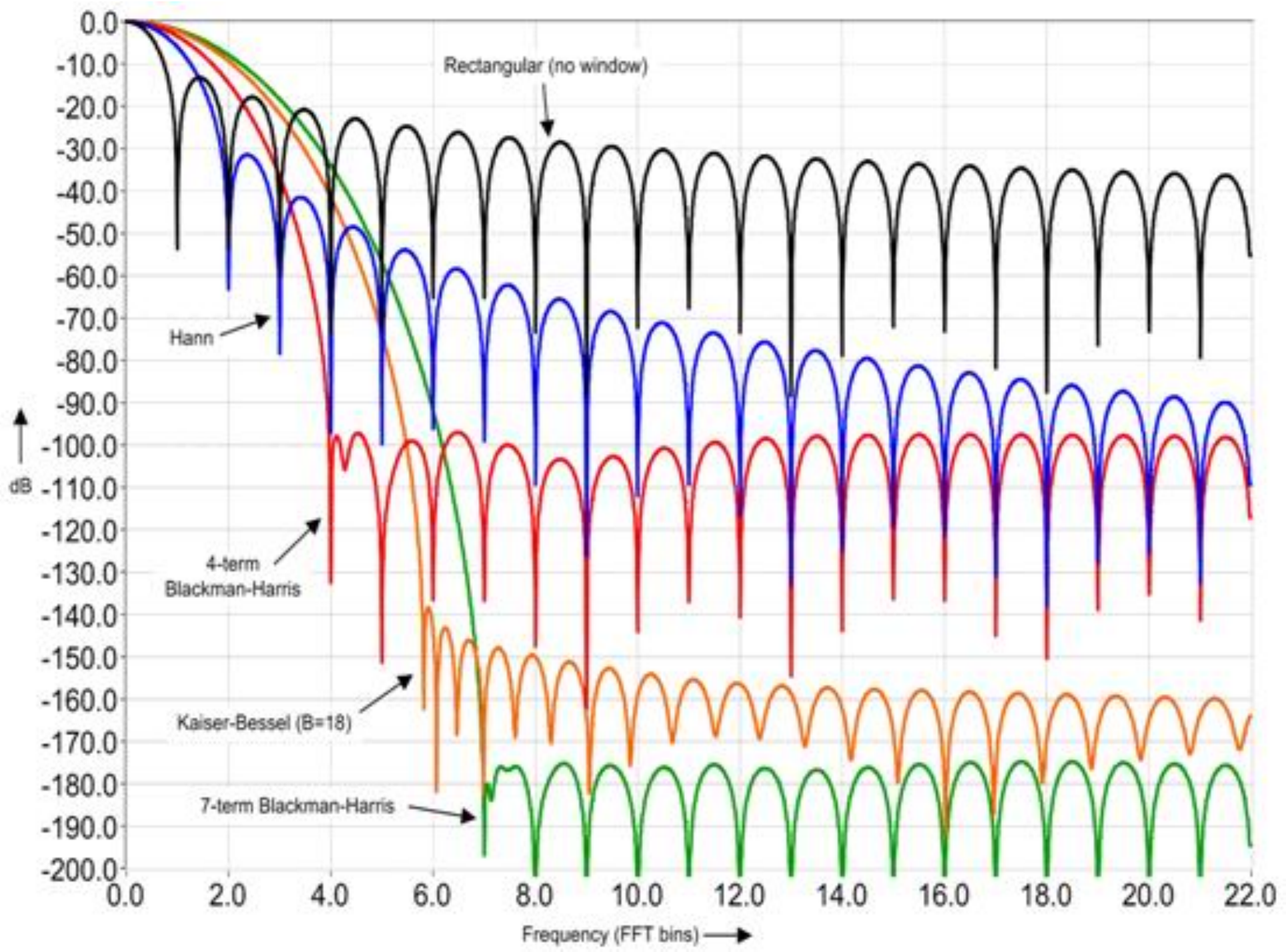}{fig:fftWndw} {Typical window functions to suppress strong sidelobes}{[t]}

\subsubsection{Gating}
The gating option of the VNA allows to eliminate or select parts of the time-domain signal,
provided they are reasonably well separated in the time-domain trace. 

For example, the already mentioned
cable, connecting to the VNA port, is assumed to have an internal irregularity at a certain position.
By suitable selection of a time-domain gate (highlighted in \refFig{fig:fftGate} from $t\approx 18$~ns
to $t\approx 26$~ns), the desired portion of the time domain trace (here, the total reflection at the open cable end)
can be separated from the the rest of the trace (set to zero).
This allows an analysis, e.g.\ by transformation back to the frequency domain, of the interesting part of the
circuit without influence of multiple reflections and perturbations from discontinuities, etc. (de-embedding). 
For transmission
measurements, usually the \emph{first} arriving pulse in the time domain is selected, thus suppressing the
effect of all following reflections and related signals. For reflection measurements, the first, but also following
pulse response in the time-domain trace may be selected. 

\Figure{0.6}{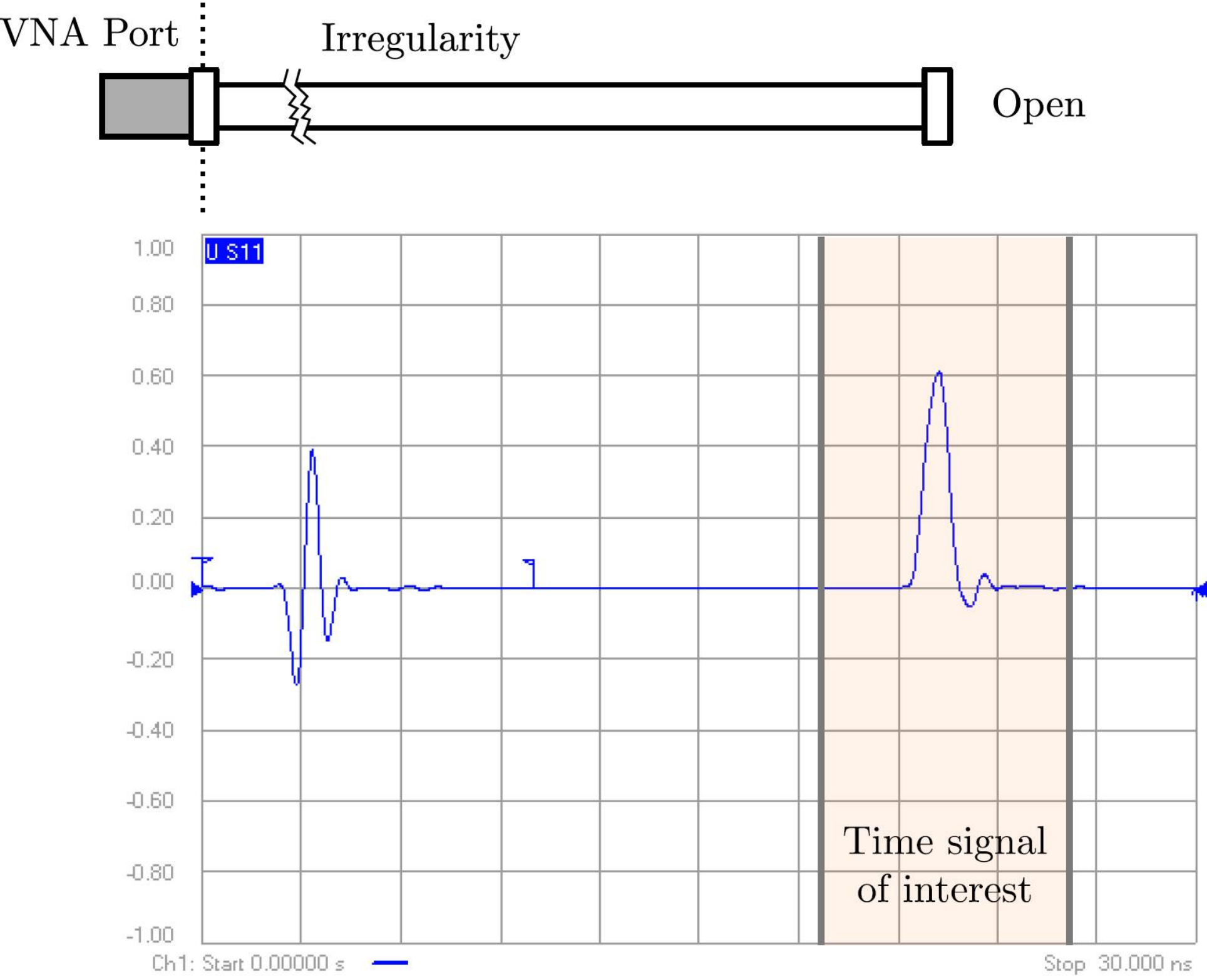}{fig:fftGate}{Only the signal in a certain time window is of interest. After selection, the FFT of this window will be calculated. Here the real values of the synthetic impulse response are shown on a linear scale.}{[t]}

The implemented time-domain gating function is not a ``brick wall'', but a soft switch applying a weighting function
similar to the iDFT window function.
As it is a \emph{non-linear} operation, it may generate additional frequency components which were not present
in the original signal. As general practical guide line, the gate should not cut into a signal trace different
from zero.


\begin{figure}[p]
\centerline{
\scalebox{1.0}{
\includegraphics*[width=1.0\textwidth]{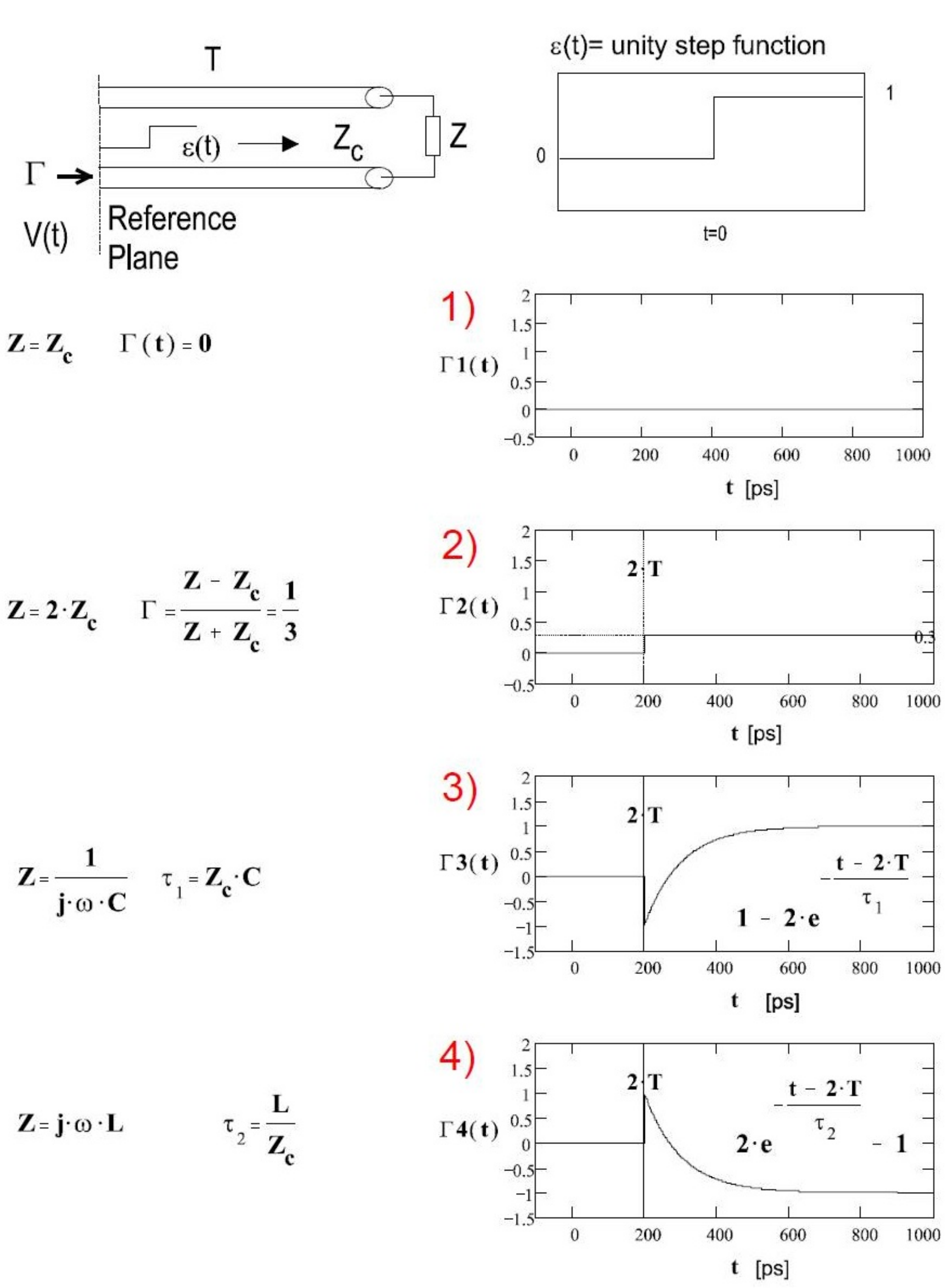}
} %
}
\caption{Examples of an arbitrary impedance, measured in TDR} %
\label{fig:fftExample} %
\end{figure} %

\subsubsection{Examples of synthetic pulse time-domain measurements}
A collection of measurement examples of simple DUTs are shown in \refFig{fig:fftExample}. For all
cases depicted, the VNA is set up in step response operation. The traces from top to bottom show:
\begin{enumerate}
  \item Matched load ($Z = Z_C$). As $\Gamma$ is equal to zero, the response is zero everywhere.
  \item Moderate (resistive) mismatch ($Z = 2 Z_C$, e.g.\ $100~\Omega$ in a $50~\Omega$ system). During the first
  200~ps the trace displays the well impedance-matched cable, following the reflection coefficient jumps to a positive,
  constand value due to the impedance mismatch.
  \item Capacitor. The TDR displays the capacitive load for a moment as a short circuit, and resumes with an
  exponential function, as the capacitor is charged. The final state is equivalent to an open circuit, as expected.
  \item Inductor. In the TDR the inductive load appears at $t = 200$~ps as an open circuit, followed by an exponential
  decay function. The steady state results in a short circuit, as the inductor is fully conducting.
\end{enumerate}
\subsubsection{Comparison to true time-domain measurements}
\label{ch:compare}
There is a wide range of applications for the discussed synthetic pulse time-domain technique. 
A VNA in
time-domain low-pass step mode has a very similar range of applications as a 
TDR sampling oscilloscope.
However, the synthetic pulse method is limited to strictly linear systems, 
therefore the analysis
of transient or non-linear systems, e.g.\ settling response of a microwave oscillator 
after power up would not give very meaningful results. 
In other words, for highly
non-linear and time-varying DUTs true time-domain measurements, based on pulse generators and oscilloscopes 
are still indispensable, e.g.\ an air traffic radar system, where we have linear but time-varying conditions.

The dynamic range of a typical
sampling oscilloscope is limited to about 60 to 80~dB with a maximum input signal of 1~V and a noise floor
around 0.1 to 1~mV (typical broadband oscilloscope). 
The dynamic range of the VNA is $>100$~dB,
allowing similar maximum input
levels of approximately +10~dBm (some VNAs allow +20~dBm). 
Both instruments are using
basically the same kind of detector, either a balanced mixer (four diodes) 
or a sampling head (two, four or six diodes),
but the essential difference lies in the noise floor and the average signal power 
arriving at the receiver input. 
In
case of the VNA the measurement is based on a continuous-wave (CW) signal 
with bandwidth of a few Hz, 
and thus can obtain with appropriate filtering a very good signal-to-noise ratio\footnotemark.
\footnotetext{Remember the thermal noise is proportional to measurement bandwidth. Its density at room
temperature is --174 dBm/Hz.}

A traditional sampling oscilloscope acquires the data during a short time with a rather low repetition rate 
(typically around 100~kHz up to a few MHz), with all the thermal noise power spread over the entire 
frequency range (typically 20--50 ~GHz bandwidth). 
With this low average signal power (around a microwatt) the signal spectral density
is orders of magnitude lower compared to the VNA measurement procedure (it acquires signals continuously), 
which explains the large difference in dynamic range (even without gain switching).

A more detailed discussion about time-domain reflectometry with vector network analysers can be found in
\cite{src:TDA}.

\subsection{Calibration methods}
The hardware of even an ``ultra-modern'' VNA is not perfect, e.g.\ the internal source
is not perfectly impedance matched to $50~\Omega$ (over the entire frequency range), 
its internal directional couplers have a finite directivity, since there exists no ideal (infinite) directivity in practice,
and finally the coaxial cables between VNA and DUT ports have frequency-dependent attenuation (dispersion) 
effects. 

This calls for a calibration to compensate all these unwanted effects, to guarantee a precise, instrument independent
analysis of the DUT.
There are several calibration procedures to eliminate some, or all of the mentioned deficiencies. The
easiest is called the ``response calibration'', typically applied for transmission, rarely for reflection measurements.
It basically is a $S_{21}$ (or $S_{12}$) transmission measurement of a quasi ``zero length'' ideal transmission-line, 
by connecting the two cable ends of the two-port VNA with each other.
For the given VNA setting, i.e.\ start / stop frequency, \# of freq.\ points, resolution bandwidth, power level, etc., magnitude and phase are acquired and stored as $S_{21 \text{reference}}$ 
in the non-volatile memory for each frequency point.   
Now, a DUT can be connected between the cable ports, with the connectors serving as \emph{reference planes}
of the calibrated system (VNA plus cables).
In calibrated mode the VNA performs: 
\Equation{equ:respCal}{S_{21 \text{DUTcal.}} = \frac{S_{21 \text{DUTmeas.}}}{S_{21 \text{reference}}},}

However, this simple calibration procedure
eliminates essentially the frequency-dependent losses and phase-transfer functions 
of the test cables only.
But, the mismatch between cable and generator, and the impact of the finite directivity 
are still present.
A more sophisticated, and widely popular calibration technique for the 
reflection measurements needs to
be performed: 
the open, short and match technique. \newline
This technique covers the three independent error sources mentioned above: 
finite directivity, generator mismatch and the transfer function of the cables.


\Figure{0.4}{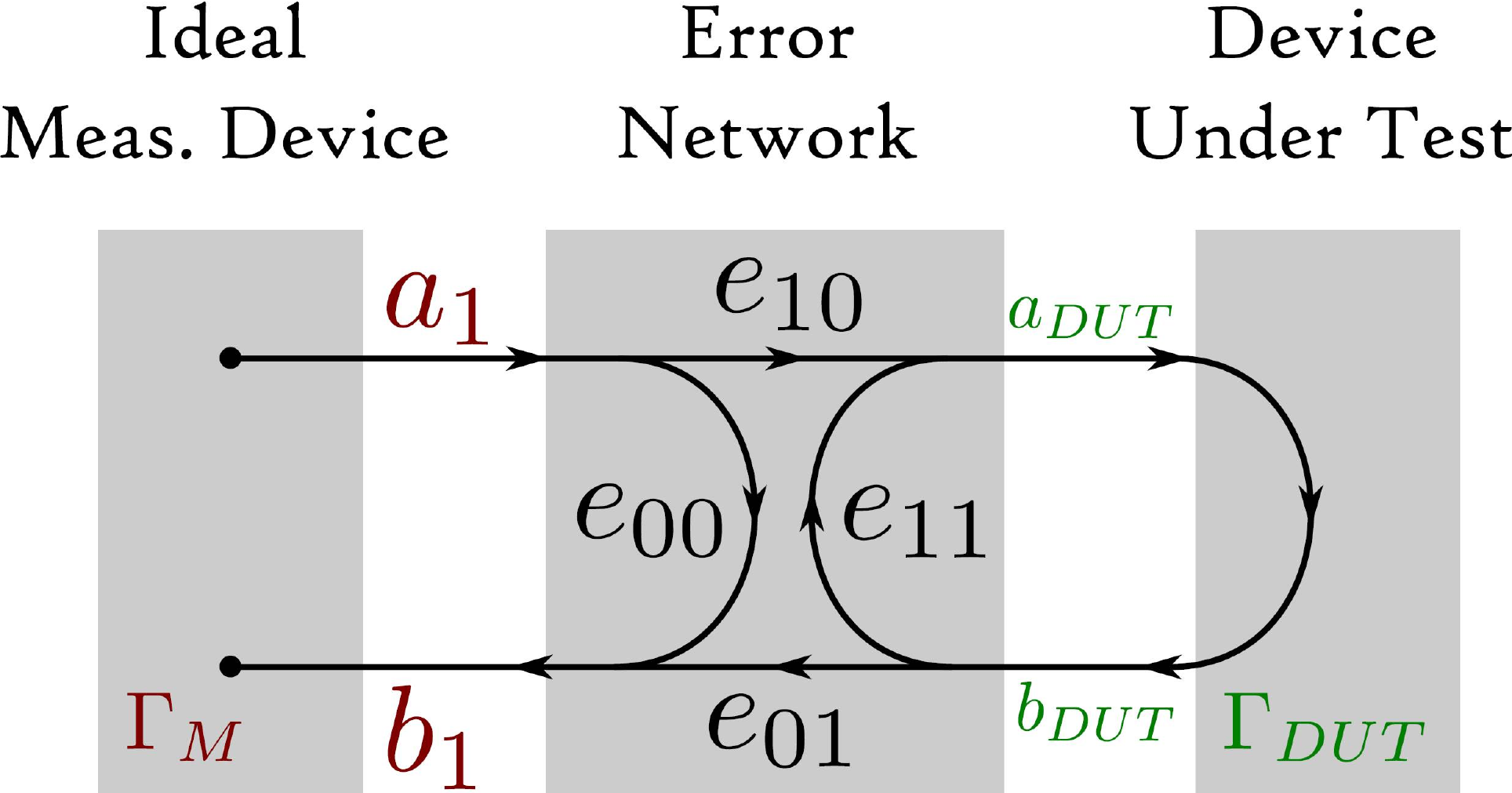}{fig:calErrorNw}{Error model of a VNA. The parameters $e_{xx}$ of the error network are determined by the calibration procedure and used to determine the true (corrected) result ($\Gamma_{\rm DUT}$) 
based on the measured result ($\Gamma_{\rm M}$).}{[t]}

The VNA applies an internal error model, shown in \refFig{fig:calErrorNw}. The measured raw data acquired
by the instrument ($\Gamma_{\rm M}$) is distorted by certain systematic errors. 
These errors are 
modeled via four parameters: $e_{10}$, $e_{00}$, $e_{01}$, $e_{11}$, based on the error network model
of \refFig{fig:calErrorNw}.
$e_{nn}$ are in general complex
and frequency dependent parameters, furthermore $e_{10} = e_{01}$.
The error parameters are extracted and stored when performing a suitable calibration method, i.e.\ open, short, match, 
such that the true value of the DUT ($\Gamma_{\rm DUT}$) is calculated and presented accordingly.
In simple terms, we need to carry out three independent measurements for each frequency point, 
to solve three coupled equations with three complex unknowns.
These error terms represent the above-mentioned effects as listed in \refTbl{tbl:eTerm}.

\begin{table}[t]
\caption{Interpretation of VNA error terms}
\begin{center}
\begin{tabular}{ll}
\hline
\hline
\bfseries Error term & \bfseries Interpretation\\
\hline
$e_{10}$ & Reflection tracking\\
$e_{00}$ & Directivity\\
$e_{11}$ & Test-port match\\
\hline
\hline
\end{tabular}
\end{center}
\label{tbl:eTerm}
\end{table}

The unknowns of the error network are determined applying a calibration measurement with three different, 
but known, calibration DUTs. These
calibration DUTs do not need to be perfect, only the electromagnetic properties need to be known with great
precision. The tabulated complex, frequency-dependent S-parameters of the calibration standards are
provided by the manufacturer of the calibration hardware (they are often referred as calibration kit), and
are stored in the VNA memory as calibration kit reference data.
Usually the calibration DUTs represent an open circuit, a short circuit and a matched load (termination), 
enabling the VNA to
determine the frequency-dependent error model. 
This is altered if different test cables are used, or if the
VNA settings are modified, and would require a re-calibration under those circumstances.
Now the VNA continuously applies the error correction during the DUT measurement, 
and the \emph{reference plane} is ``moved'' to the end of the test cables.
Only the DUT networks ``behind'' the reference plane are taken into account for the measurement.

The impact of the VNA calibration is demonstrated
in \refFig{fig:calib}, which presents a $S_{11}$ measurement of a high-quality $50~\Omega$
termination, with and without VNA calibration. 
For an ideal termination, no reflection should be present, i.e.\ $S_{11}=0\equiv -\infty$~dB. 
In this example the calibration of the VNA improves the measurement quality by 20~dB! 
In case of a short (total reflection, $S_{11}=1\equiv 0$~dB), a non-calibrated $S_{11}$ response 
typically displays a residual with values of a fraction of a dB, up to a few dB below the 0 dB line (same for an
open); after calibration these error reduces to a few millidecibels.

\Figure{0.7}{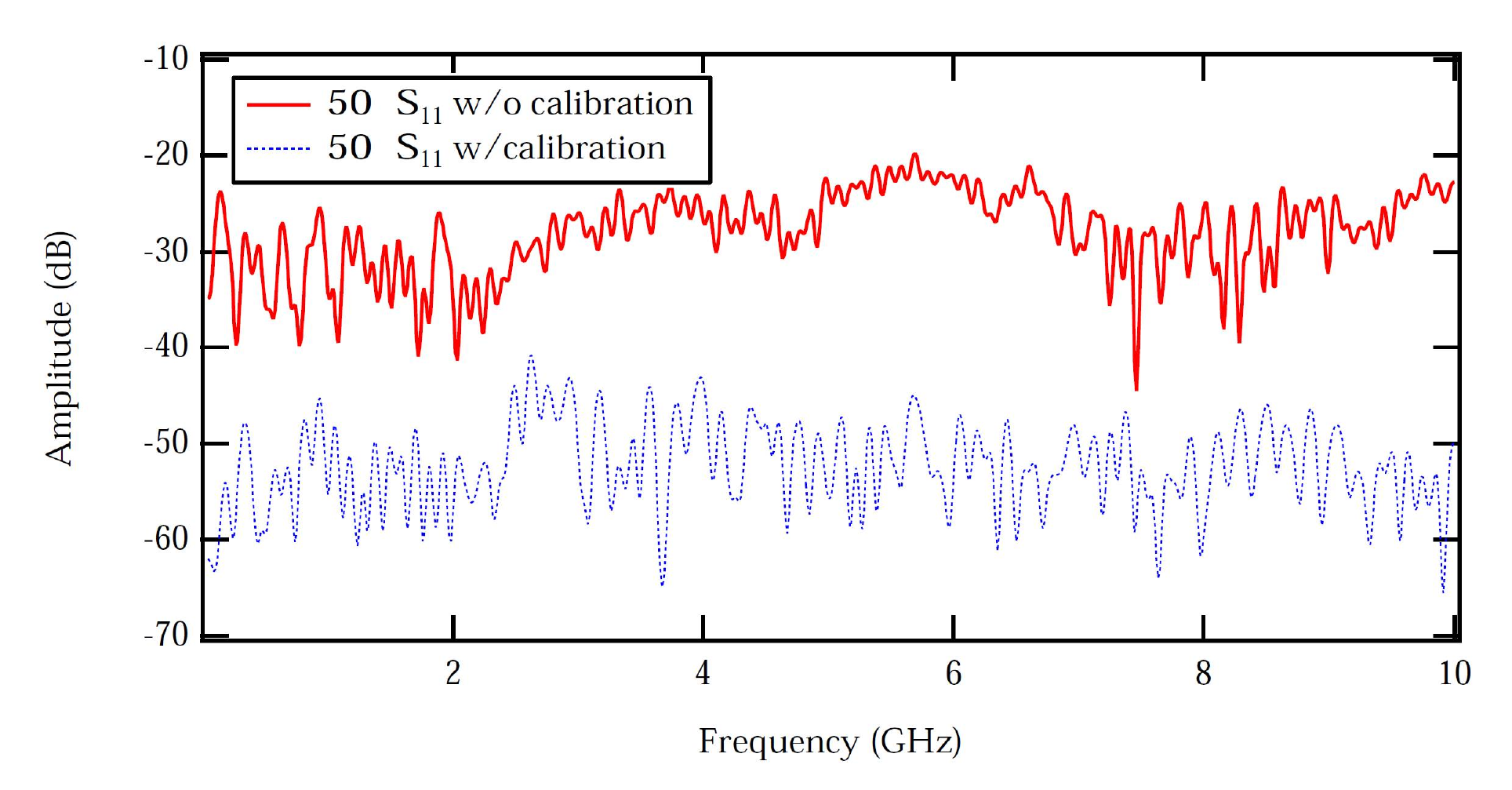}{fig:calib}{$S_{11}$ measurement of a $50~\Omega$ termination with and without calibration. The calibration provides 20 dB improvement for this frequency range.}{[t]}

So far, we have covered the ``response calibration'' and the ``complete one-port calibration''. 
To perform completely error-corrected transmission measurements, the ``full two-port calibration''
procedure has to be applied.
Therefore, the error model is expanded to include the errors from the receiving port,
requiring a calibration of each port based on the just discussed ``complete one-port calibration'' method.
Also, for transmission, we need two
standards, i.e.\ the ``response calibration'' and the ``isolation calibration'', however, latter often may be omitted.
In summary, the ``full two-port calibration'' consists out of a ``complete one-port calibration'' procedure for each port,
which requires open, short and match standards, plus the  ``response calibration'' and eventually the
``isolation calibration''.
In total eight calibration measurements have to be performed to bring the VNA into the desired \emph{CAL} status.

\begin{figure}[t]
\begin{center}
\subfloat[ ]{
\includegraphics[width=0.35\textwidth]{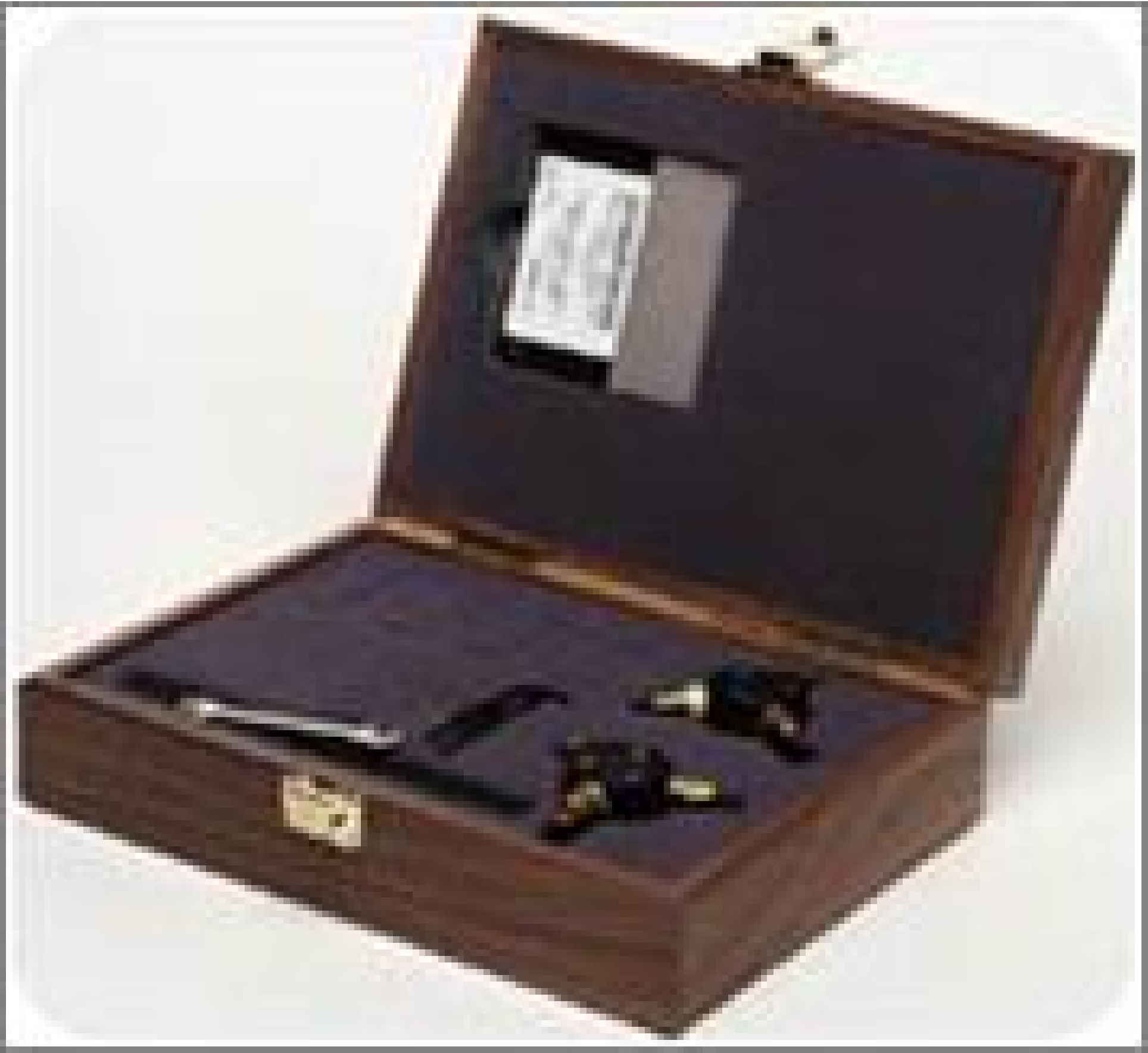}
\label{fig:calKitsa}
}
\subfloat[ ]{
\includegraphics[width=0.3\textwidth]{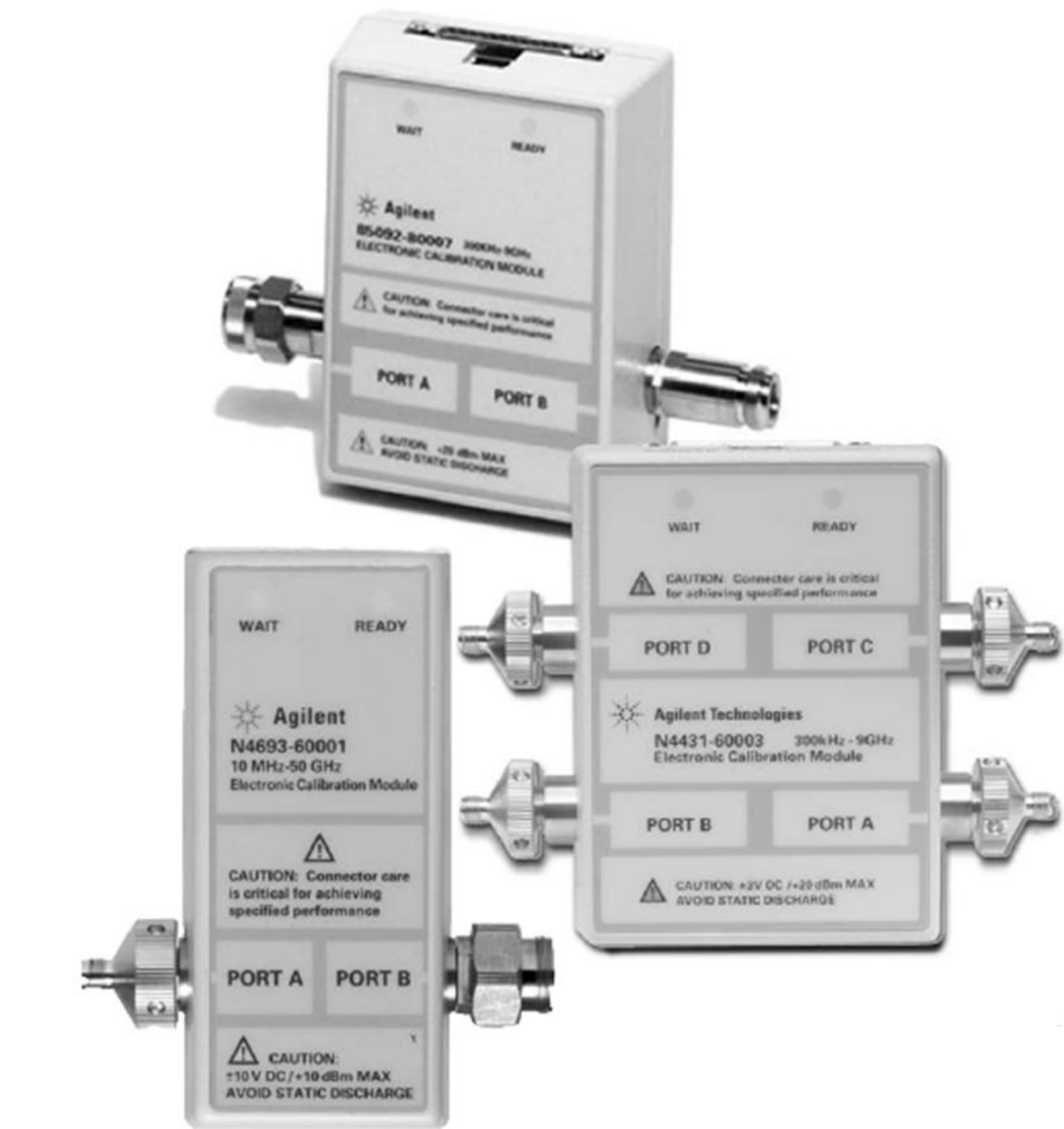}
\label{fig:calKitsb}
}
\caption{Typical calibration kits for a VNA: (a) manual (open, short, match); (b) electronic}
\label{fig:calKits}
\end{center}
\end{figure}
For measurements on devices with standard coaxial connectors, e.g.\ SMA or N-type,
calibration standards such as a termination, an open and a short circuit are available (shown in
\refFig{fig:calKitsa}). 
As mentioned, to successfully perform the calibration procedure for the reflection coefficient, the tabulated
values, representing the electromagnetic properties of the calibration standards, 
has to be present in the VNA.
Obviously, the tabulated parameters of the calibration kit does not have an infinite frequency resolution.
The instrument applies an interpolation procedure if the selected frequency points are not exactly at
the tabulated values of the calibration kit.

The calibration technique described so far is a well established industry standard for RF and microwave
VNA measurements.
However, it has a substantial disadvantage for the user: it is tedious and time consuming, 
in particular if a calibration of a multiport VNA is required.\newline
Already for the full two-port calibration requires eight calibration measurements to satisfy the
eight-term error model. The manual procedure of connection and de-connection of the calibration
standards is time consuming, boring, and prone to errors. The situation becomes even worse when performing
a full four-port calibration (32 connections and de-connections of standards). For this reason, the
electronic calibration kit method is available and now very popular. For this procedure, each port is
connected via the measurement cable to the electronic calibration box (shown in \refFig{fig:calKitsb}), 
which holds the different calibration standards, and switches them automatically controlled by the VNA. 
This method enables to perform a full four-port calibration in less than a minute.
Again, like for the manual calibration method, the standards do not need to be perfect, but well known,
reproducible (switching) and stable. More details are found in \cite{src:fundVNA,src:fundVNA2}.


\subsection{1 \mbox{dB} compression point measurement}
\label{1dBsect}
A single tone sine-wave source is connected to the input of an amplifier and its amplitude level is gradually increased versus time.
Monitoring the output of this amplifier, we notice a proportional dependence between input and output powers for
small signal levels. 
This proportionality is referred as the linear gain factor. 
For higher input signal
levels, this relationship does not hold any more, since the amplifier is not a perfectly linear system,
and suffers from ``saturation'' effects. 
A fraction of the
output power will appear at other frequencies, which are higher order harmonics of the input signal.
Typically the second and third harmonics are dominant, and the related signal distortion is referred as harmonic distortion.
In parallel, we observe a \emph{compression} of the gain for the fundamental signal.
The actual gain falls off below the small-signal gain response (\refFig{fig:1dBcompr}). 
If this deviation amounts to
1~dB, we have reached the ``1~dB compression point''.
Typically the industry refers to the output power, when specifying the 1~dB compression point for their RF products.
\Figure{0.45}{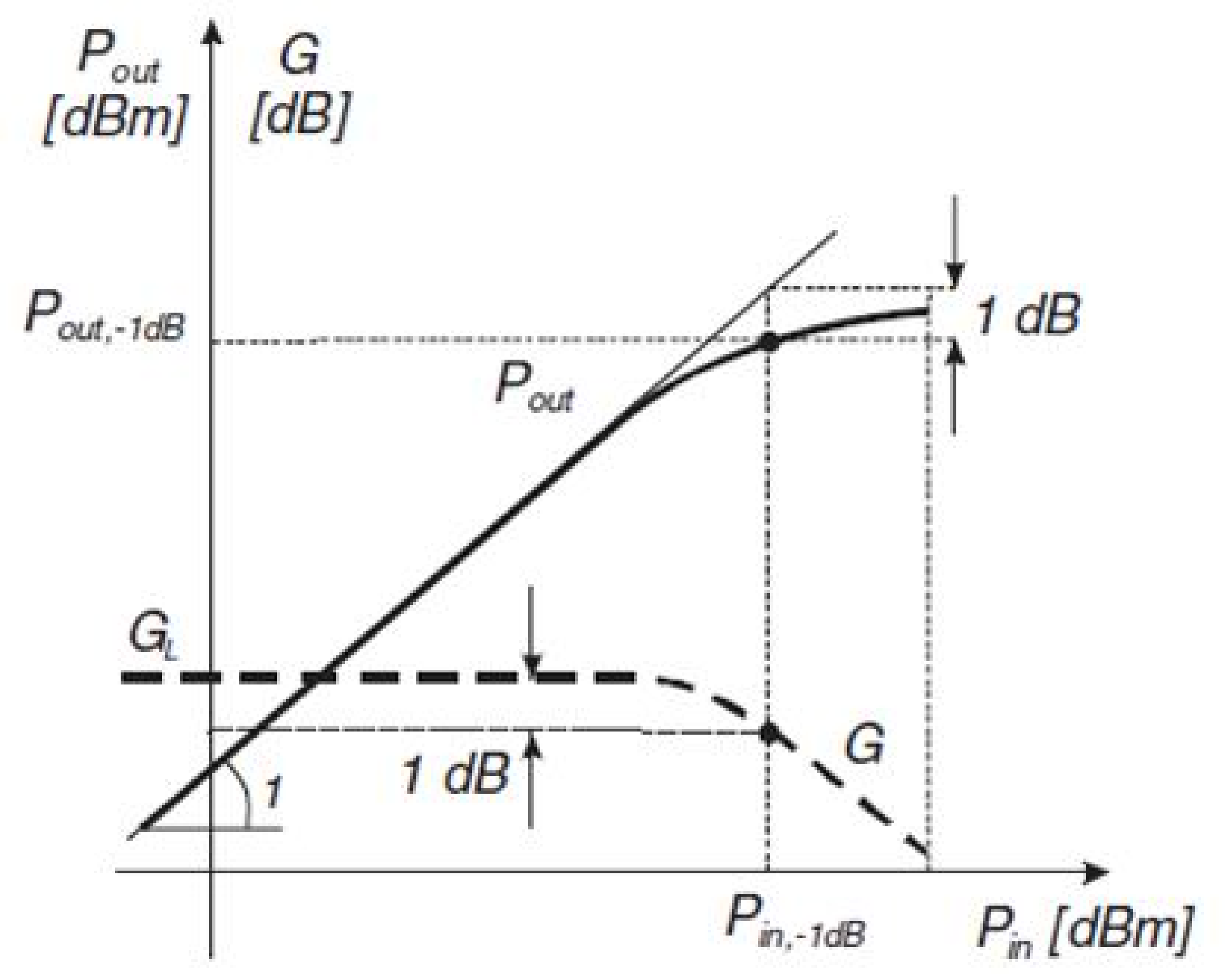}{fig:1dBcompr}{Definition of the 1~dB compression point for an amplifier: 
input vs.\ output power at the point where the power level falls below 1~dB from its (linearly) predicted value.}{[htb]}

The 1~dB compression point is an important figure of merit, used to characterize the linearity of
a RF system, in particular the performance of small-signal and power amplifiers. 
It can be comfortably 
measured with most VNAs in CW mode, i.e.\ choosing a single frequency
and performing a power sweep.
In power sweep mode, the instrument displays a trace similar as shown in \refFig{fig:1dBcompr}.

\section{Introduction to the Smith chart}

Even with today's availability of computer-aided simulation and circuit simulation software suites, 
the \textit{Smith} chart is still a very valuable and important tool that facilitates an 
interpretation of the (half) complex impedance plane with respect to the S-parameters, 
and the related calculations and measurements. 
This section gives a brief overview of the concept, and more importantly, of how to use the chart. 
Its definition, as well as an introduction of how to navigate on the chart are illustrated. 
Some typical examples illustrate the broad range of applications of the \textit{Smith} chart.
\begin{figure}[t]
\centering
\includegraphics[width=0.6\textwidth]{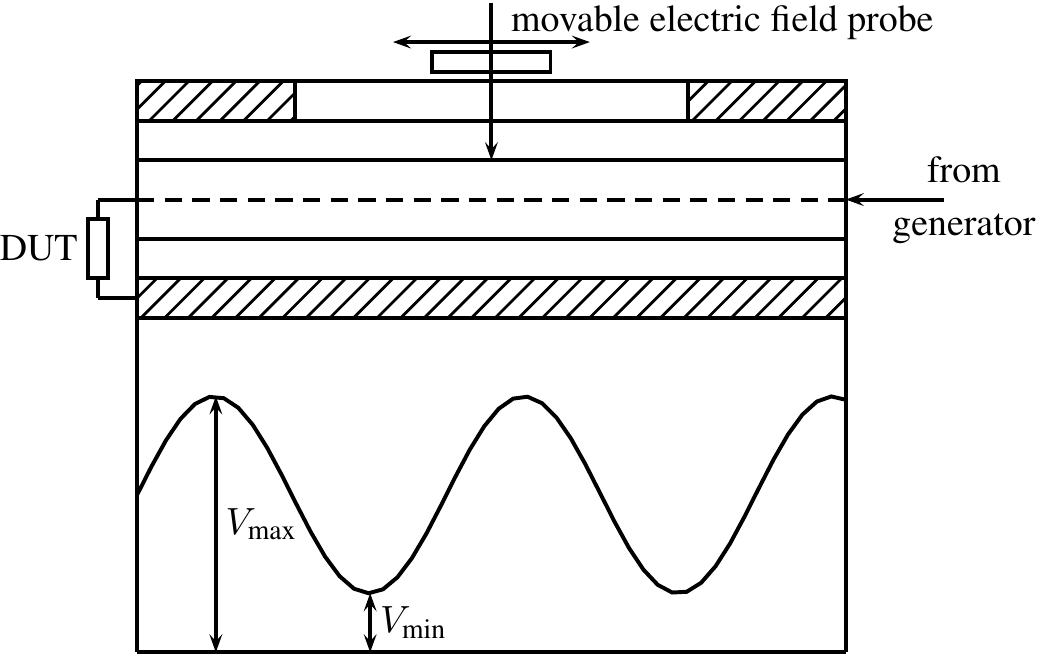}
\caption{Schematic view of a measurement set-up used to determine the reflection coefficient as well as the voltage standing wave ratio of a device under test (DUT) \cite{meinkegundlach}.}
\label{coax}
\end{figure}
\subsection{Voltage standing wave ratio ($\mathit{VSWR}$)}
\label{VSWRsect}
With modern RF measurement equipment available today it is rather easy to precisely measure 
the reflection factor $\Gamma$, even for complicated networks. 
In the ``good old days'' though, this was performed by measuring the electrical field
strength\footnote{The electrical field strength was used, since its measurement was 
considerably easier than that of the magnetic field.} along a slotted coaxial line, 
which has a longitudinal slit to allow a small field probe to be slided to any location 
along the line (Fig. \ref{coax}).
This electric field probe, protruding into the field region of the coaxial line near the outer conductor, picked up an E-field signal, which was displayed on a microvoltmeter after rectification 
via a microwave diode. 
While moving the probe, field maxima and minima, as well as their position and spacing 
where recorded. 
From this information the reflection factor $\Gamma$ and the voltage standing wave 
ratio ($\mathit{VSWR}$ or $\mathit{SWR}$) were determined:
\begin{itemize}
	\item $\Gamma$ is defined as the ratio of the electrical field strength $E$ of the reflected wave versus the forward-traveling wave:
\begin{equation}
	\Gamma = \frac{E\text{ of reflected wave}}{E \text{ of forward-traveling wave}}.
\label{eq:1}
\end{equation}
	\item The $\mathit{VSWR}$ is defined as the ratio of maximum to minimum measured voltages:
\begin{equation}
	\mathit{VSWR} = \frac{V_\text{max}}{V_{\text{min}}} = \frac{1 + |\Gamma|}{1 - |\Gamma|}.
\label{eq:2}
\end{equation}
\end{itemize}
Although today these measurements are far easier to conduct, the definitions of the aforementioned quantities are still valid. On top, their importance has not diminished in the field of microwave engineering, both reflection coefficient as well as $\mathit{VSWR}$ are still a vital part of the everyday life of a microwave engineer performing simulations or measurements.

\subsection{Definition of the \textit{Smith} chart}
\begin{figure}[t]
\centering
\includegraphics[width=0.8\textwidth]{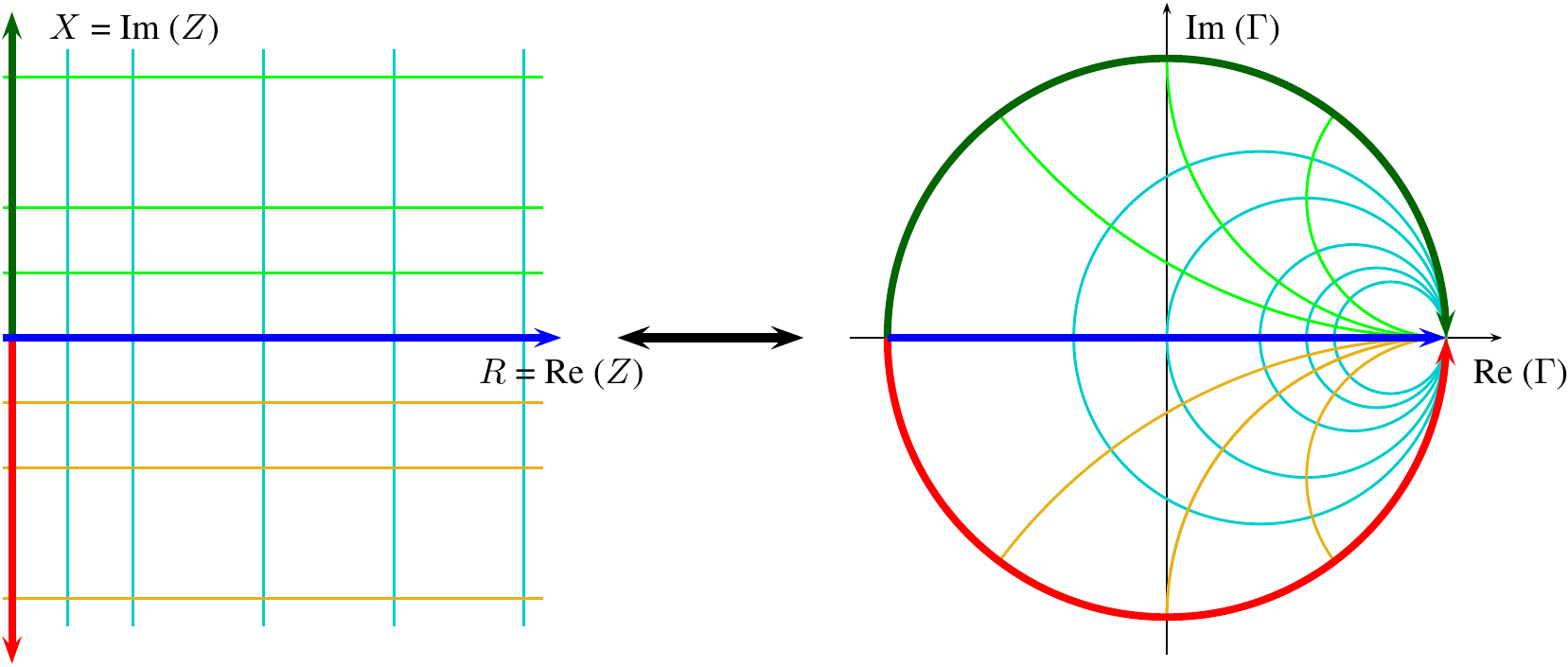}
\caption{Illustration of the \textit{Moebius} transformation from the complex impedance plane to the $\Gamma$ plane, commonly known as \textit{Smith} chart.}
\label{scbasic}
\end{figure}

The \textit{Smith} chart~\cite{smith2000} provides a graphical representation of $\Gamma$ that permits the determination of quantities like the $\mathit{VSWR}$, or the impedance of a device under test (DUT). It uses the bilinear \textit{Moebius} transformation, projecting the complex impedance plane on the complex $\Gamma$ plane:
\begin{equation}
	\Gamma = \frac{Z - Z_{\text{0}}}{Z + Z_{\text{0}}} \hspace{0.5cm}\text{ with }\hspace{0.5cm} Z = R + \text{j}\,X.
\label{eq:3}
\end{equation}
As shown in Fig. \ref{scbasic}, the half--plane with positive real part of impedance $Z$ is mapped to the interior of the unit circle of the $\Gamma$ plane.
\subsubsection{Properties of the transformation}
\begin{figure}[t]
\centering
\includegraphics[width=0.75\textwidth]{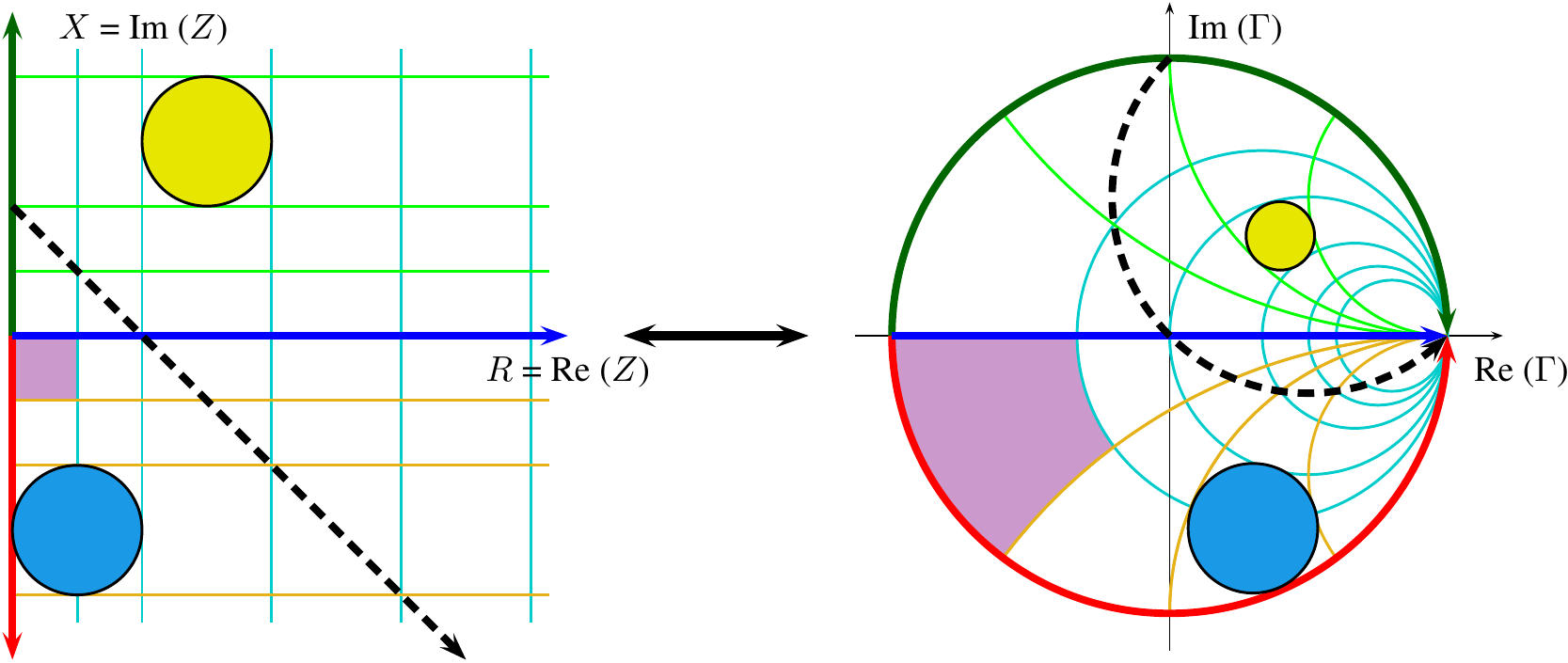}
\caption{Illustration of the transformation of basic shapes from the $Z$ to the $\Gamma$ plane.}
\label{prop}
\end{figure}

In general, this transformation has two main properties:
\begin{itemize}
	\item generalized circles are transformed to generalized circles (note that a straight line is nothing else than a circle with infinite radius and is therefore mapped as circle to the \textit{Smith} chart);
	\item angles are preserved locally.
\end{itemize}
Figure \ref{prop} illustrates how certain basic shapes transform between impedance and $\Gamma$ planes.
\begin{figure}[H]
\centering\includegraphics[width=0.75\linewidth]{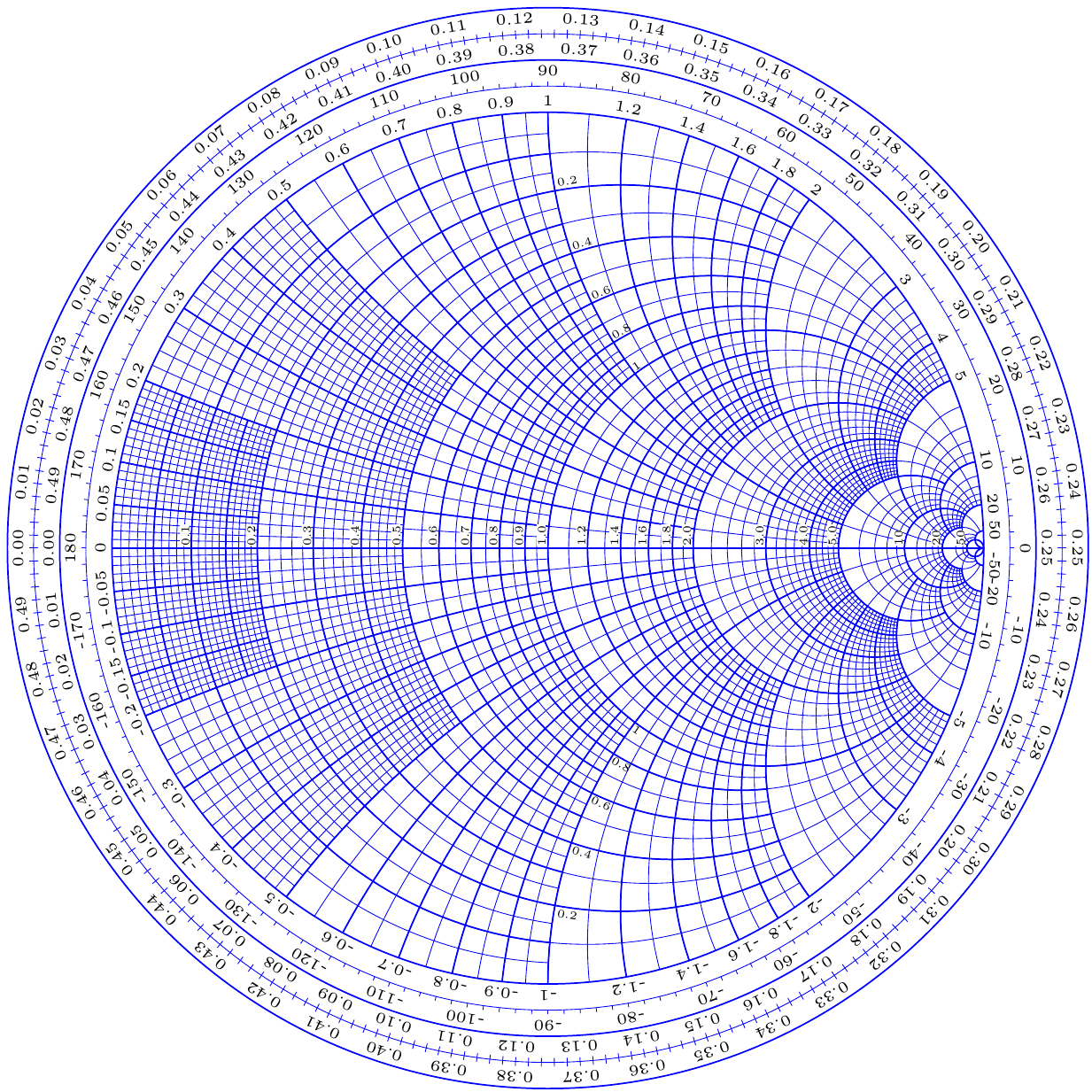}
\caption{Example of a typical \textit{Smith} chart}
\label{smith}
\end{figure}
\subsubsection{Normalization}
The Smith chart is usually normalized to a reference impedance $Z_{\text{0}}$ (= real):
\begin{equation}
	z = \frac{Z}{Z_{\text{0}}}.
\label{eq:4}
\end{equation}
This simplifies the transformation:
\begin{equation}
	\Gamma = \frac{z - 1}{z + 1} \hspace{0.5cm} \Leftrightarrow \hspace{0.5cm} z = \frac{1 + \Gamma}{1 - \Gamma}.
\label{eq:5}
\end{equation}
Although $Z_0 = 50~\Omega$ is the most common reference impedance (typical characteristic impedance of coaxial cables) and many applications use this normalization, any other real, positive value is valid. \textit{Therefore, it is crucial to check the normalization assumed, before using any chart.}

Being unfamiliar, the \textit{Smith} charts appears confusing at a first look, 
with a fine grid from the $Z$-plane mapped
to a dense grid of many circles on the chart (Fig. \ref{smith}).

\subsubsection{Admittance plane}
\begin{figure}[t]
\centering
\includegraphics[width=0.85\linewidth]{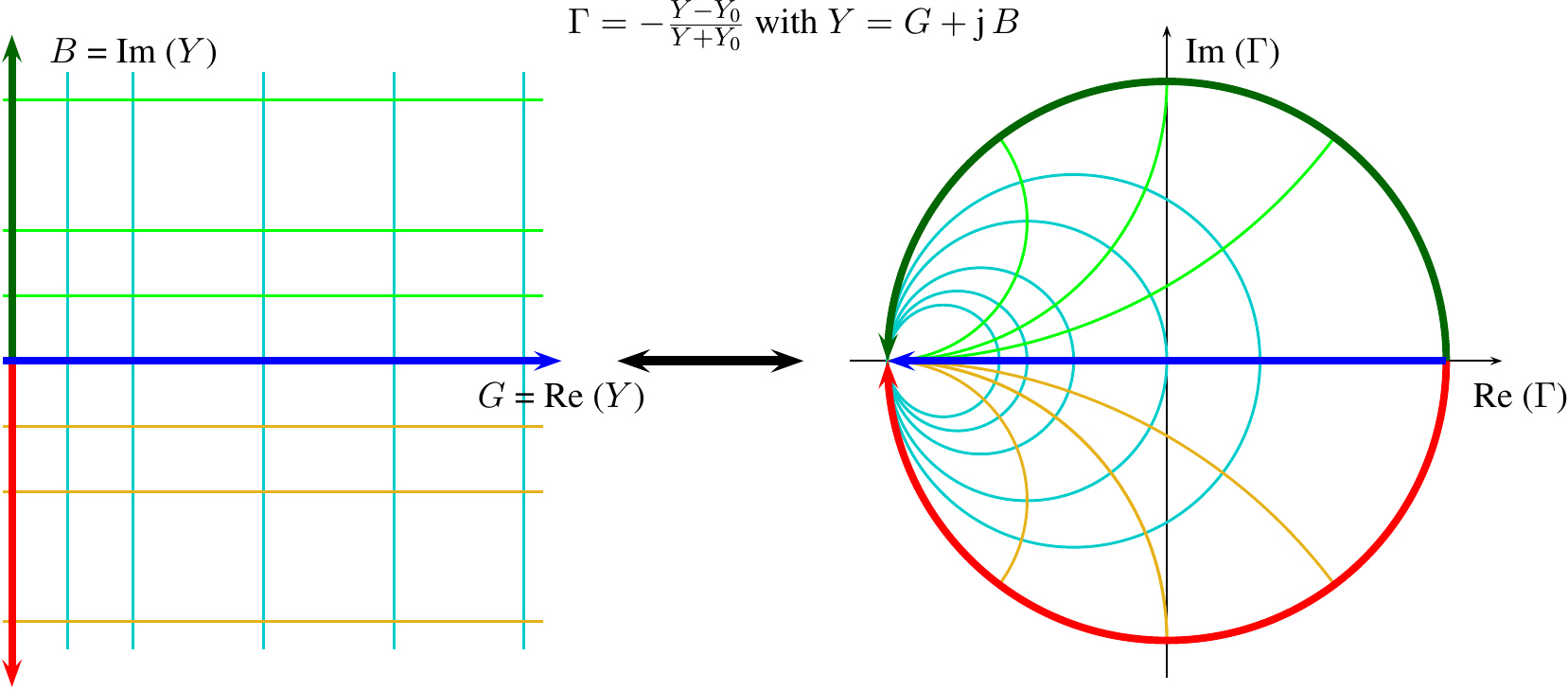}
\caption{Mapping of the admittance plane into the $\Gamma$ plane}
\label{admit}
\end{figure}

The \textit{Moebius} transformation which generates the \textit{Smith} chart also provides a mapping of the complex admittance plane ($Y = 1/Z$, or normalized $y = 1/z$) into the same chart:
\begin{equation}
	\Gamma = -\frac{y - \text{1}}{y + \text{1}} = -\frac{Y - Y_{\text{0}}}{Y + Y_{\text{0}}} = - \frac{1/Z - 1/Z_{\text{0}}}{1/Z + 1/Z_{\text{0}}} = \frac{Z - Z_{\text{0}}}{Z + Z_{\text{0}}} = \frac{z - \text{1}}{z + \text{1}}.
\label{eq:6}
\end{equation}
Using this transformation results in the same chart, but mirrored at the center of the \textit{Smith} chart (Fig. \ref{admit}).
Often both mappings, the admittance and the impedance plane are combined into one chart, which then looks even more overwhelming. For reasons of simplicity all illustrations in this article use only the mapping from the impedance to the $\Gamma$ plane.
%
\subsection{Navigation in the \textit{Smith} chart}
The representation of circuit elements in the \textit{Smith} chart is discussed in this section, starting with some important points inside the chart. The following examples of circuit elements illustrate their representation in the chart.
\subsubsection{Important points}
\begin{figure}[t]
\centering
\includegraphics[width=0.6\linewidth]{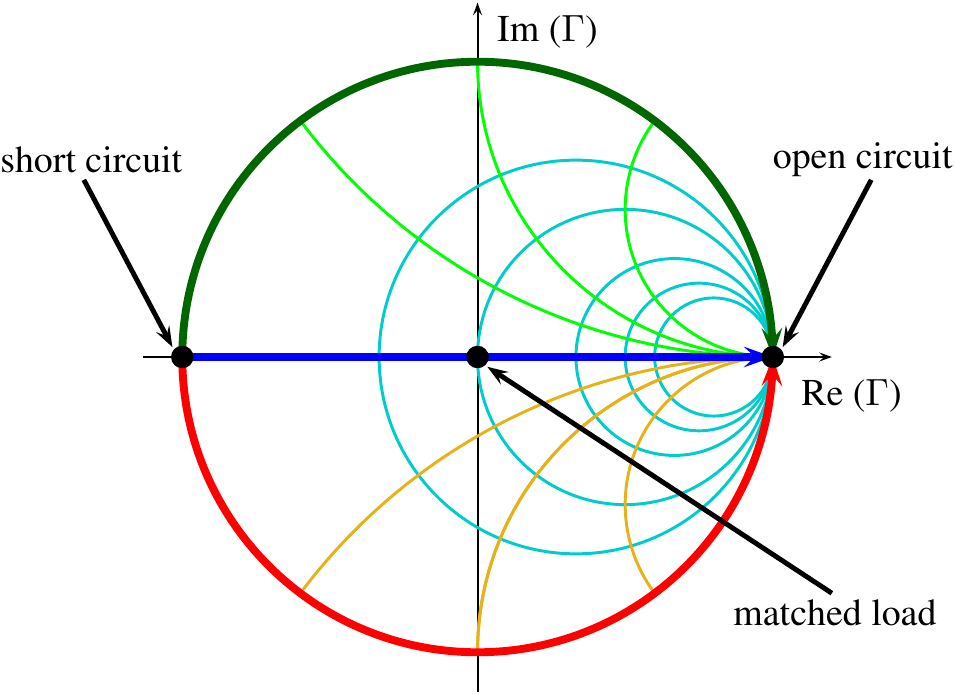}
\caption{Important points in the Smith chart}
\label{points}
\end{figure}

There are three important points in the chart:
\begin{enumerate}
	\item Open circuit with $\Gamma = 1, z \rightarrow \infty$.
	\item Short circuit with $\Gamma = -1, z = 0$.
	\item Matched load with $\Gamma = 0, z = 1$.
\end{enumerate}
They all are located along the real axis at the beginning and the end, which are also
on the outer circle (imaginary axis), and at the center of the \textit{Smith} chart (Fig. \ref{points}).
The upper half of the chart is ``inductive'', since it corresponds to the positive imaginary part of the impedance. The lower half is ``capacitive'', as it is corresponding to the negative imaginary part of the impedance.

\begin{figure}[t]
\centering
\includegraphics[width=0.4\linewidth]{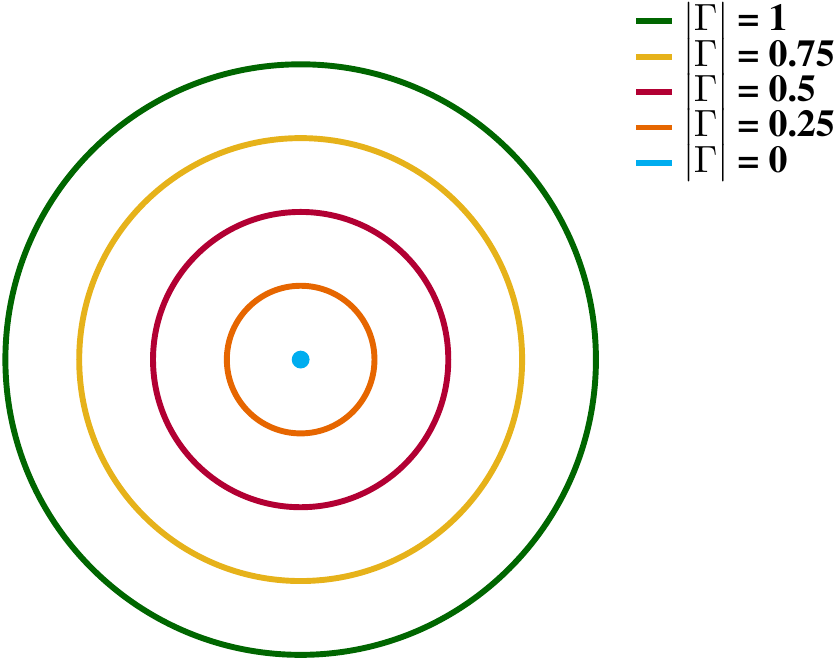}
\caption{Illustration of circles representing a constant reflection factor}
\label{concentric}
\end{figure}

Concentric circles around the center represent constant reflection factors (Fig. \ref{concentric}). Their radius is directly proportional to the magnitude of $\Gamma$; therefore, a radius of 0.5 corresponds to reflection of 3 dB (half of the signal is reflected), whereas the outermost circle (radius = 1) represents total reflection.
Evidently, matching problems are clearly visualized in the Smith chart, since a mismatch will lead to a reflection coefficient larger than 0, see Eq. (\ref{eq:7}).
\begin{equation}
	\text{Power into the load = forward power - reflected power: }P = \frac{1}{2}\left(\left|a\right|^{2} - \left|b\right|^{2}\right) = \frac{\left|a\right|^{2}}{2}\left(1 - \left|\Gamma\right|^{2}\right).
\label{eq:7}
\end{equation}
In Eq. (\ref{eq:7}) the European notation is used%
\footnote{The commonly used notation in the USA: power = $\left|a\right|^{2}$. These conventions have no impact on the S-parameters, but they are relevant for absolute power calculations. Since this is rarely used in context with \textit{Smith} chart gymnastics, the actual power definition used is not critical.}: power $= \left|a\right|^{2}/2$. Furthermore it should be noted, $(1 - \left|\Gamma\right|^{2})$ corresponds to the losses due to the impedance mismatch.

\begin{figure}[t]
\centering
\includegraphics[width=0.3\linewidth]{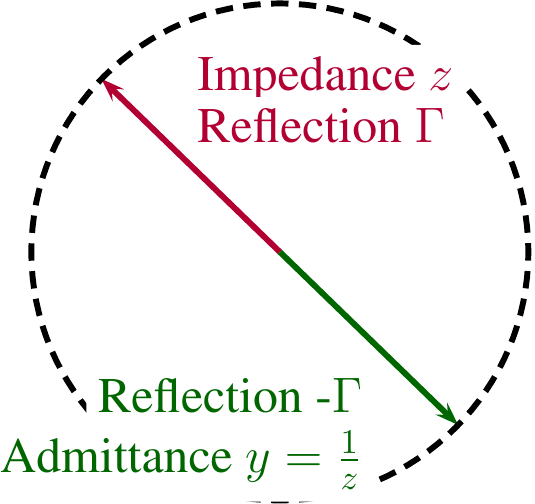}
\caption{Conversion of an impedance to the corresponding admittance in the \textit{Smith} chart}
\label{imptoad}
\end{figure}

Even though here we limit to the mapping of the impedance plane to the $\Gamma$ plane, 
The admittance is simple to determine, since
\begin{equation}
	\Gamma(\frac{1}{z}) = \frac{1/z - 1}{1/z + 1} = \frac{1 - z}{1 + z} = \left(\frac{z - 1}{z + 1}\right)\text{ or } \Gamma(\frac{1}{z}) = - \Gamma(z).
\label{eq:8}
\end{equation}
In the \textit{Smith} chart this fact is visualized as a 180$^{\circ}$ rotation of the vector of a given impedance (Fig. \ref{imptoad}).

\subsubsection{Impedance of simple, passive lumped element circuits}
\begin{figure}[t]
\centering
\includegraphics[width=0.4\linewidth]{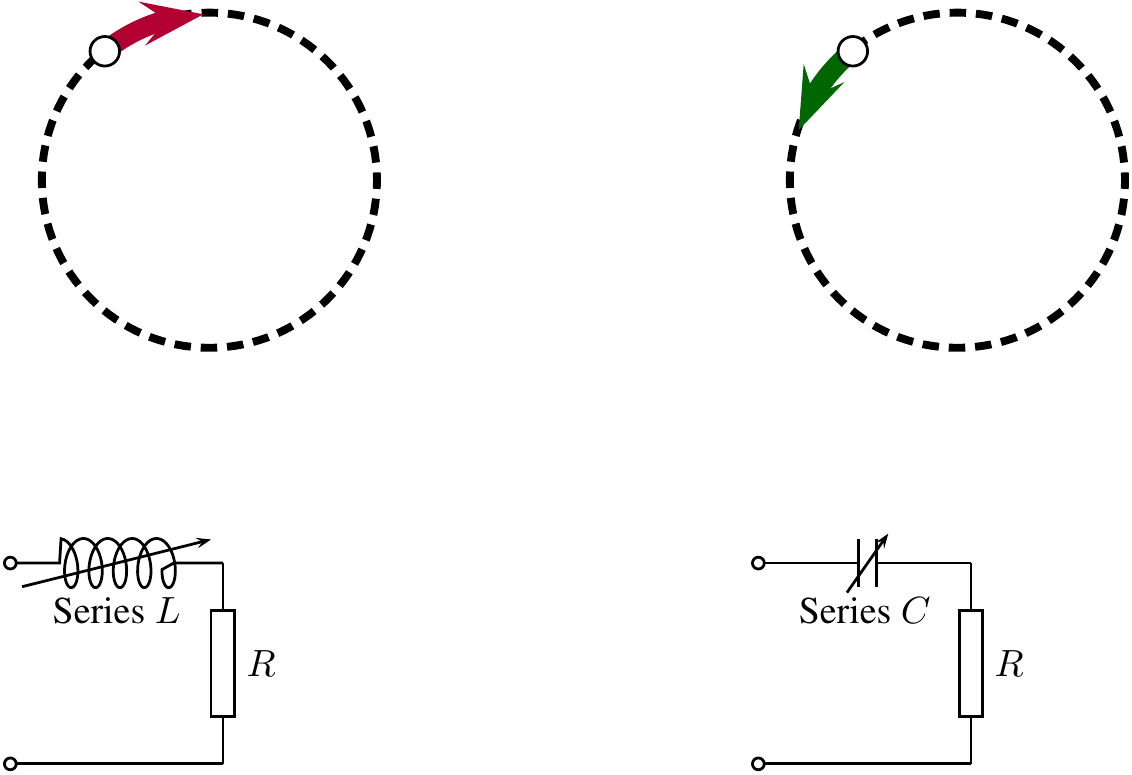}
\caption{Circular traces of reactances with varying value connected in series to a fixed impedance}
\label{series}
\includegraphics[width=0.4\linewidth]{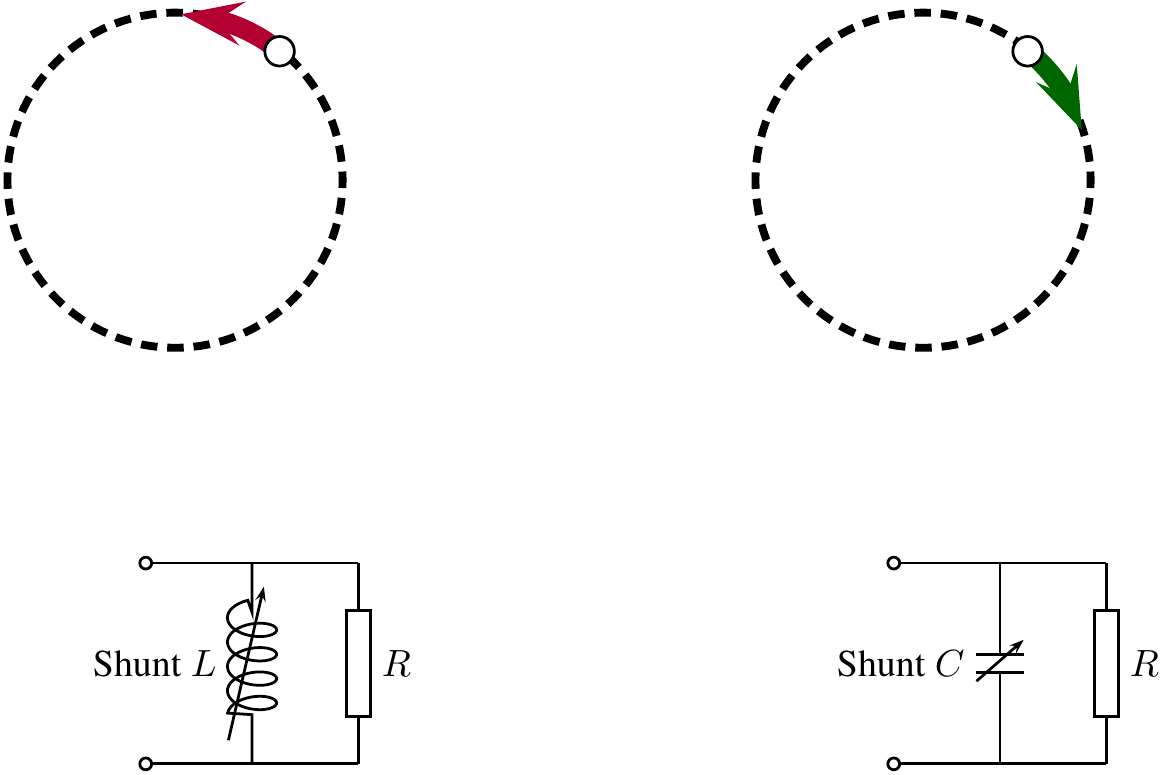}
\caption{Circular traces of reactances with varying value connected in parallel to a fixed impedance}
\label{para}
\end{figure}
\begin{figure}[t]
\centering
\includegraphics[width=0.4\linewidth]{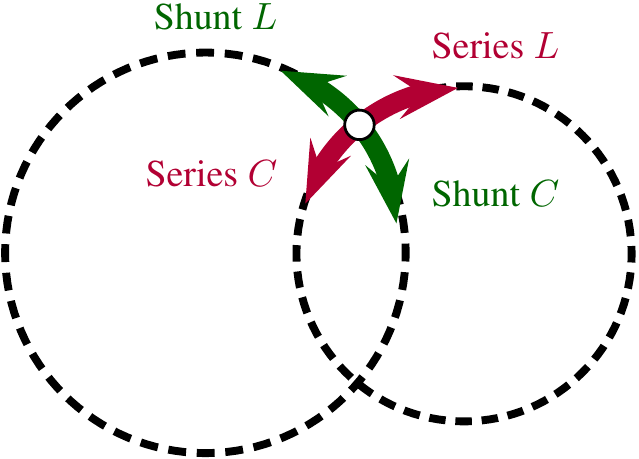}
\caption{Navigation in the \textit{Smith} chart when connecting reactive elements.}
\label{rule}
\end{figure}

Consider a simple passive circuit: a lumped, reactive element (inductance $L$, or capacitance $C$) 
of arbitrary value connected in series to an resistance $R$.
The corresponding signature of this circuit in the \textit{Smith} chart, varying the 
inductance resp.\ capacitance, 
is a circle. 
For a given type of impedance, the trace of this circle follows a clockwise (inductance), or anticlockwise
(capacitance) movement (Fig. \ref{series}).
If a lumped, reactive element is connected in parallel to $R$, the pattern is basically the same, 
but rotated by 180$^{\circ}$ (Fig. \ref{para}). 
It is equivalent to the discussed admittance mapping.
%
%
Summarizing both cases, results in a simple rule for the navigation in the \textit{Smith} chart: \\
\\
%
\textit{Reactive elements connected in series follow the trajectory of a circle in the impedance plane. Inductances move clockwise, capacitances move anticlockwise when increasing their value. 
Reactive elements connected in parallel follow a circular trajectory in the admittance plane, 
clockwise for capacitances, anticlockwise for inductances.}\\
\\
%
This rule is illustrated in Fig. \ref{rule}.
\subsubsection{Impedance transformation using a transmission-line}
The S-matrix of an ideal, lossless transmission-line of physical length $l$ is given by
\begin{equation}
	S = \left[
\begin{array}{cc}
	0 & e^{-j\beta l} \\
	e^{-j\beta l} & 0 \\
\end{array}
\right],
\label{eq:9}
\end{equation}
where $\beta = 2\pi/\lambda$ is the propagation coefficient at the wavelength $\lambda$ ($\lambda = \lambda_{\text{0}}$ for $\epsilon_{\text{r}} = 1$).

\begin{figure}[H]
\centering
\includegraphics[width=0.3\linewidth]{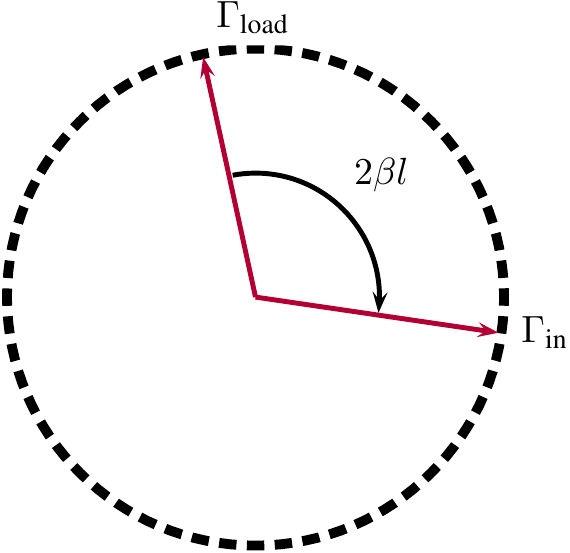}
\caption{Adding a lossless transmission-line of physical length $l$ to an impedance $Z_{\text{load}}$}
\label{transmissionline}
\end{figure}

The lossless transmission-line changes only the phase between its ports. 
Adding a short piece of, e.g.\  of coaxial cable in front of a load impedance, will turn the corresponding circle of $Z_{\text{load}}$ clockwise, which is effectively a transformation of the reflection factor $\Gamma _{\text{load}}$ (without line) to the new reflection factor $\Gamma _{\text{in}} = \Gamma _{\text{load}}e^{-j2\beta l}$. Graphically speaking, the vector corresponding to $\Gamma_{\text{in}}$ is rotated clockwise by an angle of 2$\beta l$ (Fig. \ref{transmissionline}).

The input impedance of a lossless transmission-line of characteristic impedance $Z_0$, terminated with $Z_{\text{load}}$ is given by:
\begin{equation}
Z_{\text{in}} = Z_0 \frac{Z_{\text{load}}+j Z_0 \tan(\beta l)}{Z_0+j Z_{\text{load}}\tan(\beta l)}
\label{eq:11}
\end{equation}
and the corresponding reflection coefficient follows as mentioned:
\begin{equation}
\Gamma _{\text{in}} = \Gamma _{\text{load}}e^{-j2\beta l}
\label{eq:refl}
\end{equation}
Depending on the values of $\beta$, $Z_0$, $Z_{\text{load}}$, and $l$, the input impedance will be quite different from the load impedance $Z_{\text{load}}$.
Special cases are:
\begin{itemize} 
\item $l=\lambda/2$: $Z_{\text{in}}=Z_{\text{load}}$
\item $l=\lambda/4$: $Z_{\text{in}}=Z^2_0/Z_{\text{load}}$ (impedance transformer)
\item $Z_{\text{load}}=Z_0$: $Z_{\text{in}}=Z_0$ (matched termination)
\item $Z_{\text{load}}=j X_{\text{load}}$: $Z_{\text{in}}=j X_{\text{in}}$ (reactive load $\Rightarrow$ reactive input impedance)
\item $l\ll\lambda$: $Z_{\text{in}}=Z_{\text{load}}$ (basically no line present)
\end{itemize}
Terminating a transmission-line with a short circuit, $Z_{\text{load}}=0$, simplifies Eq.~\ref{eq:11} to
\begin{equation}
Z_{\text{in}} = j Z_0\tan(\beta l)
\label{eq:tlshort}
\end{equation}
which results in an ``inductive'' or ``capacitive'' impedance behavior at the input, depending on the length of the line (see Fig.~\ref{tangens}).

Adding a transmission-line of length $\lambda/4$ interestingly results in a change of $\Gamma$ by a factor $-1$:
%
%
%
\begin{figure}[t]
\centering
\includegraphics[width=0.4\linewidth]{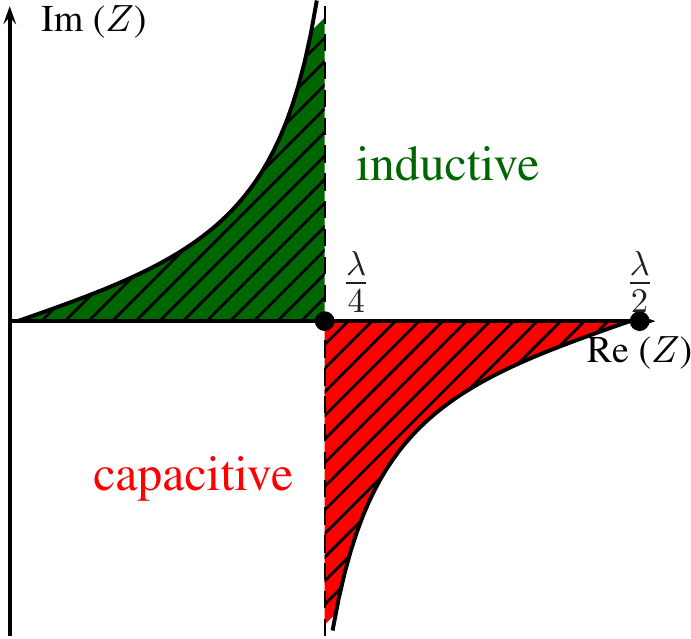}
\caption{Impedance of a transmission line as a function of its length $l$}
\label{tangens}
\end{figure}
%
%
\begin{equation}
	\Gamma_{\text{in}} = \Gamma_{\text{load}} e^{-j 2\beta l} = \Gamma_{\text{load}} e^{-j 2(\frac{2\pi}{\lambda}) l} \stackrel{l=\frac{\lambda}{4}}{=} \Gamma_{\text{load}} e^{-j\pi} = -\Gamma_{\text{load}}.
\label{eq:12}
\end{equation}
Again, this is equivalent to inverting an impedance $z$ to its admittance $1/z$, or the clockwise rotation of the impedance vector by 180$^{\circ}$. Especially when starting with a short circuit ($Z_{\text{load}}=0\,\Rightarrow\, -1$ in the \textit{Smith} chart), adding a transmission line of length $\lambda/4$ transforms it into an open circuit ($+1$ in the \textit{Smith} chart), and vice versa.

\subsubsection{Two-port examples}
\begin{figure}[t]
\centering
\includegraphics[width=0.4\linewidth]{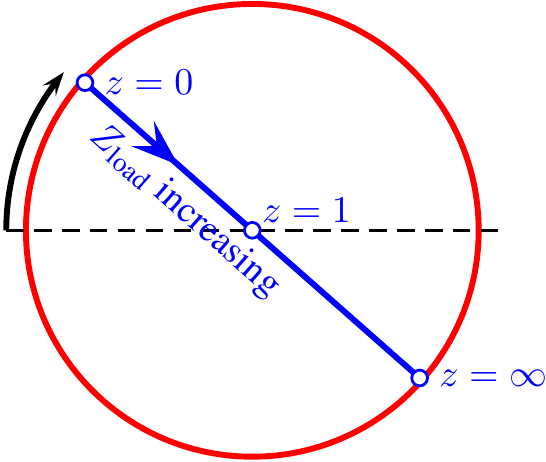}
\caption{Rotation of the  real axis, therefore the reference plane of the \textit{Smith} chart 
when adding a transmission-line}
\label{tlsimple}
\end{figure}

The general form of Eq.~\ref{eq:refl} returns the input reflection coefficient $\Gamma_{\text{in}}$ 
for a 2-port network terminated with $Z_{\text{load}}$, i.e.\ a reflection coefficient 
$\Gamma_{\text{out}}$ at the output port:
%
\begin{equation}
	\Gamma_{\text{in}} = {S}_{11} + \frac{{S}_{12} {S}_{21} \Gamma_{\text{load}}}{1 - {S}_{22} \Gamma_{\text{load}}}.
\label{eq:13}
\end{equation}
%
Lets evaluate some examples, defined by their S-matrix, which map their impedance to particular characteristic lines and circles on the \textit{Smith} chart. 
For this illustration, a very simplified \textit{Smith} chart, consisting just of the outermost circle (imaginary axis) and the real axis is used.
\subsubsubsection{Transmission-line of length $\lambda/16$}
\label{tl}
The S-matrix of a $\lambda/16$ transmission-line is
\begin{equation}
	\text{S} = \left[
		\begin{array}{cc}
			0 & \text{e}^{-\text{j}\frac{\pi}{8}} \\
			\text{e}^{-\text{j}\frac{\pi}{8}} & 0 \\
		\end{array}
\right]
\label{eq:14}
\end{equation}
has a input reflection coefficient of
\begin{equation}
	\Gamma_{\text{in}} = \Gamma_{\text{load}} \text{e}^{-\text{j}\frac{\pi}{4}}
\label{eq:15}
\end{equation}
This corresponds to a rotation of the real axis of the \textit{Smith chart} by an angle of 45$^{\circ}$ (Fig. \ref{tlsimple}) and hence a change of the reference plane of the chart (Fig. \ref{tlsimple}). Consider, for example, a transmission-line terminated by a short and hence $\Gamma_{\text{load}} = -1$.  The resulting reflection coefficient is then equal to $\Gamma_{\text{in}} = \text{e}^{-\text{j}\frac{\pi}{4}}$.
\begin{figure}[b]
\centering
\includegraphics[width=0.4\linewidth]{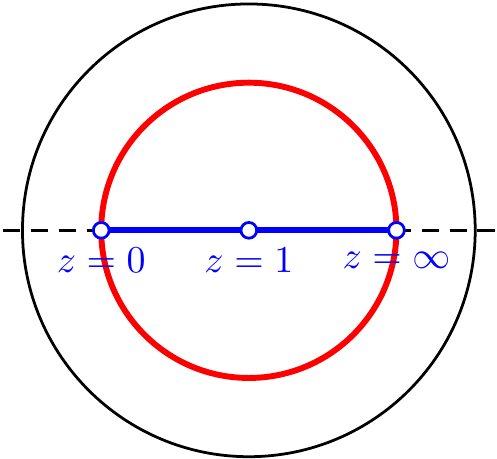}
\caption{Effect of an attenuator in the \textit{Smith} chart}
\label{att}
\end{figure}
\subsubsubsection{3~dB attenuator}
The S-matrix of a 3~dB attenuator is given by
\begin{equation}
	\text{S} = \left[
		\begin{array}{cc}
			0 & \frac{\sqrt{2}}{2} \\
			\frac{\sqrt{2}}{2} & 0 \\
		\end{array}
\right].
\label{eq:16}
\end{equation}
The resulting reflection coefficient is
\begin{equation}
	\Gamma_{\text{in}} = \frac{\Gamma_{\text{load}}}{2}
\label{eq:17}
\end{equation}
In the Smith chart, the connection of such an attenuator causes the outermost circle to shrink to a radius 
of 0.5, see Fig. \ref{att}%
\footnote{An attenuation of 3 dB corresponds to a reduction by a factor 2 in power.}.
\subsubsection{Resistive load}
\begin{figure}[t]
\centering
\includegraphics[width=0.4\linewidth]{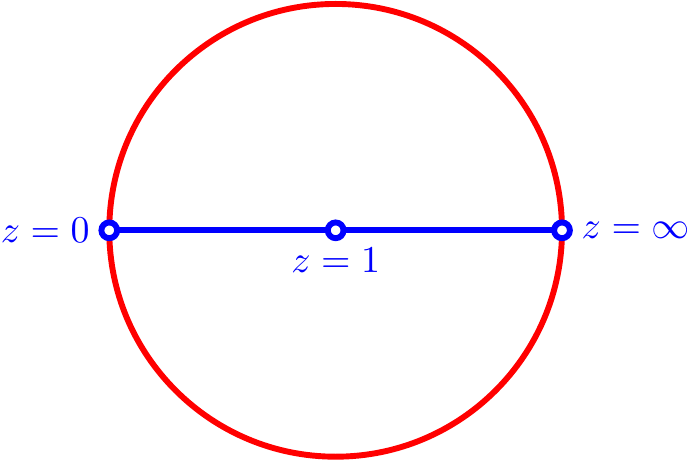}
\caption{A load resistor of variable value in the simplified \textit{Smith} chart. Since the impedance has a real part only, the trace remains on the real axis of the $\Gamma$ plane.}
\label{res}
\end{figure}

Fig. \ref{res} illustrates how the real axis is passed, if a resistive load changes its value 0 $< z < \infty$.

\subsection{Examples for applications of the Smith chart}
In this section two examples of typical RF problems demonstrate how the \textit{Smith} chart greatly facilitates their solutions.
\subsubsection{A step in the characteristic impedance}
\begin{figure}[b]
\centering
\includegraphics[width=0.4\linewidth]{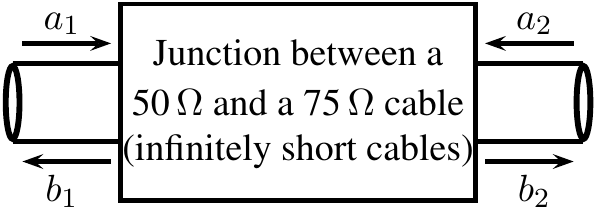}
\caption{Junction between two coaxial cables, one with with $Z_1=50\Omega$, the other with $Z_2=75\Omega$ characteristic impedance. Infinitely short cables are assumed -- 
only the junction is considered.}
\label{junct}
\end{figure}

Consider a junction between two infinitely short cables, an incoming with a characteristic impedance 
of $Z_1=50 \Omega$, the outgoing with $Z_2=75\Omega$ (Fig. \ref{junct}). 
Both ports are matched in their characteristic impedance.

The incident waves are denoted with $a_{i}$ ($i = 1,2$), the reflecting waves with $b_{i}$. 
The reflection coefficient at port 1 follows as
\begin{equation}
	\Gamma_{1} = \frac{Z_2 - Z_1}{Z_2 + Z_1} = \frac{75 - 50}{75 + 50} = +0.2.
\label{eq:18}
\end{equation}

\begin{figure}[t]
\centering
\includegraphics[width=0.4\linewidth]{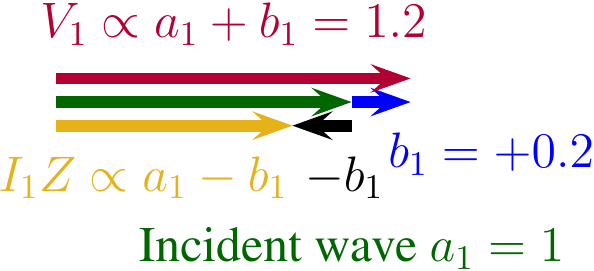}
\caption{Visualization of the two-port formed by the two cables of different characteristic impedances}
\label{tlsolution}
\end{figure}
Thus, the voltage of the reflected wave at port 1 is 20\% of the incident wave ($b_{1} = a_{1}$ $\cdot$ $0.2$), and the reflected power at port 1 is $\Gamma^2_1=0.04\equiv$ 4\%. 
From conservation of energy, the transmitted power has to be 96\%, i.e.\ $b_{2}^{2} = 1-\Gamma^2_1=0.96$.
%

The voltage transmission coefficient in this particular case computes $t = 1 + \Gamma$, and the output voltage of the transmitted wave at port 2 is \emph{higher} than the voltage of the incident wave at port 1:
$V_{\text{transmitted}} = V_{\text{incident}} + V_{\text{reflected}} =1+0.2=1.2$.
Also, note that this structure is not symmetric ($S_{11}=+0.2 \neq  S_{22}=-0.2$), but reciprocal 
($S_{21} = S_{12}=\sqrt{1-\Gamma^2_1}$).
As all impedances are real, the corresponding vectors show up in the \textit{Smith} chart on the real axis (Fig. \ref{tlsolution}).




\subsubsection{Quality ($Q$) factor of a cavity}
The second example shows the calculation of the quality factor of a cavity resonator with help of the \textit{Smith} chart.
%

\begin{figure}[b]
\centering
\includegraphics[width=0.7\linewidth]{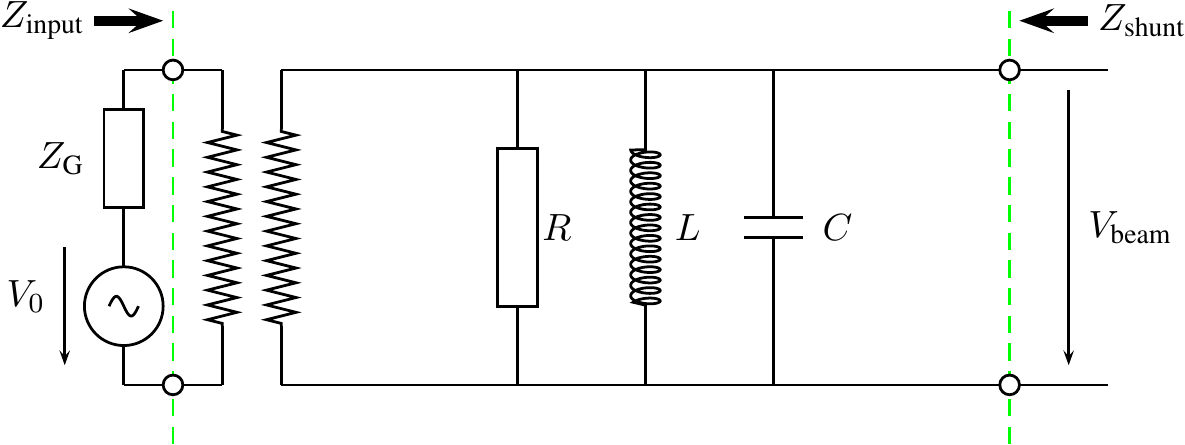}
\caption{Equivalent circuit of a cavity near resonance. The transformer describes the coupling of the cavity 
(typically $Z_{\text{shunt}} \approx 1$ M$\Omega$, as seen by the beam) to the generator (often $Z_G = 50~\Omega$).}
\label{rlc}
\end{figure}

A cavity at or near to one of its eigenmode resonances can be approximated by a parallel $RLC$ equivalent circuit (Fig. \ref{rlc}).
The resonance condition is given as
\begin{equation}
	\omega L = \frac{1}{\omega C}
\label{eq:19}
\end{equation}
from which the resonance frequency follows
\begin{equation}
	\omega_{\text{res}} = \frac{1}{\sqrt{LC}} \text{\hspace{1cm} or \hspace{1cm} }f_{\text{res}} = \frac{1}{2 \pi}\frac{1}{\sqrt{LC}}.
\label{eq:20}
\end{equation}

The impedance $Z$ of the cavity equivalent circuit is simply
\begin{equation}
	Z(\omega) = \frac{1}{\frac{1}{R} + \text{j}\omega C + \frac{1}{\text{j}\omega L}}.
\label{eq:24}
\end{equation}

\begin{figure}[t]
\centering
\includegraphics[width=0.7\linewidth]{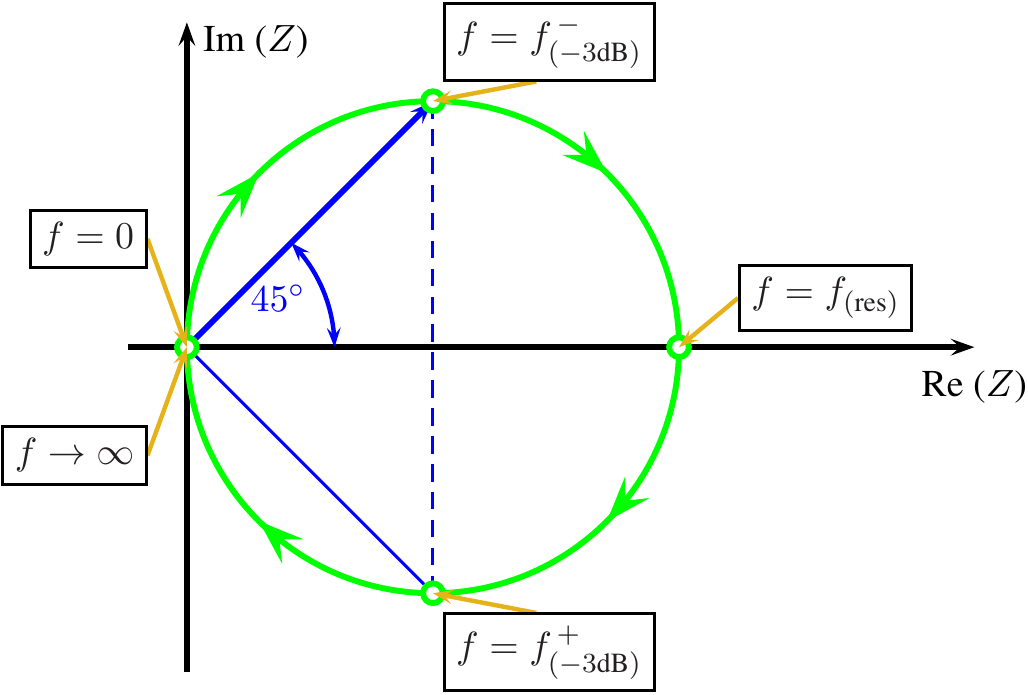}
\caption{Schematic drawing of the 3 dB bandwidth in the impedance plane}
\label{3db}
\end{figure}

The 3 dB bandwidth $\Delta f$ refers to the points where Re($Z$) = Im($Z$), which correspond to two vectors with an argument of 45$^{\circ}$ (Fig. \ref{3db}) and an impedance of $|Z_{(-3~\text{dB})}| = 0.707 R = R/\sqrt{2}$.

In general, the quality factor $Q$ of a resonant circuit is defined as the ratio of the stored energy $W$ over the energy dissipated $P$ in one oscillation cycle:
\begin{equation}
	Q = \frac{\omega W}{P}.
\label{eq:21}
\end{equation}
However, the $Q$ factor for a resonance can also be calculated using the 3~dB bandwidth and the resonance frequency:
\begin{equation}
	Q = \frac{f_{\text{res}}}{\Delta f}.
\label{eq:22}
\end{equation}
For a cavity, three different quality factors are defined:
\begin{itemize}
	\item $Q_{0}$ (unloaded $Q$): $Q$ factor of the unperturbed system, i.e. the stand-alone cavity;
	\item $Q_{\text{L}}$ (loaded $Q$): $Q$ factor of the cavity when connected to a generator and/or measurement circuits;
	\item $Q_{\text{ext}}$ (external $Q$): $Q$ factor that describes the degeneration of $Q_{0}$ due to the generator and/or diagnostic impedances.
\end{itemize}
All these $Q$ factors are linked via a simple relation:
\begin{equation}
	\frac{1}{Q_{\text{L}}} = \frac{1}{Q_{0}} + \frac{1}{Q_{\text{ext}}}.
\label{eq:23}
\end{equation}
The coupling coefficient $\beta$ is then defined as
\begin{equation}
	\beta = \frac{Q_{0}}{Q_{\text{ext}}}.
\label{eq:25}
\end{equation}
\emph{This coupling coefficient has not to be confused with the propagation coefficient of transmission lines, which is also denoted as $\beta$.}

In the \textit{Smith} chart, a resonant circuit shows up as a circle (Fig. \ref{qfactor}, dashed, red circle shown in the ``detuned short'' position). The larger the circle, the stronger is the coupling. Three types of coupling are distinguished, depending on the range of $beta$ (= size of the circle, assuming the circle is in the ``detuned short'' position):
\begin{figure}[t]
\centering
\includegraphics[width=0.5\linewidth]{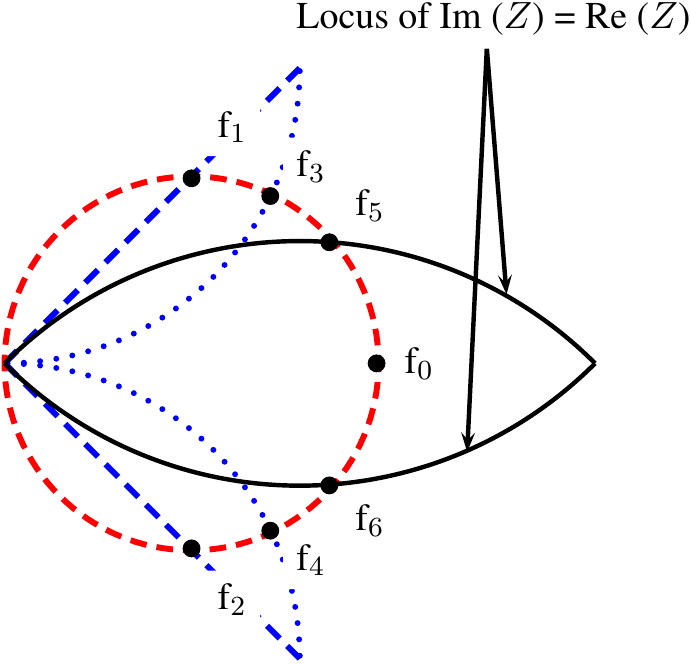}
\caption{Evaluation of the different $Q$ factors of a resonant cavity with help of the \textit{Smith} chart}
\label{qfactor}
\end{figure}
\begin{itemize}
	\item Undercritical coupling ($0 < \beta < 1$): the radius of the resonance circle is smaller than 0.25. Hence, the center of the chart ($\Gamma=0$) lies outside the circle.
	\item Critical coupling ($\beta = 1$): the radius of the resonance circle is exactly 0.25. Hence, the circle crosses $\Gamma=0$ at the resonance frequency $f_{\text{res}}$.
	\item Overcritical coupling ($1 < \beta < \infty$): the radius of the resonance circle is larger than 0.25. Hence, the center of the chart lies inside the circle.
\end{itemize}
In practice, the circle may be rotated around the origin due to the transmission lines between the resonant circuit and the measurement device.

From the different marked frequency points in Fig. \ref{qfactor} the 3~dB bandwidth, and thus the quality factors $Q_{0}$, $Q_{\text{L}}$ and $Q_{\text{ext}}$ are determined as follows:
\begin{itemize}
\item The unloaded $Q$ is determined from $f_{5}$ and $f_{6}$. The condition for these points is Re($Z$) = Im($Z$), with the resonance circle in the ``detuned short'' position.
\item The loaded $Q$ is determined from $f_{1}$ and $f_{2}$. The condition to find these points is $\left|\text{Im}(S_{11})\right| \rightarrow$ max.\ in ``detuned short'' position.
\item The external $Q$ is calculated from $f_{3}$ and $f_{4}$. The condition to determine these points is $Z$ = $\pm \text{j}$ in ``open short'' position, which is equivalent to $Y$ = $\pm \text{j}$ in ``detuned short'' position
\end{itemize}

To determine the points $f_{1}$ to $f_{6}$ with a network analyzer, the following steps are applicable:
\begin{itemize}
	\item $f_{1}$ and $f_{2}$: set the marker format to Re($S_{11}$) + j Im($S_{11}$) and determine the two points where Im($S_{11}$) = max.
	\item $f_{3}$ and $f_{4}$: set the marker format to $Z$ and find the two points where $Z = \pm$ j.
	\item $f_{5}$ and $f_{6}$: set the marker format to $Z$ and locate the two points where Re($Z$) = Im($Z$).
\end{itemize}

\section{Summary}

Some fundamental concepts on RF devices, instruments, and signal processing techniques
have been presented in this introduction to RF measurement techniques. 
Advantages of various measurement methods using spectrum and network analyzers 
have been emphasized. 
In the last section the definition of the \textit{Smith} chart, 
and its usage were illustrated with several examples. 
This article supports the practical part of the CAS RF course and CAS 2018 special topic on beam instrumentation, 
and serves as background information. 

\section{Acknowledgments}
Greatest respect and many thanks go to \emph{Fritz Caspers}, who started this CAS RF training initiative! 
Many contributions, ideas and concepts in this text, and in the lecture root to him, 
and to the numerous contributions and support of his former Ph.D.\ students!

\section{Appendix: Meaning of the rulers below the Smith chart}

How to use the rulers that are often plotted below the Smith chart? 

\begin{figure}[h]
\centerline{
\scalebox{1.0}{
\includegraphics*[width=1.0\textwidth]{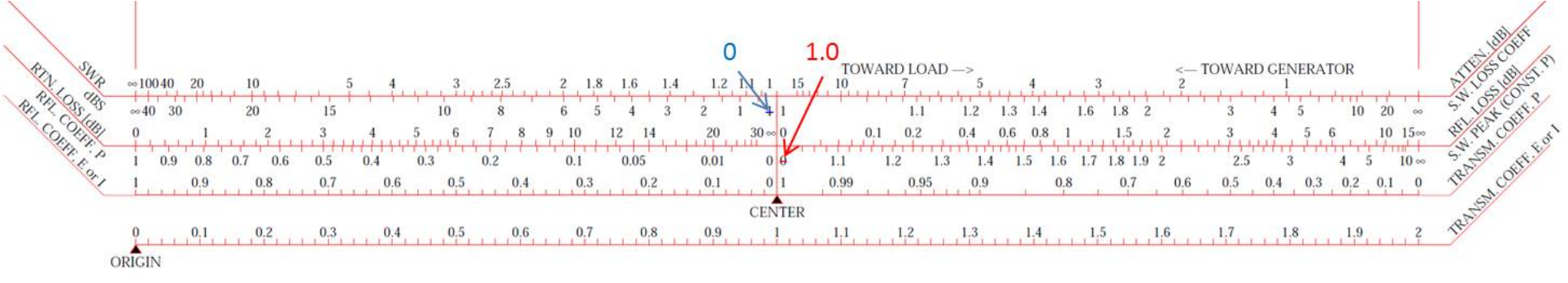}
} %
}
\caption{Example for a set of rulers that can be found underneath the Smith chart (please note corrections in respect to the RF-course printouts)} %
\label{ruler0} %
\end{figure}

A commonly used set of rulers is usually found below the Smith chart, see Fig.~\ref{ruler0}. 
There are four rulers, some with an upper and lower part, to quickly estimate and compare some important
properties in terms of modulus values.
For the following discussion lets split the upper three rulers at the line marked \emph{CENTER} to a left and right part,
each to be discussed separately. These rulers start at the \emph{CENTER}, referring to the center of the Smith chart, and end at the left or right
boundary, referring to the circular boundary of the Smith chart.
The 4th ruler at the bottom is different, it starts at the left boundary \emph{ORIGIN} and ends at the right boundary.

\begin{figure}[h]
\centerline{
\scalebox{1.0}{
\includegraphics*[width=1.0\textwidth]{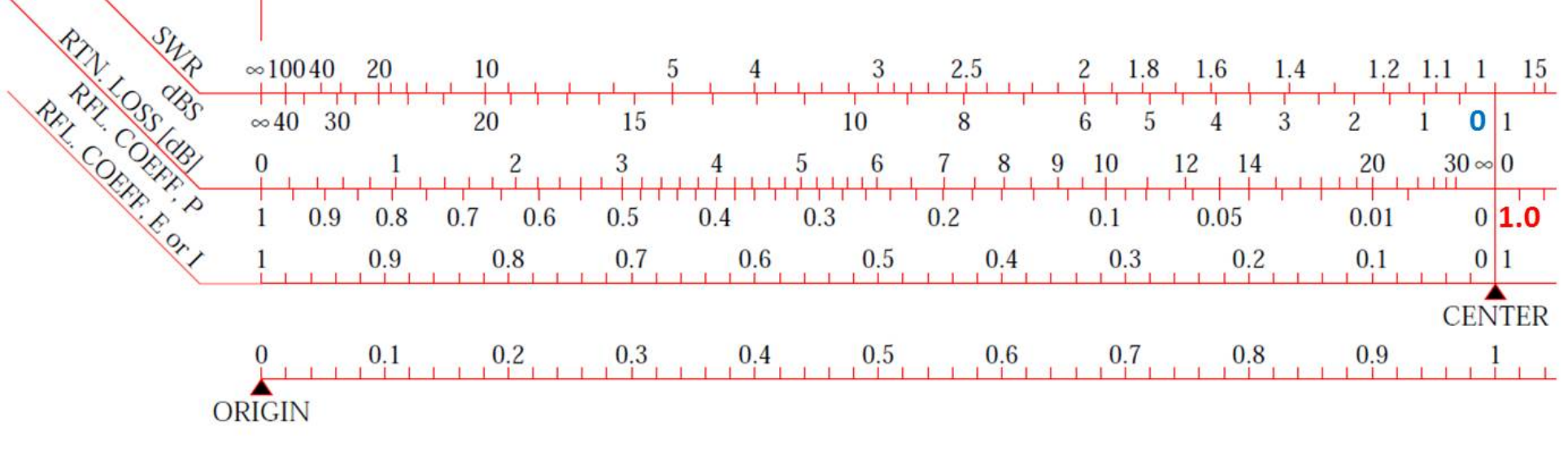}
} %
}
\caption{Left part of the rulers usually plotted underneath the Smith Chart} %
\label{ruler1} %
\end{figure} %

First ruler, left/upper part in Fig.~\ref{ruler1} is marked as \emph{SWR} which mean actually VSWR,\ i.e. voltage standing wave ratio. It ranges between one -- for the matched case (center of the Smith chart) and infinity -- for total reflection (boundary of the Smith chart), respectively. The upper part is in linear scale, the lower part of this ruler is in dB, noted as dBS (dB referred to Standing Wave Ratio). Example: SWR = 10 corresponds to 20 dBS, SWR = 100 corresponds to 40 dBS [voltage ratios, not power ratios].

Second ruler, left/upper part, marked as \emph{RTN.LOSS} i.e. return loss in dB. This indicates the amount of reflected wave expressed in dB. Thus, in the center of SC nothing is reflected and the return loss is infinite. At the boundary we have full reflection, thus return loss is 0 dB. The lower part of the scale denoted as \emph{RFL.COEFF.P} is a reflection coefficient in terms of POWER (proportional $| \Gamma | ^2$).
If there is no reflected power for the matched case locus is in the center of the Smith chart (SC). On the contrary, if normalized reflected power is equal to 1 locus is at the boundary. 

Third ruler, left, marked as \emph{RFL.COEFF,E or I} gives us the absolute value of the reflection coefficient in linear scale. Note that since we have the modulus we can refer it both to voltage or current as we have omitted the sign. Obviously in the center the reflection coefficient is zero, at the boundary it is one. 

The fourth is a Voltage transmission coefficient. Note that the modulus of the voltage (and current) transmission coefficient has a range from zero, i.e. short circuit, to +2 (open = 1+$\Gamma$ with $\Gamma$ = 1). This ruler is only valid for $Z_{load}$ = real, i.e. the case of a step in characteristic impedance of the coaxial line.

\begin{figure}[h]
\centerline{
\scalebox{1.0}{
\includegraphics*[width=1.0\textwidth]{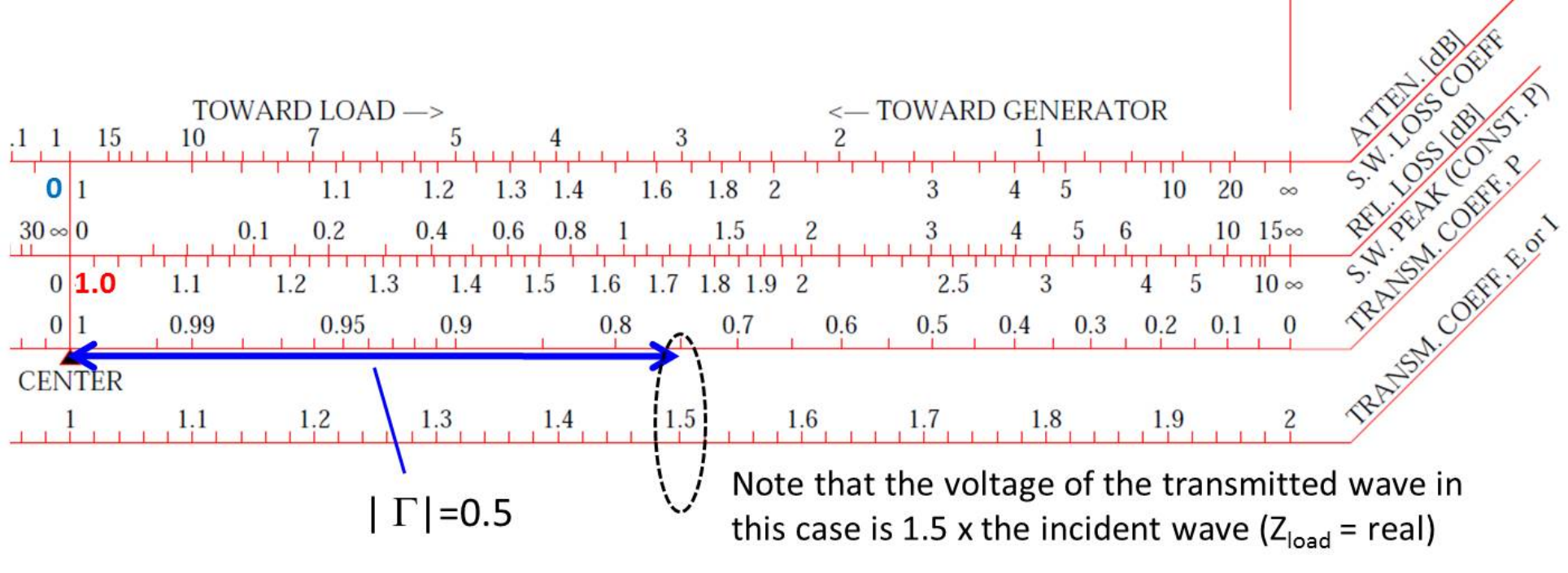}
} %
}
\caption{Right part of the rulers usually plotted underneath the Smith Chart} %
\label{ruler2} %
\end{figure}

Third ruler, right (see Fig.~\ref{ruler2}) marked as \emph{TRANSM.COEFF.P} refers to the transmitted power as a function of mismatch and displays essentially the relation $P_{t}=1-|\Gamma|^2$. Thus, in the center of the SC full match, all the power is transmitted. At the boundary we have total reflection and e.g. for a $\Gamma$ value of 0.5 we see that 75 \% of the incident power is transmitted.

Second ruler, right/upper part, denoted as \emph{RFL.LOSS in dB} denotes reflection loss. This ruler refers to the loss in the transmitted wave, and should not be confounded with the return loss referring to the reflected wave. It displays the relation  $P_{t}=1-|\Gamma|^2$ in dB.
This ruler is nowadays rather not more in use.

\begin{figure}[h]
\centerline{
\scalebox{1.0}{
\includegraphics*[width=1.0\textwidth]{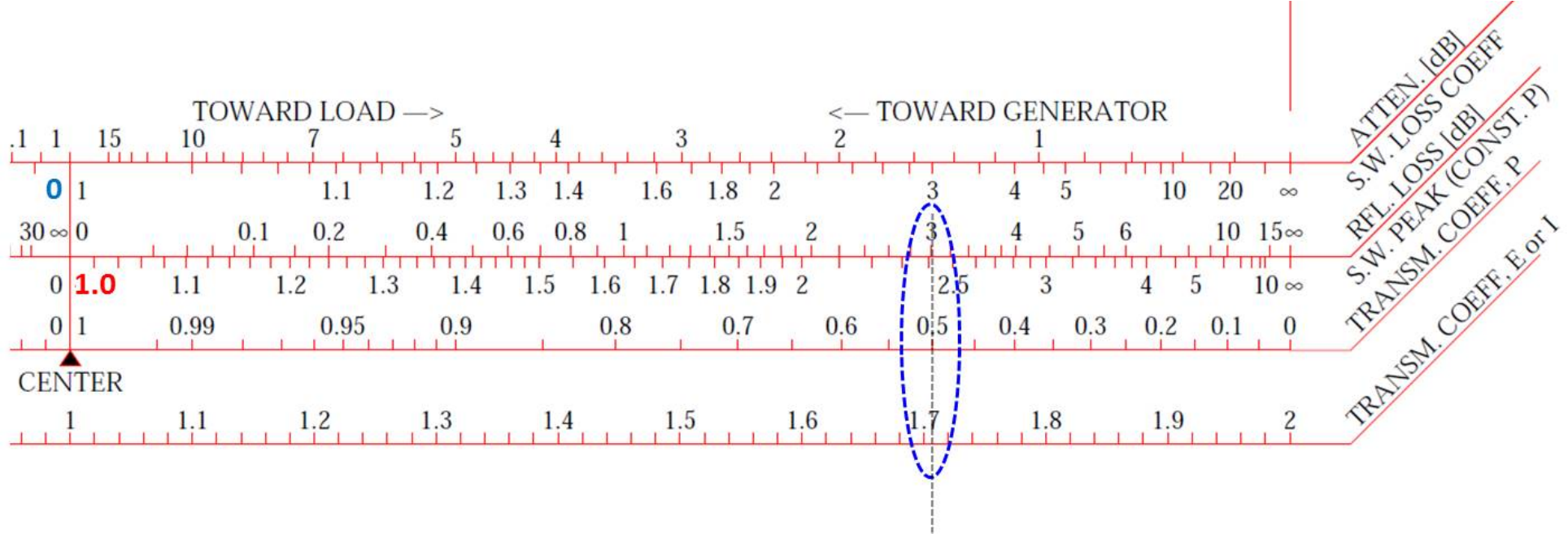}
} %
}
\caption{Example for $|\Gamma| = 1/\sqrt{2}=0.707$ and 50 \% of transmitted power (i.e. 3 dB loss), see description in text} %
\label{ruler3} %
\end{figure}

Let us analyse an example from Fig.~\ref{ruler3}: $|\Gamma| = 1/\sqrt{2}=0.707$  , transmitted power = 50 \% thus loss = 50 \% = 3 dB. 
Note that in the lowest ruler the voltage of the transmitted wave ($Z_{load}$ = real) would be $V_{t} = 1.707 = 1+1/\sqrt{2}$ if referring to the voltage.

Finally, the First ruler, right/upper part, denoted as \emph{ATTEN. in dB} assumes that one is measuring an attenuator or a lossy line which itself is terminated by an open or short circuit (full reflection). Thus the wave is traveling twice through the attenuator (forward and backward). The value of this attenuator can be between zero and some very high number corresponding to the matched case. 
The lower scale of first ruler displays the same situation just in terms of VSWR.

\begin{figure}[h]
\centerline{
\scalebox{1.0}{
\includegraphics*[width=1.0\textwidth]{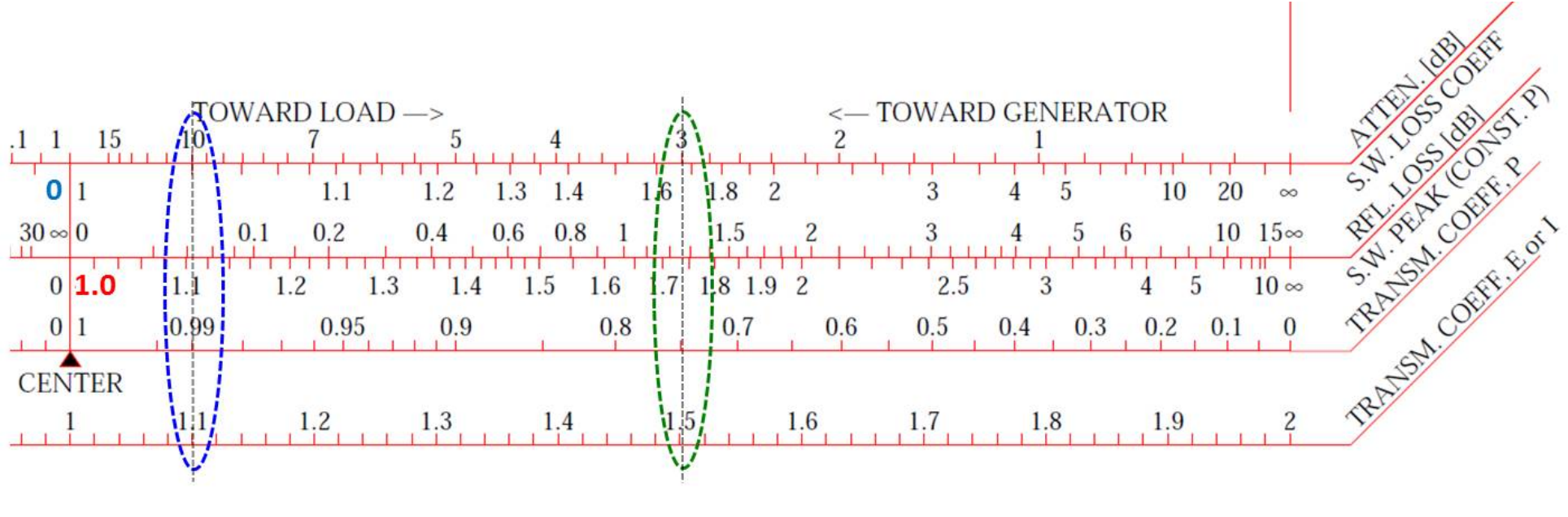}
} %
}
\caption{Example for 10 dB and 3 dB attenuator, see description in text} %
\label{ruler4} %
\end{figure}

For the next example see Fig.~\ref{ruler4}: an 10 dB attenuator attenuates the reflected wave by 20 dB going forth and back and we get a reflection coefficient of $\Gamma$ = 0.1. This correspond to the reflection of 10 \% in voltage.
Another example is 3 dB attenuator: for the forth and back transmission it gives 6 dB which correspond to half of the voltage. Table~\ref{tab:formulas} is reprinted from an original paper of Phillip~H.~Smith~\cite{smith2000} and summarizes reflection formulas discussed above.

\begin{table}[h]
\caption{Reflection formulas}
\label{tab:formulas}
\centering
\begin{tabular}{|l|c|c|c|}
\hline
\bfseries function & \bfseries	traveling waves & \bfseries reflection coefficient & \bfseries standing waves	\\
\hline
& & & \\
VOLTAGE REFL.  COEFF.	&	$r \over i$	&	$\Gamma$															&	${S-1 \over S+1}$ \\
& & & \\
\hline
& & & \\
POWER   REFL.  COEF.	&	$({r \over i})^2$	& 	$\Gamma^2$ 													&	$({S-1 \over S+1})^2$ \\
& & & \\
\hline
& & & \\
RETURN  LOSS  [dB] 		&	$10 \cdot \log({i \over r})^2$	& 		$-10 \cdot \log(\Gamma^2)$					& 	$-10 \cdot \log({S-1 \over S+1})^2$ \\
& & & \\
\hline
& & & \\
REFLECTION  LOSS [dB]	&	$10 \cdot \log({i^2 \over i^2-r^2})$		& 	$-10 \cdot \log(1-\Gamma^2)$		&	$-10 \cdot \log[1-({S-1 \over S+1})^2]$ \\
& & & \\
\hline
& & & \\
STDG. WAVE LOSS COEF.	&	$1- {[(i+r)/(i-r)]^2 \over 2[(i+r)/(i-r)]}$	& 	${1-\Gamma+\Gamma^2-\Gamma^3}\over {1-\Gamma-\Gamma^2+\Gamma^3}$ 	& 	${1+S^2 \over 2S}$ \\
& & & \\
\hline
& & & \\
STDG. WAVE RATIO [dB]	&	$20 \cdot \log({i+r \over i-r})$	& 	$20 \cdot \log({1+\Gamma \over 1-\Gamma})$ 							&	$20 \cdot \log(S)$ \\
& & & \\
\hline
& & & \\
MAX. OF STDG. WAVE 		&	$({i+r \over i-r})^{1/2}$	& 	$({1+\Gamma \over 1-\Gamma})^{1/2}$ 								& 	$\sqrt S$ \\
& & & \\
\hline
& & & \\
MIN. OF STDG. WAVE  	&	$({i-r \over i+r})^{1/2}$	& 	$({1-\Gamma \over 1+\Gamma})^{1/2}$									& 	$1 \over \sqrt S$ \\
& & & \\
\hline
& & & \\
STANDING WAVE RATIO		&	${i+r \over i-r}$	& 	${1+\Gamma \over 1-\Gamma}$ 										& 	$S$ \\
& & & \\
\hline
& & & \\
ATTENUATION [dB]		&	$-10 \cdot \log({r \over i})$	& 	$-10 \cdot \log(\Gamma)$ 											& 	$-10 \cdot \log({S-1 \over S+1})$ \\
& & & \\
\hline
\end{tabular}

\vspace{0,2cm}
whereas: $i$ = incident wave amplitude, 
$r$ = reflected wave amplitude, 
$\Gamma$ = reflection coefficient,
$S \equiv  \mbox{SWR}$ = voltage standing wave ratio.

\end{table}


\begin{thebibliography}{99}
\bibitem{vendelin}G.D. Vendelin, A.M. Pavio and U.L. Rohde, \textit{Microwave Circuit Design Using Linear and Nonlinear Techniques}, second ed. (Wiley-Interscience, New Jersey, 2005), ISBN-10 0-471-41479-4.
\bibitem{src:oxford} F. Caspers, Proc. CERN Accelerator School, RF Engineering for Particle Accelerators, Oxford, UK, 1991, p.181.
\bibitem{thumm}M. Thumm, W. Wiesbeck and S. Kern, \textit{Hochfrequenzmesstechnik} (Teubner, Stuttgart/Leipzig, 1998), ISBN 3-519-16360-8.
\bibitem{witte}R.A. Witte, \textit{Spectrum and Network Measurements} (Prentice-Hall, New Jersey , 1991), ISBN 0-13-826959-9.
\bibitem{Schleifer}W.O. Schleifer, \textit{Hochfrequenz und Mikrowellenmesstechnik in der Praxis} (H\"{u}thig, Heidelberg,  1981), ISBN 3-7785-0675-7.
\bibitem{Schiek}B. Schiek and H.J. Sieveris, \textit{Rauschen im Hochfrequenzschaltungen} (H\"{u}thig, Heidelberg, 1984), ISBN 3-7785-2007-5.
\bibitem{Yip}P.C.L. Yip, \textit{High Frequency Circuit Design and Measurement} (Chapman and Hall, London, 1990), ISBN 0-412-34160-3.
\bibitem{Evans}G. Evans and C.W. McLeisch, \textit{RF-Radiometer Handbook} (Artech, Dedham,  1977), ISBN 0-89006-055-X.
\bibitem{Connor}F.R. Connor, \textit{Noise} (Edward Arnold, London, 1973), ISBN 0-7131-3306-6.
\bibitem{Landstorfer}F. Landstorfer and H. Graf, \textit{Rauschprobleme der Nachrichtentechnik} (Oldenbourg,  M\"unchen, 1981), ISBN 3-486-24681-X.
\bibitem{Zinke}O. Zinke and H. Brunswig, \textit{Lehrbuch der Hochfrequenztechnik, Zweiter Band} (Springer, Berlin, 1974), ISBN 3-540-06245-9.
\bibitem{HP}Agilent Technologies, Inc., \textit{Fundamentals of RF and microwave noise figure measurements}, Agilent Application Note 57-1, 2010.
\bibitem{Schiek2}B. Schiek, \textit{Messysteme der Hochfrequenztechnik} (H\"{u}thig, Heidelberg, 1984), ISBN 3-7785-1045-2.
\bibitem{FritzSpara}F. Caspers, RF engineering basic concepts: S-parameters, CAS Proc., 2010, CERN Yellow Report CERN-2011-007, pp. 67-93.
\bibitem{src:xPar} J. Verspecht and D. Root, \textit{Polyharmonic Distortion Modeling}, IEEE Microwave Magazine, Vol. 7, Issue 3, June 2006, pp. 44-57.
\bibitem{src:thumm} M. Thumm, W. Wiesbeck and S. Kern, \textit{Hochfrequenzmesstechnik, Verfahren und Messsysteme} (Teubner, Stuttgart, 1998),  ISBN 978-3519163602.
\bibitem{src:fundVNA} M. Hiebel, \textit{Fundamentals of Vector Network Analysis} (Rohde \& Schwarz, M\"unchen, 2007), ISBN 3939837067.
\bibitem{src:fundVNA2} Agilent Technologies, Inc., \textit{Understanding the fundamental principles of vector network analysis}, Agilent Application Note AN 1287-1, 2000.
\bibitem{src:kruschdPaper} Anritsu Company,  \textit{Time domain measurements using vector network analyzers}, Anritsu Application Note No. 11410-00206, R , 2009.
\bibitem{src:TDA} Agilent Technologies, Inc., \textit{Time domain analysis using a network analyzer}, Agilent Application Note 1287-12, 2012.
\bibitem{meinkegundlach} H. Meinke and F.-W. Gundlach, \textit{Taschenbuch der Hochfrequenztechnik} (Springer, Berlin, 1992).
\bibitem{smith2000} P. Smith, \textit{Electronic Applications of the Smith Chart} (Noble Publishing, Atlanta, 2000), ISBN 1-884932-39-8.
%
%
%
%
\end{thebibliography}
\end{document}